\documentclass[preprint,5p,twocolumn,authoryear]{elsarticle}

\usepackage{amsmath,amssymb,amsfonts,dsfont,amsthm}
\usepackage{cmap}
\usepackage[noTeX]{mmap}   
\usepackage[utf8]{inputenc} 
\usepackage{resizegather}
\usepackage{subcaption}
\usepackage{graphbox}
\usepackage{float}
\usepackage{array}
\usepackage{color}
\usepackage{todonotes}
\usepackage{hyperxmp}
\usepackage[pdfdisplaydoctitle=true,
      colorlinks=true,
      urlcolor=blue,
      citecolor=blue,
      linkcolor=blue,
      pdfstartview=FitH,
      pdfpagemode=None,
      bookmarksnumbered=true]{hyperref} 

\newtheorem{thm}{Theorem}[section]
\newtheorem{cor}[thm]{Corollary}

\theoremstyle{definition}

\theoremstyle{remark}
\newtheorem{rem}[thm]{Remark}
\newtheorem{ex}[thm]{Example}
\numberwithin{equation}{section}
\newcommand{\CC}{\mathbb{C}}                
\newcommand{\RR}{\mathbb{R}}                

\newcommand{\Ela}{\mathbb{E}\mathrm{la}}    
\newcommand{\HH}{\mathbb{H}}                

\newcommand{\OO}{\mathrm{O}}                
\newcommand{\SO}{\mathrm{SO}}               
\newcommand{\DD}{\mathbb{D}}                

\newcommand{\eee}{\pmb{e}}                   
\newcommand{\nn}{\pmb{n}}                   
\newcommand{\ww}{\pmb{w}}                   
\newcommand{\xx}{\pmb{x}}                   

\newcommand{\Idd}{\mathbf{1}}               
\newcommand{\rh}{\mathrm{h}}                

\newcommand{\ba}{\mathbf{a}}
\newcommand{\bb}{\mathbf{b}}

\newcommand{\bd}{\mathbf{d}}
\newcommand{\be}{\mathbf{e}}
\newcommand{\bh}{\mathbf{h}}
\newcommand{\bk}{\mathbf{k}}
\newcommand{\bn}{\mathbf{n}}

\newcommand{\bt}{\mathbf{t}}
\newcommand{\bu}{\mathbf{u}}
\newcommand{\bv}{\mathbf{v}}

\newcommand{\bepsilon}{\pmb \epsilon}
\newcommand{\bsigma}{\pmb \sigma}
\newcommand{\bomega}{\pmb \omega}

\newcommand{\bC}{\mathbf{C}}                
\newcommand{\bA}{\mathbf{A}}                
\newcommand{\bH}{\mathbf{H}}                
\newcommand{\bI}{\mathbf{I}}                
\newcommand{\bJ}{\mathbf{J}}                
\newcommand{\bS}{\mathbf{S}}                
\newcommand{\bT}{\mathbf{T}}                
\newcommand{\bD}{\mathbf{D}}                
\newcommand{\bOmega}{\pmb{\Omega}}          

\DeclareMathOperator{\tr}{tr}
\renewcommand\Re{\operatorname{\mathfrak{Re}}}

\newcommand{\norm}[1]{\left\Vert#1\right\Vert}  

\renewcommand{\vec}{\pmb}

\newcommand{\otimesbar}{\; \underline{\overline{\otimes}} \;}

\newcommand{\beq}{\begin{equation}}
\newcommand{\eeq}{\end{equation}}

\journal{}

\hypersetup{
  pdfauthor={C. Oliver-Leblond, R. Desmorat and B. Kolev}, 
  pdftitle={Anisotropic damage modelling as a numerical twin of bi-dimensional discrete elements}, 
  pdfsubject={}, 
  pdfkeywords={crack density \sep damage tensor \sep harmonic decomposition}, 
  pdflang=en, 
  }
\begin{document}

\begin{frontmatter}



  \title{Continuous anisotropic damage as a twin modelling \protect\\ of discrete bi-dimensional fracture}


  \author{C. Oliver-Leblond\corref{cor}}
  \ead{cecile.oliver@ens-paris-saclay.fr}
  \author{R. Desmorat}
  \ead{rodrigue.desmorat@ens-paris-saclay.fr}
  \author{B. Kolev}
  \ead{boris.kolev@math.cnrs.fr}

  \address{Université Paris-Saclay, ENS Paris-Saclay, CNRS,  LMT - Laboratoire de Mécanique et Technologie, 91190, Gif-sur-Yvette, France}

  \begin{abstract}

    In this contribution, the use of discrete simulations to formulate an anisotropic damage model is investigated. It is proposed to use a beam-particle model to perform numerical characterization tests. Indeed, this discrete model explicitly describes cracking by allowing displacement discontinuities and thus capture crack induced anisotropy.

    Through 2D discrete simulations, the evolution of the effective elasticity tensor for various loading tests, up to failure, is obtained. The analysis of these tensors through bi-dimensional harmonic decomposition is then performed to estimate the tensorial damage evolution. As a by-product of present work we obtain an upper bound of the distance to  the orthotropic symmetry class of bi-dimensional elasticity.

  \end{abstract}


\end{frontmatter}

\section*{Introduction}
\label{sec:introduction}

For most quasi-brittle materials, the initial mechanical behaviour of the uncracked material can be considered as isotropic. On the other hand, degradation induced by mechanical loading generally leads to anisotropy. Indeed, cracks are naturally oriented and therefore their appearance will not affect the material properties in an isotropic manner. We note that the orientation of these cracks for an initially isotropic material can be determined by the direction and sign of the loading \citep{mazars1990unilateral,ramtani1992orthotropic}.

To obtain an explicit representation of cracking -- and thus of its impact on material properties -- it is possible to use discrete models of the lattice or particle type. The first lattice models were historically introduced by \cite{poisson1828memoire} and more recently by \cite{hrennikoff1941solution} to solve classical elasticity problems. The elastic material is discretized using 1D elements -- springs or beams -- which allow for the transfer of forces between the nodes of the lattice. The development of numerical simulations has allowed for its extension to the study of fracture behaviour by considering a brittle behaviour for the elements forming this lattice \citep{herrmann1989fracture}. The approach was then applied to the quasi-brittle fracture of concrete subjected to tension \citep{schlangen1992simple}. However, these models do not allow for the representation of compression cracking nor of cyclic loading. Particulate models were proposed in 1979 to study the behaviour of granular \citep{cundall1979discrete} joints. For these applications, contact forces alone were sufficient to correctly describe the behaviour. A cohesive version has been proposed latter \citep{meguro1989fracture} but does not offer the simplicity and rapidity of the lattice models. In the present work, we propose to use a beam-particle model combining the lattice approach and the particulate approach \citep{daddetta2002application,delaplace2008modelisation,vassaux2016beam}. Concrete is represented via an assembly of polygonal particles bounded together by brittle beams. After the beams break, frictional contact forces are introduced between the particles. This model allows for a fine and explicit description of the cracking and the associated mechanisms (initiation, propagation, closure with stiffness recovery, friction). Its application to structural calculations is not yet common but can be envisaged with the implementation of high-performance calculation techniques. These discrete methods can be used as numerical experimentation tools at the scale of the Representative Elemental Volume or at the scale of a laboratory specimen to establish and identify the constituent equations of a continuous model \citep{vassaux2015regularised} or simply a part of these equations \citep{delaplace2007discrete}.

In order to obtain a representative, robust and efficient numerical model, it is common to use macroscopic laws that account for the formation, propagation and coalescence of micro-cracks by introducing an internal damage variable. One of the first damage models for quasi-brittle materials \citep{mazars1984application} made it possible to reproduce the degradation of the concrete material with loading, via a scalar damage variable, by considering concrete as a homogeneous material at the scale of the volume element of continuum mechanics. This isotropic damage model is commonly used to study the behaviour of concrete structures but its isotropic nature does not allow for it to take into account complex multi-axial loading (as observed by Mazars and his collaborators, \cite{ramtani1992orthotropic}, see also the work of \cite{HD1998}). To tackle this issue, damage models with induced anisotropy -- such as the one proposed by \cite{desmorat2007nonlocal} and made more robust later \citep{desmorat2016anisotropic} -- can be used. In this model, developed within the framework of the thermodynamics of irreversible processes, the damage is represented by a tensor variable of order 2. It should be noted that these macroscopic models are globally phenomenological because their formulation and identification are based on experimental observations of the behaviour of quasi-brittle materials. It is the same for their laws of evolution \citep{ramtani1992orthotropic,lemaitre2005engineering} or their non-local character. Their representativeness is thus limited by the capacity to carry out experiments on the degraded material.

Discrete cracking models and continuous damage models both have advantages and limitations, and it makes sense to combine these advantages by combining the two modelling methods. In this paper, we investigate the relevance to use the beam-particle model to perform virtual tests in order to formulate an anisotropic damage model. Let us recall that in \citep{delaplace2007discrete} the methodology was limited to the identification of a single parameter of the continuous model (certainly delicate to measure on real tests because it was the one governing the so-called "shear-bulk" coupling of a more important effect of the damage on compressibility than on shear). More recent studies have proposed the formulation of continuous damage models from a discrete analysis: by introducing scalar damage variables calculated from the macroscopic loss of stiffness in a lattice simulation \citep{rinaldi2007statistical,rinaldi2013bottom,jivkov2014structure}, or by continualization of discrete equations \citep{challamel2015discrete}.

In this paper, the goal is to derive a tensorial damage variable from discrete simulations. The tools for the intrinsic analysis of tensors, introduced by \cite{backus1970geometrical} in elasticity and by \cite{leckie1981tensorial} in damage mechanics, mixing harmonic analysis and the notion of covariants (generalizing that of invariants, \cite{kolev2018characterization}), are used here in order to analyze the effective elasticity tensors obtained through discrete simulations, without reference to a particular basis, and to achieve a general tensorial representation of damage. Discrete simulations will be used to obtain the evolution of the effective elasticity tensor during the rupture of a numerical specimen under different mechanical loads. The analysis of the tensors via the harmonic decomposition and their covariant reconstruction \citep{olive2018harmonic} will then allow us to estimate the evolution of the tensor damage.

In section~\ref{sec:def}, definitions relating to the properties of fully symmetrical tensors are recalled.  The harmonic decomposition of a bi-dimensional elasticity tensor is then performed in section~\ref{sec:harmonic} and it is reminded that the fourth order harmonic part of this elasticity tensor is always an \emph{harmonic square}. In section~\ref{sec:covariant}, a covariant reconstruction of 2D orthotropic elasticity tensors is used to derive an upper bound of the distance to the orthotropic symmetry class in bi-dimensional elasticity. The beam-particle model, used in this contribution as a numerical testing tool, is presented in section~\ref{sec:discrete} as well as the methodology to extract the effective elasticity tensor from the discrete simulations. A first analysis is realised in section~\ref{sec:isotropy} to check whether the initial elasticity tensor, of the uncracked medium, can be considered as isotropic. In section~\ref{sec:telas_analysis}, the analysis of the effective elasticity tensors -- and more precisely of their harmonic part -- of cracked media for various loading test up to failure is achieved.  Finally, the analysis of those tensors through harmonic decomposition/covariant reconstruction is then performed in section~\ref{sec:damage} to estimate the tensorial damage evolution.

\section{Definitions}
\label{sec:def}

We next make use of the Euclidean structure of $\RR^{d}$, $d=2,3$, and do not make difference between covariant, contravariant or mixed tensors.

\subsection{Symmetric tensor product}

We denote by $\bT^\mathrm{s}$ the totally symmetric part of a possibly non symmetric tensor $\bT$. More precisely, if $\bT\in\otimes^n \RR^d$ is of order $n$,
\begin{equation*}
  \bT^\mathrm{s}(\xx_{1}, \dotsc , \xx_{n}) := \frac{1}{n!} \sum_{\sigma \in \mathfrak{S}_{n}} \bT(\xx_{\sigma(1)}, \dotsc , \xx_{\sigma(n)}),
\end{equation*}
where $\mathfrak{S}_{n}$ is the permutation group of $n$ elements.

The \emph{symmetric tensor product} of two tensors $\bT_{1}$ and $\bT_{2}$, of respective orders $n_1$ and $n_2$, is the symmetrization of $\bT_{1} \otimes \bT_{2}$, defining a totally symmetric tensor of order $n=n_1+n_2$:
\begin{equation*}
  \bT_{1} \odot \bT_{2} := ( \bT_{1} \otimes \bT_{2} )^\mathrm{s}.
\end{equation*}

\subsection{Traces -- Harmonic tensors}

Contracting two subscripts $i,j$ of a tensor $\bT$ of order $n$ defines a new tensor of order $n-2$ denoted as $\tr_{ij} \bT$. For a totally symmetric tensor $\bT$, this operation does not depend on a particular choice of the pair $i,j$. Thus, we can refer to this contraction just as the \emph{trace} of $\bT$ and we will denote it as $\tr \bT$. It is a totally symmetric tensor of order $n-2$. Iterating the process, we define
\begin{equation*}
  \tr^{k} \bT = \tr(\tr(\dotsb (\tr \bT))),
\end{equation*}
which is a totally symmetric tensor of order $n-2k$.

Harmonic tensors $\bH$  are by definition totally symmetric traceless tensors, \emph{i.e.} such as $\bH=\bH^s$, $\tr^k \bH=0$. Their vector spaces are denoted $\HH^n(\RR^d)$, with $n$ the order of considered tensor, $d$ the dimension. In  2D ($d=2$), $\dim \HH^n(\RR^2)=2$, while in 3D ($d=3$), $\dim \HH^n(\RR^3)=2n+1$. The harmonic tensors of order two (\emph{i.e.} the deviatoric tensors) will be denoted by lowercase letter $\bh$.

\subsection{Harmonic product}

Let $\bH_1$ and $ \bH_2 $ be two harmonic tensors of orders $n_1$ and $n_2$ respectively.
The harmonic product $\bH_1 \ast \bH_2$, defining an harmonic tensor of order $n=n_1+n_2$, has been introduced in~\citep{olive2018harmonic} as the leading harmonic part of the symmetric tensor product $\bH_1 \odot \bH_2 $,
\begin{equation*}
  \bH_1 \ast \bH_2:= \left( \bH_1 \odot \bH_2 \right)' \in \HH^{n_1+n_2}(\RR^d), \qquad d=2,3.
\end{equation*}
In previous works, the leading harmonic part was sometimes denoted $ \left( \bH_1 \odot \bH_2 \right)_{0}$. As it is a generalization of the deviatoric part, the notation $(\cdot)'$ is here preferred to $(\cdot)_0$. It is computed at any order thanks to the harmonic decomposition (see section~\ref{sec:harmonic}). The harmonic product is \emph{associative} and \emph{commutative},
\begin{equation*}
  \bH_1 \ast (\bH_2\ast \bH_3) = (\bH_1 \ast \bH_2) \ast \bH_3,
  \quad
  \bH_1 \ast \bH_2=\bH_2 \ast \bH_1.
\end{equation*}

Let us particularize the harmonic product for specific cases of bi-dimensional (2D) vectors and harmonic second-order tensors (belonging thus to $\HH^n(\RR^2)$ for $n=1,2$).

\begin{ex}
  For two vectors $\vec w_1, \vec w_2\in \HH^1(\RR^2)$, we have
  \begin{equation*}
    \begin{split}
      \ww_1 \ast \ww_2 = &(\ww_1 \odot \ww_2)'
      \\
      = &\frac{1}{2}\left(\ww_1 \otimes \ww_2 + \ww_2 \otimes \ww_1 \right) - \frac{1}{2} (\ww_1 \cdot \ww_2) \,\Idd,
    \end{split}
  \end{equation*}
  where $\ww_1 \cdot \ww_2=\ww_1^T \ww_2$ is the scalar product.
\end{ex}

\begin{ex}
  For two second-order harmonic (deviatoric) tensors  $\bh_1, \bh_2 \in \HH^2(\RR^2)$, we have \citep{olive2018harmonic}
  \begin{equation*}
    \begin{split}
      \bh_1 \ast \bh_2 = &( \bh_1 \odot \bh_2)'
      \\
      = & \bh_{1}\odot \bh_{2}-\frac{1}{4}\tr (\bh_{1}\bh_{2})\Idd \odot \Idd.
    \end{split}
  \end{equation*}
\end{ex}

\begin{ex}\label{ex:hasth}
  The harmonic square of a 2D second order harmonic (deviatoric) tensor $\bh \in \HH^2(\RR^2)$ writes  \citep{olive2018harmonic}:
  \begin{equation}\label{eq:H2astH2}
    \begin{aligned}
      \bh \ast \bh = & ( \bh \odot \bh)'
      \\
      =              & \bh \odot \bh-\frac{1}{4}(\bh:\bh) \Idd \odot \Idd
      \\
      =              & \bh\otimes \bh -\frac{1}{2} (\bh:\bh) \, \bJ,
    \end{aligned}
  \end{equation}
  where
  \begin{equation}\label{eq:J}
    \bJ=\bI-\frac{1}{2} \Idd \otimes \Idd.
  \end{equation}
  with $I_{ijkl}=\frac{1}{2}(\delta_{ik}\delta_{jl}+\delta_{il}\delta_{jk})$.
\end{ex}

\subsection{Covariants of a tensor $\bT$ (for the rotation group)}

The action $g\star \bT$ of a rotation $g\in \SO(d)$ on a tensor $\bT$ of order $2$ or $4$ is
\begin{equation*}
  (g\star \bT)_{ij}=g_{ik}g_{jl} T_{kl}, \quad
  (g\star \bT)_{ijkl}=g_{ip}g_{jq}g_{kr}g_{ls}  T_{pqrs}.
\end{equation*}

A tensor $\bA(\bT)$ is said to be a covariant of tensor $\bT$ for $\SO(d)$ if
\begin{equation*}
  g\star \bA(\bT)=\bA(g \star \bT), \quad \forall g\in  \SO(d),
\end{equation*}
and dimension $d$ is taken next as $d=2$.

The algebra of polynomial covariants of the elasticity tensor has been defined (and studied) in~\citep{kolev2018characterization} for $d=3$ (three-dimensional case) and in~\citep{desmorat2020computation}, for $d=2$ (bi-dimensional case).

\section{Harmonic fourth order part $\bH$ of the 2D elasticity tensor as an harmonic square}
\label{sec:harmonic}

\subsection{Harmonic decomposition}
\label{ssec:harmonic-decomposition}

The harmonic decomposition of tensors is a powerful mathematical tool \citep{schouten1989tensor,spencer1970note}, that has first been applied to three-dimensional elasticity tensors $\bC \in \Ela(\RR^3)$ by \cite{backus1970geometrical}.
Formally in 2D, it is the equivariant decomposition
\begin{equation*}
  \bC= (\mu, \kappa, \bd', \bH)\in\Ela(\RR^2),
\end{equation*}
into two scalars (invariants) $\mu, \kappa \in \HH^0(\RR^2)\simeq \RR$, $\mu$ being the shear modulus, $\kappa$ the bi-dimensional bulk modulus, one harmonic (deviatoric) second order covariant $\bd'=\bd'(\bC)\in\HH^2(\RR^2)$ and one harmonic fourth order covariant $\bH=\bH(\bC)\in\HH^4(\RR^2)$, such as
\begin{equation*}
  g\star \bC= (\mu, \kappa, g\star \bd', g\star \bH)\, \quad \forall g\in \SO(2).
\end{equation*}

An explicit harmonic decomposition of $\bC\in\Ela(\RR^2)$ is:
\begin{equation}\label{eq:HarmDecomp}
  \bC= 2 \mu \bJ + \kappa \Idd \otimes \Idd +\frac{1}{2} \left( \Idd \otimes \bd' +\bd' \otimes \Idd\right) + \bH,
\end{equation}
with $\bJ$ defined by \eqref{eq:J}. The closed form expressions of harmonic components $\mu$, $\kappa$, $\bd'$ and $\bH$ are gained thanks to the definition of dilatation and Voigt second order tensors
$\bd=\tr_{12} \bC$ and of $\bv=\tr_{13} \bC$,
\begin{equation}\label{eq:HarmDecompComponents}
  \begin{split}
    \mu = & \frac{1}{8}(2\tr \bv- \tr \bd),
    \\
    \kappa = & \frac{1}{4} \tr \bd,
    \\
    \bd' =& \, \bd-\frac{1}{2} (\tr \bd)\, \Idd,
    \\
    \bH=&\, \bC- 2 \mu \bJ - \kappa \Idd \otimes \Idd -\frac{1}{2} \left( \Idd \otimes \bd' +\bd' \otimes \Idd\right) .
  \end{split}
\end{equation}

In the isotropic (initial) case, one has $\bH=0$ and
\begin{equation*}
  \bd=2\kappa \Idd, \qquad \bv=(2\mu+\kappa) \Idd ,
\end{equation*}
the Young's modulus and Poisson's ratio being
\begin{equation*}
  E=\frac{4 \kappa  \mu }{\kappa +\mu },
  \qquad
  \nu=\frac{\kappa -\mu }{\kappa +\mu }.
\end{equation*}

\begin{rem}\label{rem1}
  In 2D the deviatoric parts of dilatation and of Voigt second order tensors are equal, $\bv'=\bd'$, and the term $\Idd \otimes \bd' +\bd' \otimes \Idd$ is also equal to $\Idd\, \underline{\overline \otimes} \, \bd' +\bd'\,  \underline{\overline \otimes} \, \Idd$, where
  $(\ba \underline{\overline \otimes} \bb)_{ij}=\frac{1}{2}(a_{ik} b_{jl}+a_{il} b_{jk})$.
\end{rem}

One can also perform the harmonic decomposition of the compliance tensor $\bS=\bC^{-1}$, and gets then
\begin{equation*}
  \bd'(\bS)=\bv'(\bS)=(\tr_{12} \bS)',
\end{equation*}
where
\begin{equation*}
  \bS= \frac{1}{2 \mu} \bJ + \frac{1}{4\kappa} \Idd \otimes \Idd +\frac{1}{2} \left( \Idd \otimes \bd'(\bS) +\bd'(\bS) \otimes \Idd\right)+\bH(\bS) ,
\end{equation*}
and $\bH(\bS)$ is the harmonic (totally symmetric and traceless) fourth order part of $\bS$.

Recall finally that a minimal generating set (a minimal integrity basis) of the invariant algebra of the elasticity tensor in 2D, under the action of the orthogonal group $\OO(2)$, consists in the following 5 invariants~\citep{vianello1997integrity},
  \begin{equation}\label{eq:invC2D}
    \begin{aligned}
      \mu    & =  \frac{1}{8}(2\tr \bv- \tr \bd),
      \\
      \kappa & = \frac{1}{4} \tr \bd,
      \\
      I_2    & = \norm{\bd'}^2 = \bd':\bd',
      \\
      J_2    & =\norm{\bH}^2 = \bH :: \bH,
      \\
      K_3    & = \bd:\bH:\bd  .
    \end{aligned}
  \end{equation}

\subsection{The harmonic part $\bH \in \HH^4(\RR^2)$ as an harmonic square}

In 2D, it has been shown that any fourth order harmonic tensor $\bH\neq 0$ is of the form \citep{desmorat2015tensorial}
\begin{equation}\label{eq:HLambda}
  \bH= 2 \Lambda\, \be \ast \be, \quad \tr \be=0, \quad  \| \be \|=1,
\end{equation}
where $\be$ is a unit deviatoric (second order) eigentensor associated with a non zero eigenvalue $\Lambda$ of the Kelvin representation of $\bH$ (\emph{i.e.} such that $\bH:\be=\Lambda \be$ in an orthonormal basis).

\begin{rem}
  Since $\bH\neq 0$ has two opposite eigenvalues $\Lambda_{+}>0$ and $\Lambda_{-} = -\Lambda_{+}<0$, this implies that any bi-dimensional fourth order harmonic tensor is always an \emph{harmonic square}~\citep{desmorat2015tensorial,desmorat2016second}. The corresponding explicit formula is as follows,
  \begin{equation}\label{eq:Hhsquare}
    \bH= \bh \ast \bh, \qquad \bh=\pm \sqrt{2 \Lambda_{+}}\, \be_{+} ,
  \end{equation}
  where the unit deviatoric second order tensor $\be=\be_{+}$ is the eigentensor of $\bH$ associated with the strictly positive eigenvalue $\Lambda_{+}$. Observe that there are two opposite \emph{harmonic square roots} $\bh$ of $\bH$, which are opposite to each other. A new proof of this result, more conceptual, is provided in~\ref{sec:harmonic-square-root}.
\end{rem}

\section{Covariant reconstruction of 2D orthotropic elasticity tensors}
\label{sec:covariant}

Bi-dimensional elasticity tensors $\bC\in \Ela(\RR^2)$ have Kelvin matrix representation
\begin{equation*}
  [\bC]=\left[
    \begin{array}{ccc}
      C_{1111}          & C_{1122}          & \sqrt{2} C_{1112}
      \\
      C_{1122}          & C_{2222}          & \sqrt{2} C_{1222}
      \\
      \sqrt{2} C_{1112} & \sqrt{2} C_{1222} & 2C_{1212}
    \end{array}\right].
\end{equation*}
Bi-dimensional harmonic fourth order tensors $\bH\in \HH^4(\RR^2)$ have Kelvin representation
\begin{equation}\label{eq:KelvH}
  [\bH]=\left[
    \begin{array}{ccc}
      H_{1111}          & -H_{1111}          & \sqrt{2} H_{1112}
      \\
      -H_{1111}         & H_{1111}           & - \sqrt{2} H_{1112}
      \\
      \sqrt{2} H_{1112} & -\sqrt{2} H_{1112} & -2H_{1111}
    \end{array}\right].
\end{equation}

The normal form of an orthotropic elasticity tensor corresponds to $C_{1112}=C_{1222}=0$. If $\bC = \bH$ is moreover harmonic, we get $H_{1112}=0$.

A bi-dimensional harmonic fourth order tensor $\bH$ cannot be strictly orthotropic (\emph{i.e.} with exact symmetry group $\DD_2$, the symmetry group of a rectangle). A non-vanishing harmonic fourth-order tensor $\bH$ has always the symmetry group $\DD_4$ (the symmetry group of a square)~\citep{VV2001,vannucci2005plane}. The covariants of $\bH$ inherit its symmetry \citep{olive2018harmonic,kolev2018characterization}: this implies that all the second order covariants of $\bH$ have at least the square symmetry, they are therefore isotropic (and all the deviatoric second order covariants of $\bH$ vanish). Combined with the fact that  the harmonic product $\ast$ is itself covariant, this geometrical property implies that the harmonic square root $\bh$ (\emph{i.e.} such as $\bH=\bh \ast \bh$) is not a covariant of $\bH$.

\subsection{Square symmetry case}

If an elasticity tensor itself has the (exact) \emph{square symmetry}, then $\bd'=0$ and $\bH\neq 0$ \citep{verchery1982invariants,vianello1997integrity,vannucci2005plane}, and $\bC$ has no better reconstruction formula by means of its covariants than its harmonic decomposition,
\begin{equation*}
  \bC=  2 \mu \bJ + \kappa \Idd \otimes \Idd + \bH.
\end{equation*}

\subsection{Orthotropic case}

When an elasticity tensor $\bC$ is orthotropic, its second order covariant $\bd'$ is an eigentensor of its harmonic part $\bH=\bH(\bC)$ (given by~\eqref{eq:HarmDecomp}), \emph{i.e.}
\begin{equation}\label{eq:Heigen}
  \bH:\bd'=  \Lambda \bd',
\end{equation}
but where $\Lambda=\pm \Lambda_{+}$ is a non-vanishing eigenvalue. To check this, just consider its Kelvin normal form
\begin{equation}\label{eq:Cortho2D}
  [\bC]=\left[
    \begin{array}{ccc}
      C_{1111} & C_{1122} & 0
      \\
      C_{1122} & C_{2222} & 0
      \\
      0        & 0        & 2C_{1212}
    \end{array}\right],
\end{equation}
which leads to
\begin{equation*}
  \bd'=(\tr_{12}\bC)'=\frac{C_{1111}-C_{2222}  }{2}
  \begin{pmatrix}
    1 & 0
    \\
    0 & -1
  \end{pmatrix}
  \neq 0,
\end{equation*}
and, using~\eqref{eq:HarmDecomp}, we get
\begin{equation*}
  [\bH]= H_{1111} \left[
    \begin{array}{rrr}
      1  & -1 & 0
      \\
      -1 & 1  & 0
      \\
      0  & 0  & -2
    \end{array}\right].
\end{equation*}

Contracting~\eqref{eq:Heigen} with $\bd'$, altogether with the relations
\begin{equation*}
  \bH:\Idd = \Idd:\bH=0, \qquad \bd':\bH:\bd'=K_3,
\end{equation*}
we obtain finally
\begin{equation*}
  \Lambda = \frac{K_3}{I_2},
\end{equation*}
where the invariants $I_2$ and $K_3$ are defined by~\eqref{eq:invC2D}. This means by~\eqref{eq:HLambda} that
\begin{equation*}
  \bH = \frac{2\Lambda}{\| \bd'\|^2} \bd' \ast \bd'.
\end{equation*}
Hence, we have the following result.

\begin{thm}\label{them:main}
  Any bi-dimensional orthotropic elasticity tensor $\bC=(\mu, \kappa, \bd', \bH)\in \Ela(\RR^2)$ can be reconstructed by means of its 4 invariants $\mu$, $\kappa$, $I_2$, $K_3$, and of its (deviatoric)  second order covariant $\bd'$, as
  \begin{equation}
    \begin{aligned}
      \bC=  & 2 \mu \bJ + \kappa \Idd \otimes \Idd +\frac{1}{2} \left( \Idd \otimes \bd' +\bd' \otimes \Idd\right) + \bH,
      \\
      \bH = & \frac{2 K_3}{I_2^{\, 2}} \,\bd'\ast \bd' ,
    \end{aligned}
  \end{equation}
  where
  $\bd'\ast \bd'=  \displaystyle  \bd'\otimes \bd' -\frac{1}{2} (\bd':\bd')\, \bJ$.
\end{thm}

From theorem~\ref{them:main}, one can derive an upper bound $\Delta$ for the distance to orthotropy of a bi-dimensional elasticity tensor as defined below.

\begin{cor}\label{cor:Delta}
  Let $\bC=(\mu, \kappa, \bd', \bH)\in \Ela(\RR^2)$ be a bi-dimensional elasticity tensor  with no material symmetry, let
  $I_2=\|\bd'\|^2$, $J_2=\| \bH \|^2$ and $K_3=\bd: \bH: \bd$.
  Then, the positive invariant
    {\small
      \begin{equation*}
        \begin{aligned}
          \Delta \! = &
          \Big\| \bC
          - \!  2 \mu \bJ - \!  \kappa \Idd \!  \otimes \!  \Idd
          -\! \frac{1}{2}( \Idd \!  \otimes \bd' +\!  \bd' \!  \otimes \!  \Idd)
          -\!   \frac{2 K_3}{I_2^{\, 2}}\, \bd'\ast \bd'
          \Big\|
          \\
          =           & \frac{\sqrt{J_2 I_2^2-2K_3^2}}{I_2} .
        \end{aligned}
      \end{equation*}
    }
  is an upper bound of the distance of $\bC$ to the orthotropic symmetry class.
\end{cor}

In corollary~\ref{cor:Delta}, the fact that $\bC$ has no material symmetry implies that $\| \bd' \|^2=I_2\neq 0$. The first equality corresponds to the norm of the difference between the elasticity tensor and its orthotropic reconstruction formula (in which necessarily $\bd'\neq 0$). The Harmonic decomposition formula~\eqref{eq:HarmDecomp} gives indeed
\begin{equation*}
  \Delta
  =\left\| \bH -   \frac{2 K_3}{I_2^{\, 2}}\, \bd'\ast \bd' \right\|.
\end{equation*}
The second equality in corollary~\ref{cor:Delta} is obtained by expanding the square norm, using~\eqref{eq:H2astH2} and the facts that $\bJ::\bJ=2$ and $\bH::\bJ=0$, see~\citep{desmorat2015tensorial}.

\section{Discrete elements representative volumes}
\label{sec:discrete}

\subsection{Beam-particle model}

\begin{figure}[htp]
  \centering
  \includegraphics[scale=0.8]{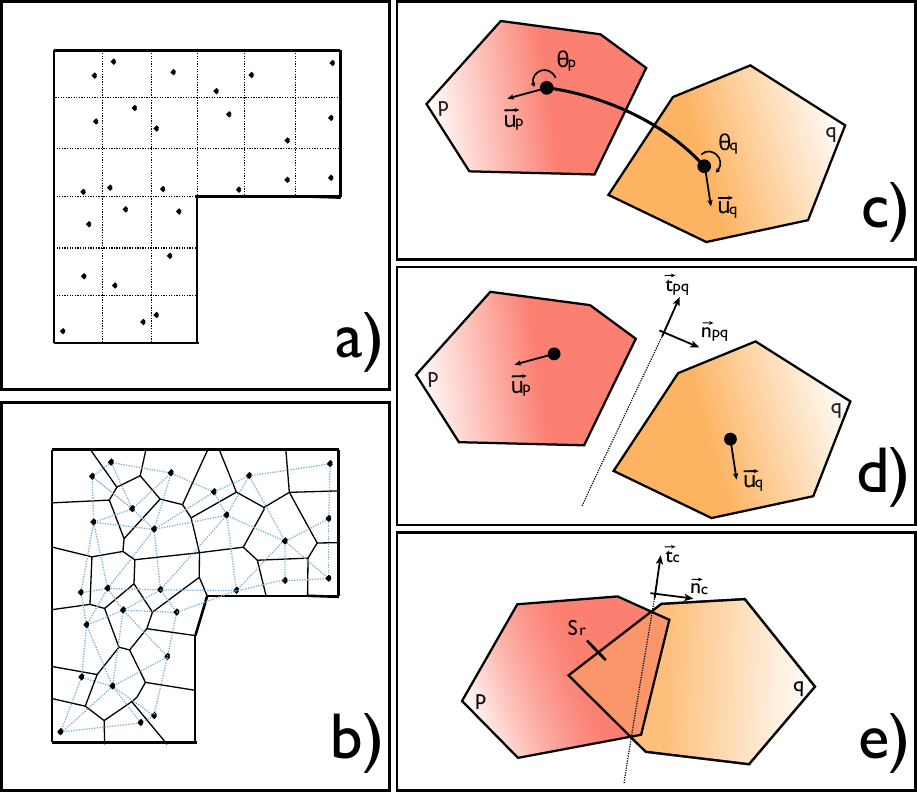}
  \caption{Description of the main ingredients in the beam-particle model (from \cite{oliver2019discontinuous}).}
  \label{fig:deap}
\end{figure}

The discrete method used here to perform the virtual testing of a quasi-brittle heterogeneous material is a beam-particle approach detailed in \citep{vassaux2016beam} and summarized in figure~\ref{fig:deap}.

The representative volume is divided into an assembly of rigid particles. The particle mesh is generated from a Voronoi tesselation of a set of randomly generated points within a grid. This operation generates polygonal particles.

Dual of the Voronoi tesselation, the Delaunay triangulation associates a segment with each pair of neighbouring particles. This segment is used as a geometric support for an elastic Euler-Bernoulli beam modelling the cohesion of the material. Each beam $b$ is parameterized by its length $l_b$, its section $A_b$, the Young modulus $E$ and the coefficient of inertia $\alpha = 64 I_b \pi / A_b^2$. The first two coefficients $l_b$ and $A_b$ are different for each beam and imposed by the geometry of the mesh. The next two, $E$ and $\alpha$, are identical for all the beams and identified in order to reproduce the macroscopic elastic behaviour.

In order to reproduce the fracture behaviour, a failure criterion $P_{pq}$ is associated with each beam connecting two particles $p$ and $q$:

\begin{equation}
  P_{pq} = \frac{\varepsilon_{pq}}{\varepsilon_{pq}^{cr}} + \frac{|\theta_{p}-\theta_{q}|}{\theta_{pq}^{cr}} > 1
\end{equation}
where the breaking threshold in extension $\varepsilon_{pq}^{cr}$ and the breaking threshold in rotation $\theta_{pq}^{cr}$ are generated for each beam according to a Weibull distribution. The Weibull probability density function adopted in this study is :
\begin{equation}
  f(x) = \frac{k}{\lambda} \left(\frac{x}{\lambda	}\right)^{k-1}e^{-(x/\lambda)^k}
\end{equation}
with the scale factor $\lambda$ and the shape factor $k$. In fact, the spatial variability for the two breaking thresholds $\varepsilon_{pq}^{cr}$, $\theta_{pq}^{cr}$, is supposed to be identical. Therefore, three parameters control the fracture behaviour:
a shape factor $k$ (common to both distributions), a scale factor in extension $\lambda_{\epsilon_{cr}}$ and a scale factor in rotation $\lambda_{\theta{cr}}$. The combination of this failure criterion and of a random generation of failure thresholds makes it possible to model a quasi-brittle behaviour in tension and compression (as proposed and shown in \cite{vassaux2016beam}). These three failure parameters are identified in such a way as to reproduce the non-linear behaviour of the material.

Finally, in order to be able to capture the mechanisms of crack closure and of crack sliding, frictional contact is introduced between the particles if they overlap while they are not connected by a beam. The detection of the soft contact between two polygonal particles, the calculation of the normal contact force as well as that of tangential contact via Coulomb's law of friction follows the proposal of \cite{tillemans1995simulating}. A rewriting is however proposed in order to use relative displacements rather than relative velocities \citep{vassaux2015lattice} (figure~\ref{fig:deap}e), the former being more suitable in the quasi-static framework in which we place ourselves. To parametrize the friction, an angle of friction $\phi$ is introduced resulting in a coefficient of friction $\tan \phi$ for Coulomb's law.

\subsection{Extraction of the effective elasticity tensor}

The definition of homogenized quantities on the discrete volume, such as stress and strain, is necessary to link the discrete description, providing detailed information on particle movements and interaction forces, to the continuous description.

The average Cauchy stress tensor is computed from the symmetrization of the definition proposed by \cite{bagi1996stress}:
\begin{equation}
  \bsigma  = \frac{1}{S} \sum_{p=1}^{N} \mathbf{f}^{(p)} \odot \mathbf{x}^{(p)}
\end{equation}
where $S$ is the area of the discrete representative volume (the so-called Representative Volume Element or RVE in 3D, the Representative Area Element or RAE in 2D), $\mathbf{f}^{(p)}$ is the resulting force on the particle $p$ and $\mathbf{x}^{(p)}$ is the vector position of the center of the particle $p$. The summation is made on the $N$ particles constituting the boundary of the RAE.

The macroscopic strain tensor is defined as the following mean value:
\begin{equation}
  \bepsilon = \frac{1}{S} \sum_{p=1}^{N} \left( \frac{\bu^{(p)}+\bu^{(p+1)}}{2} \odot \nn^{(p,p+1)} \right) l^{(p,p+1)}
\end{equation}
where $\bu^{(p)}$ is the displacement vector of the particle $p$, $l^{(p,p+1)}$ is the length of the segment linking the particles $p$ and $p+1$ and $\nn^{(p,p+1)}$ the outward pointing normal vector of this same segment.

In order to obtain the effective elasticity tensor, three measurement loadings are performed and the corresponding average stress and strain vectors are extracted. For each loading $i$, the effective elasticity tensor links the stress to the strain in the following way:
\begin{equation}
  \hat \sigma^{(i)} = [\bC] \,\hat \epsilon^{(i)}
\end{equation}
where $[\bC]$ is Kelvin matrix representation of macroscopic elasticity tensor $\bC$ and
\begin{equation*}
  \hat \sigma^{(i)} = \left(
  \begin{matrix}
      \sigma_{11}^{(i)} \\
      \sigma_{22}^{(i)} \\
      \sqrt{2} \sigma_{12}^{(i)}
    \end{matrix}\right),
  ~~~
  \hat \epsilon^{(i)} = \left(
  \begin{matrix}
      \epsilon_{11}^{(i)} \\
      \epsilon_{22}^{(i)} \\
      \sqrt{2} \epsilon_{12}^{(i)}
    \end{matrix}\right).
\end{equation*}

Therefore, if the measurement loadings are chosen to ensure that the strain vectors are linearly independent, their determinant is non-zero and the symmetric effective elasticity tensor can be computed from
\begin{equation}
  [\bC] = \left(\left(\hat \sigma^{(1)} ~ \hat \sigma^{(2)} ~ \hat \sigma^{(3)}\right) \left(\hat \epsilon^{(1)} ~ \hat \epsilon^{(2)} ~ \hat \epsilon^{(3)}\right)^{-1}\right)^s.
\end{equation}

\subsection{Definition of possible measurement loadings}

Two requirements must be met to define the measurement loadings: ensuring the linear independence of the strain vectors and maintaining the cracks open during measurement. Several sets of measurement loadings are used and compared:
\begin{itemize}

  \item DEF\_WoC : Elementary strain loading without contact
        \begin{equation}
          \hat \epsilon^{(1)} = \left(
          \begin{matrix}
              \epsilon \\
              0        \\
              0
            \end{matrix}\right),
          ~~~
          \hat \epsilon^{(2)} = \left(
          \begin{matrix}
              0        \\
              \epsilon \\
              0
            \end{matrix}\right),
          ~~~
          \hat \epsilon^{(3)} = \left(
          \begin{matrix}
              0 \\
              0 \\
              \sqrt{2} \epsilon
            \end{matrix}\right),
          \nonumber
        \end{equation}
        where $\epsilon$ is the applied strain that must be sufficiently small for the (possibly cracked) RAE to remain in the elasticity domain.
        The following displacement fields are thus applied on the boundaries of the specimen:
        \begin{equation}
          \bu^{(1)} = \left(
          \begin{matrix}
              \epsilon x \\
              0
            \end{matrix}\right),
          ~~~
          \bu^{(2)} = \left(
          \begin{matrix}
              0 \\
              \epsilon y
            \end{matrix}\right),
          ~~~
          \bu^{(3)} = \left(
          \begin{matrix}
              \epsilon y \\
              \epsilon x
            \end{matrix}\right),
          \nonumber
        \end{equation}
        with $x$ and $y$ the positions of the center of the particles on the boundaries.

  \item DEF\_BT\_C : Bi-tension strain loading with contact
        \begin{equation}
          \hat \epsilon^{(1)} = \left(
          \begin{matrix}
              \epsilon_{bt}+\epsilon \\
              \epsilon_{bt}          \\
              0
            \end{matrix}\right),
          ~~~
          \hat \epsilon^{(2)} = \left(
          \begin{matrix}
              \epsilon_{bt}            \\
              \epsilon_{bt} + \epsilon \\
              0
            \end{matrix}\right),
          ~~~
          \hat \epsilon^{(3)} = \left(
          \begin{matrix}
              \epsilon_{bt} \\
              \epsilon_{bt} \\
              \sqrt{2} \epsilon
            \end{matrix}\right),
          \nonumber
        \end{equation}
        where $\epsilon_{bt}$ is introduced to create a bi-tension loading keeping the cracks open.
        The following displacement fields are thus applied on the boundaries of the specimen:
        \begin{gather}
          \bu^{(1)} = \left(
          \begin{matrix}
              (\epsilon_{bt}+\epsilon) x \\
              \epsilon_{bt} y
            \end{matrix}\right),
          ~~~
          \bu^{(2)} = \left(
          \begin{matrix}
              \epsilon_{bt} x \\
              (\epsilon_{bt} + \epsilon) y
            \end{matrix}\right),
          ~~~
          \bu^{(3)} = \left(
          \begin{matrix}
              \epsilon_{bt} x + \epsilon y \\
              \epsilon x + \epsilon_{bt} y
            \end{matrix}\right),
          \nonumber
        \end{gather}

  \item SIG\_F\_WoC : Elementary stress loading with applied forces without contact
        \begin{equation}
          \hat \sigma^{(1)} = \left(
          \begin{matrix}
              \sigma \\
              0      \\
              0
            \end{matrix}\right),
          ~~~
          \hat \sigma^{(2)} = \left(
          \begin{matrix}
              0      \\
              \sigma \\
              0
            \end{matrix}\right),
          ~~~
          \hat \sigma^{(3)} = \left(
          \begin{matrix}
              0 \\
              0 \\
              \sqrt{2} \sigma
            \end{matrix}\right),
          \nonumber
        \end{equation}
        where $\sigma$ is the applied stress that must be sufficiently small to remain in the elasticity domain.
        The traction vectors $\bt=\bsigma \bn$ are then applied on the boundaries ($\bn$: outer unit normal) and the central particle is completely blocked to avoid rigid motions:
        \begin{equation}
          \bt^{(1)}_{xmin} = \left(
          \begin{matrix}
              - \sigma \\
              0
            \end{matrix}\right),
          ~~~
          \bt^{(1)}_{xmax} = \left(
          \begin{matrix}
              \sigma \\
              0
            \end{matrix}\right),
          \nonumber
        \end{equation}
        \begin{equation}
          \bt^{(2)}_{ymin} = \left(
          \begin{matrix}
              0        \\
              - \sigma \\
            \end{matrix}\right),
          ~~~
          \bt^{(2)}_{ymax} = \left(
          \begin{matrix}
              0      \\
              \sigma \\
            \end{matrix}\right),
          \nonumber
        \end{equation}
        \begin{equation}
          \bt^{(3)}_{xmin} = \left(
          \begin{matrix}
              0 \\
              -  \sigma
            \end{matrix}\right),
          ~~~
          \bt^{(3)}_{xmax} = \left(
          \begin{matrix}
              0 \\
              \sigma
            \end{matrix}\right),
          ~~~
          \bt^{(3)}_{ymin} = \left(
          \begin{matrix}
              -  \sigma \\
              0
            \end{matrix}\right),
          ~~~
          \bt^{(3)}_{ymax} = \left(
          \begin{matrix}
              \sigma \\
              0
            \end{matrix}\right).
          \nonumber
        \end{equation}

  \item SIG\_D\_WoC : Elementary stress loading with applied displacements without contact.

        The objective is to obtain the three stress states of the previous measurement loading by applying only kinematic conditions.

        \begin{equation}
          \bu^{(1)}_{xmin}\cdot\eee_x = 0,
          ~~~
          \bu^{(1)}_{xmax}\cdot\eee_x = u,
          ~~~
          \bu^{(1)}_{ymin}\cdot\eee_y = 0,
          \nonumber
        \end{equation}
        \begin{equation}
          \bu^{(2)}_{ymin}\cdot\eee_y = 0,
          ~~~
          \bu^{(2)}_{ymax}\cdot\eee_y = u,
          ~~~
          \bu^{(2)}_{xmin}\cdot\eee_x = 0,
          \nonumber
        \end{equation}
        \begin{equation}
          \bu^{(3)}_{xmin}\cdot\eee_y = 0,
          ~~~
          \bu^{(3)}_{xmax}\cdot\eee_y = \sqrt{2} u,
          ~~~
          \bu^{(3)}_{ymin}\cdot\eee_x = 0,
          ~~~
          \bu^{(3)}_{ymax}\cdot\eee_x = \sqrt{2} u.
          \nonumber
        \end{equation}
        where $\bu$ is the applied displacement that must be sufficiently small for the RAE to remain in the elasticity domain.

\end{itemize}

\subsection{Comparison of the different measurement loadings}

We first compare the four measurement loadings described previously on an uncracked square specimen of $100 \times 100$ particles (see first line of table~\ref{tab:mes_loads}). We observe that they give very close results. The dispersion can be attributed to the measurement noise. This result was expected because there are enough particles in the specimen to make it a Representative Area Element. Thus, the type of boundary conditions does not influence the result, as in an infinite medium.

\begin{table*}[hp]
  \begin{center}
    \begin{tabular}[c]{|c|m{3cm}|m{3cm}|m{3cm}|m{3cm}|}
      \hline
      \textbf{Crack patterns} & \textbf{DEF\_WoC} & \textbf{DEF\_BT\_C} & \textbf{SIG\_F\_WoC} & \textbf{SIG\_D\_WoC} \\
      \hline
      \hline
      \begin{minipage}{.2\textwidth}
        \centering
        \vspace{0.2cm}
        Initial state  \\
        No damage \\
        \vspace{0.2cm}
        \includegraphics[width=0.5\textwidth]{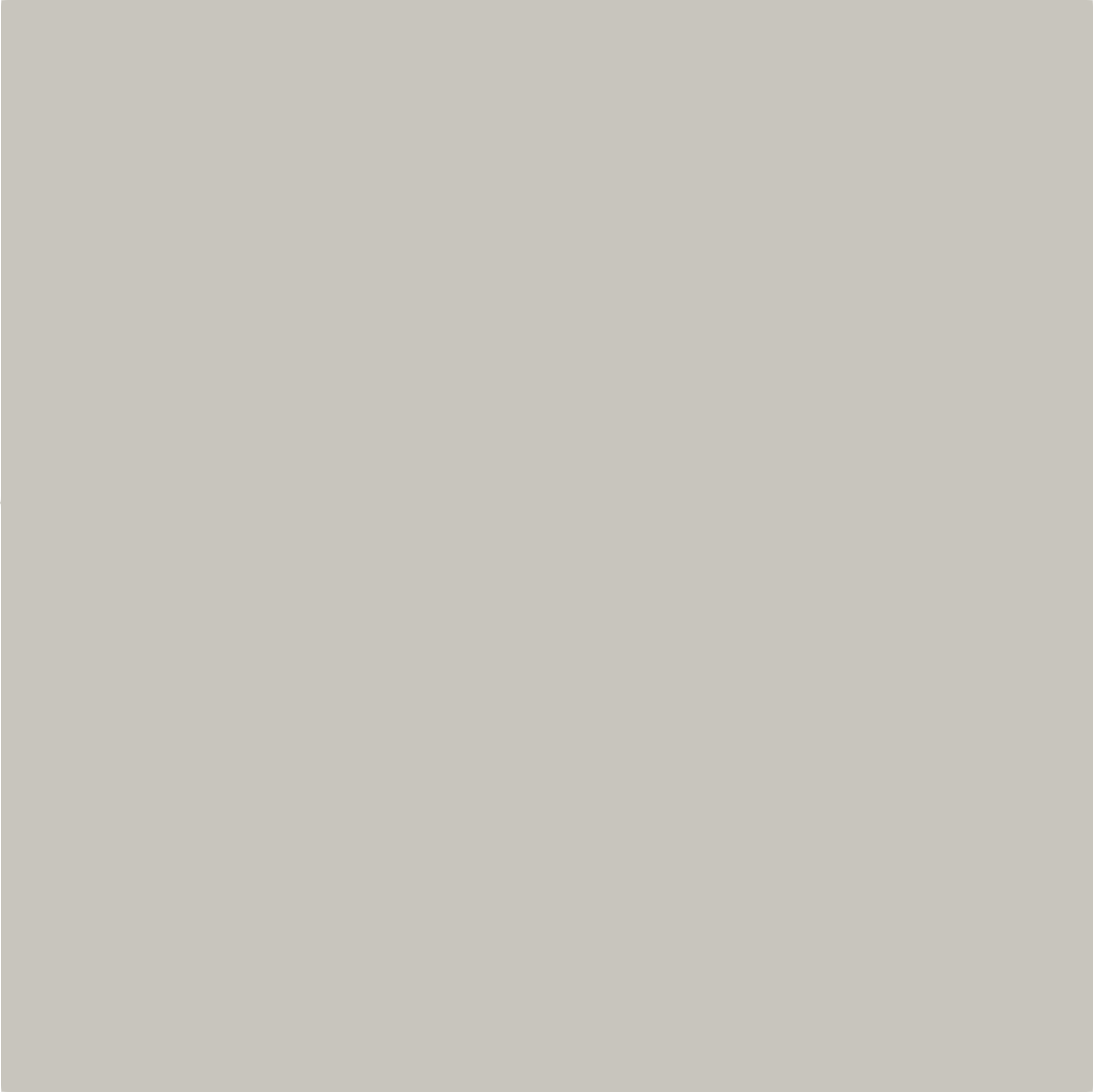}
        \vspace{0.2cm}
      \end{minipage}
                              &
      \begin{equation*} \scriptsize{\begin{bmatrix} 38.3 & 8.1 & 0.1 \\ 8.1 & 38.4 & -0.3\\ 0.1 & -0.3 & 29.2 \end{bmatrix}} \end{equation*}  
                              &
      \begin{equation*} \scriptsize{\begin{bmatrix} 38.4 & 8.0 & -0.3\\ 8.0 & 38.6 & -0.1\\ -0.3 & -0.1 & 29.4 \end{bmatrix}} \end{equation*}     
                              &
      \begin{equation*} \scriptsize{\begin{bmatrix} 38.0 & 8.1 & 0.1\\ 8.1 & 38.6 & 0.0\\ 0.1 & 0.0 & 29.4 \end{bmatrix}} \end{equation*}  
                              &
      \begin{equation*} \scriptsize{\begin{bmatrix} 38.3 & 8.1 & 0.1\\ 8.1 & 38.4 & -0.1\\ 0.1 & -0.1 & 29.2 \end{bmatrix}} \end{equation*}  
      \\
      \hline
      \begin{minipage}{.2\textwidth}
        \centering
        \vspace{0.2cm}
        Localized cracking  \\
        Low level of damage \\
        \vspace{0.2cm}
        \includegraphics[width=0.5\textwidth]{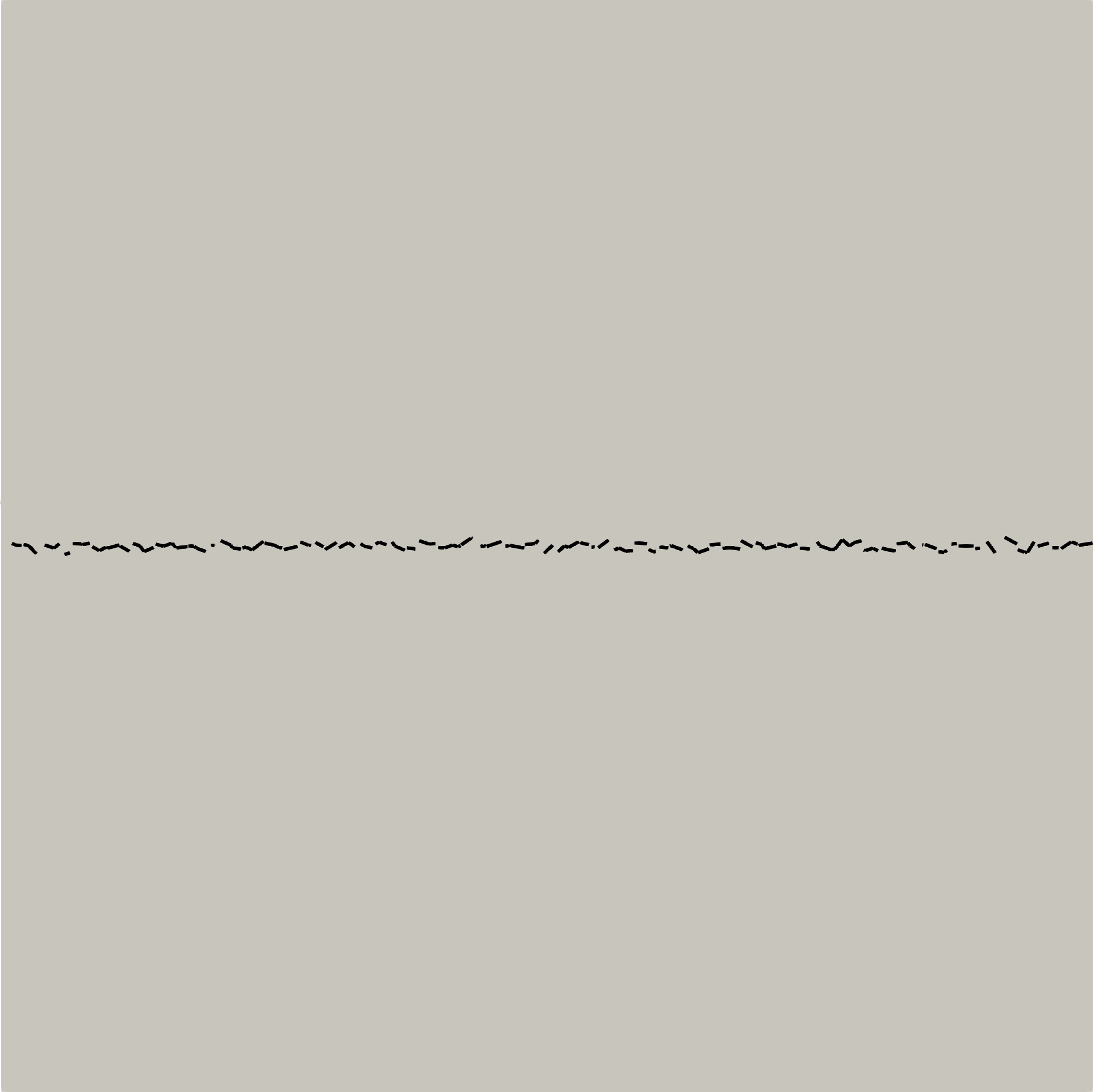}
        \vspace{0.2cm}
      \end{minipage}
                              &
      \begin{equation*} \scriptsize{\begin{bmatrix} 38.1 & 7.4 & 0.0\\ 7.4 & 34.6 & -0.4\\ 0.0 & -0.4 & 28.6 \end{bmatrix}} \end{equation*}           
                              &
      \begin{equation*} \scriptsize{\begin{bmatrix} 38.2 & 7.3 & -0.3\\ 7.3 & 34.7 & -0.4\\ -0.3 & -0.4 & 28.6 \end{bmatrix}} \end{equation*}  
                              &
      \begin{equation*} \scriptsize{\begin{bmatrix} 37.7 & 7.4 & 0.0\\ 7.4 & 34.4 & -0.3\\ 0.0 & -0.3 & 28.6 \end{bmatrix}} \end{equation*}    
                              &
      \begin{equation*} \scriptsize{\begin{bmatrix} 38.0 & 7.3 & 0.0\\ 7.3 & 34.1 & -0.4\\ 0.0 & -0.4 & 28.2 \end{bmatrix}} \end{equation*}     
      \\
      \hline
      \begin{minipage}{.2\textwidth}
        \centering
        \vspace{0.2cm}
        Localized cracking  \\
        Mid level of damage \\
        \vspace{0.2cm}
        \includegraphics[width=0.5\textwidth]{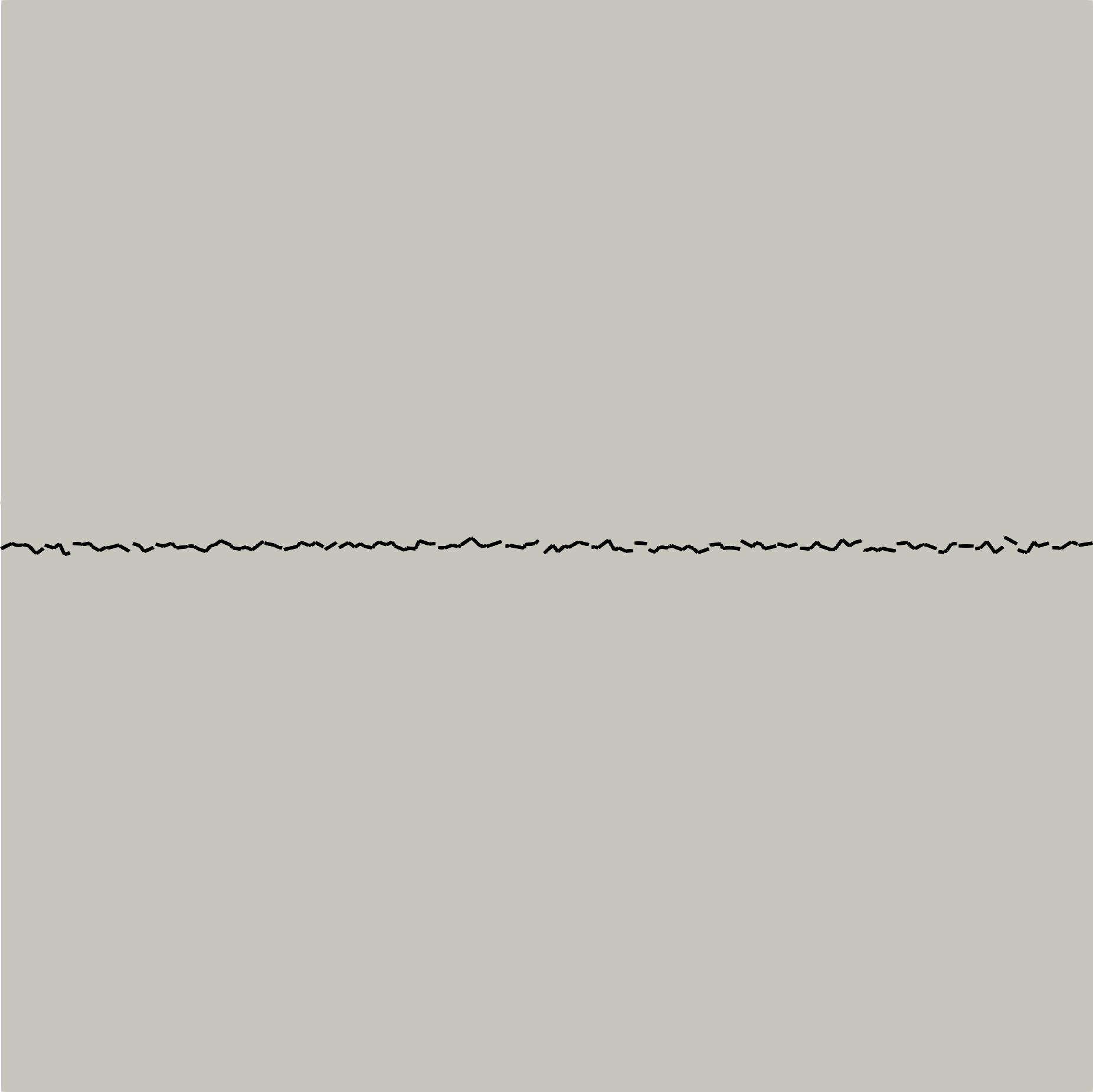}
        \vspace{0.2cm}
      \end{minipage}
                              &
      \begin{equation*} \scriptsize{\begin{bmatrix} 37.6 & 5.6 & 0.4\\ 5.6 & 25.6 & 1.1\\ 0.4 & 1.1 & 27.2 \end{bmatrix}} \end{equation*}        
                              &
      \begin{equation*} \scriptsize{\begin{bmatrix} 37.7 & 5.5 & 0.0\\ 5.5 & 25.8 & 1.3\\ 0.0 & 1.3 & 27.2 \end{bmatrix}} \end{equation*}     
                              &
      \begin{equation*} \scriptsize{\begin{bmatrix} 37.2 & 5.0 & 0.4\\ 5.0 & 23.0 & 1.6\\ 0.4 & 1.6 & 26.6 \end{bmatrix}} \end{equation*}     
                              &
      \begin{equation*} \scriptsize{\begin{bmatrix} 37.5 & 5.0 & 0.4\\ 5.0 & 22.7 & 1.6\\ 0.4 & 1.6 & 26.4 \end{bmatrix}} \end{equation*}     
      \\
      \hline
      \begin{minipage}{.2\textwidth}
        \centering
        \vspace{0.2cm}
        Localized cracking  \\
        High level of damage \\
        \vspace{0.2cm}
        \includegraphics[width=0.5\textwidth]{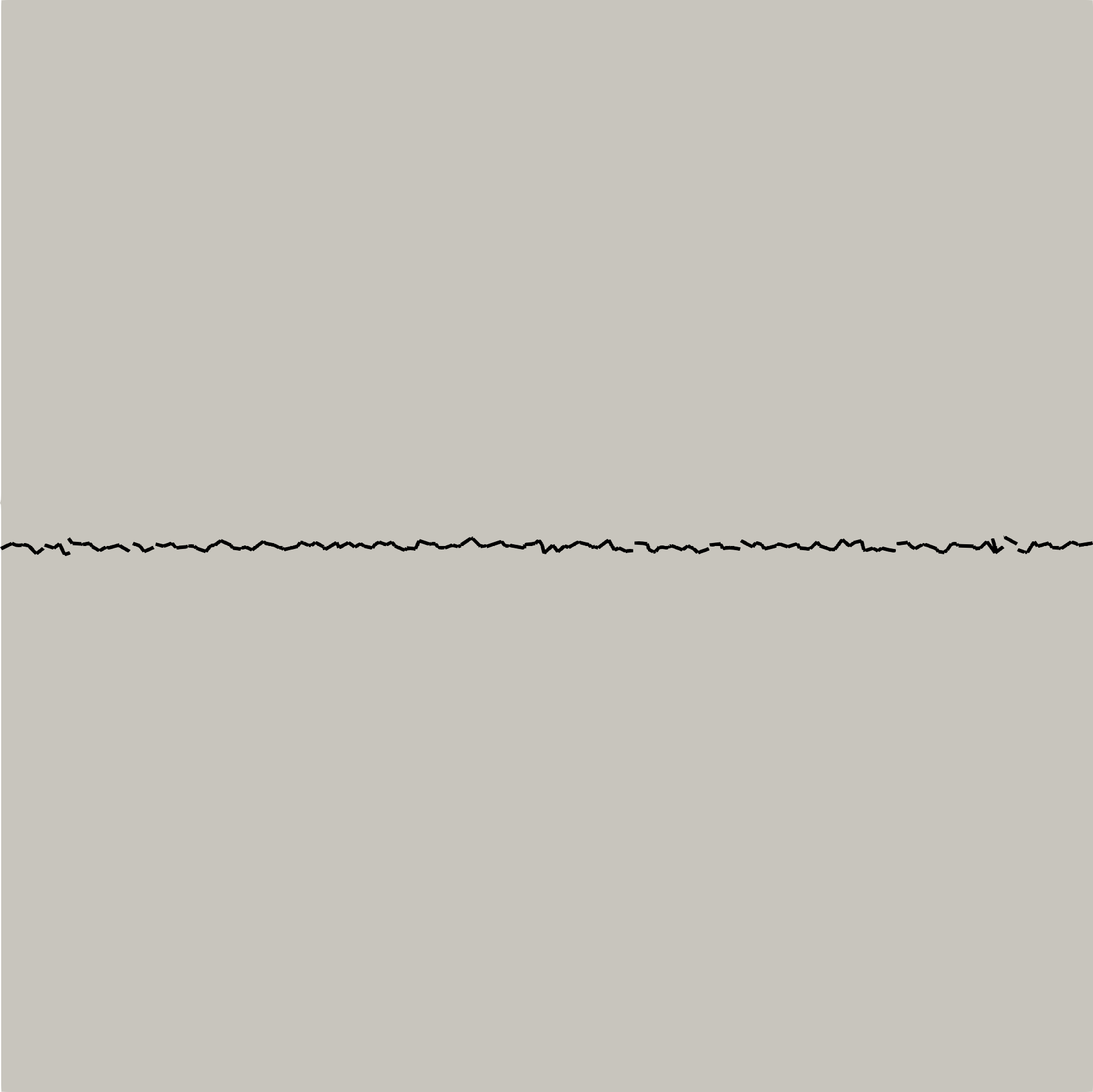}
        \vspace{0.2cm}
      \end{minipage}
                              &
      \begin{equation*} \scriptsize{\begin{bmatrix} 37.2 & 3.7 & -0.3\\ 3.7 & 17.0 & -1.6\\ -0.3 & -1.6 & 24.0 \end{bmatrix}} \end{equation*}        
                              &
      \begin{equation*} \scriptsize{\begin{bmatrix} 37.3 & 3.6 & -0.7\\ 3.6 & 17.1 & -1.4\\ -0.7 & -1.4 & 24.2 \end{bmatrix}} \end{equation*}    
                              &
      \begin{equation*} \scriptsize{\begin{bmatrix} 36.6 & 2.6 & -0.7\\ 2.6 & 12.0 & -2.1\\ -0.7 & -2.1 & 21.4 \end{bmatrix}} \end{equation*}     
                              &
      \begin{equation*} \scriptsize{\begin{bmatrix} 36.9 & 2.5 & -0.4\\ 2.5 & 11.9 & -2.1\\ -0.4 & -2.1 & 21.4 \end{bmatrix}} \end{equation*}    
      \\
      \hline
      \begin{minipage}{.2\textwidth}
        \centering
        \vspace{0.2cm}
        Diffused cracking  \\
        Low level of damage \\
        \vspace{0.2cm}
        \includegraphics[width=0.5\textwidth]{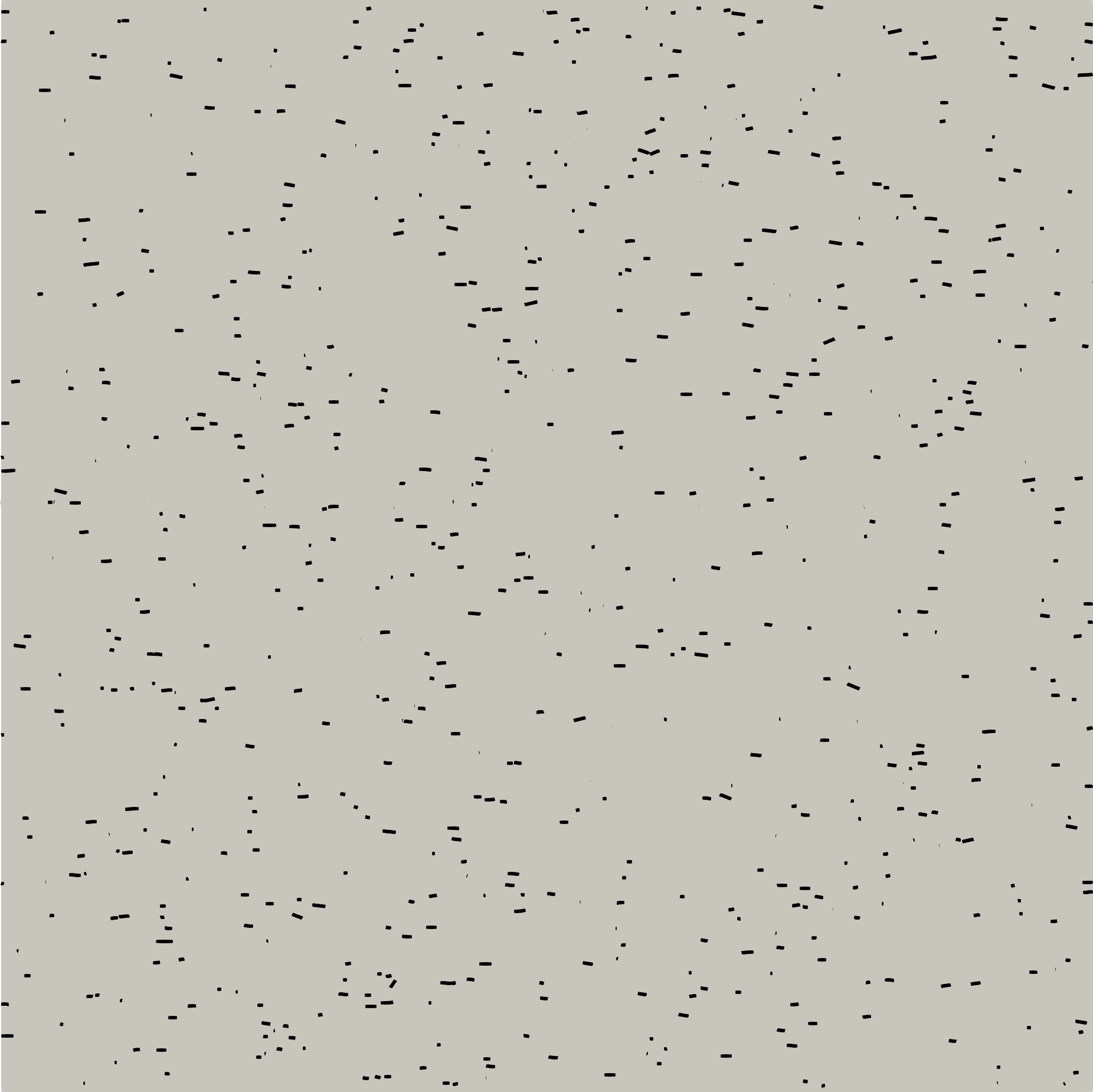}
        \vspace{0.2cm}
      \end{minipage}
                              &
      \begin{equation*} \scriptsize{\begin{bmatrix} 38.2 & 8.0 & 0.1 \\ 8.0 & 34.4 & -0.3\\ 0.1 & -0.3 & 28.4 \end{bmatrix}} \end{equation*}       
                              &
      \begin{equation*} \scriptsize{\begin{bmatrix} 38.3 & 7.9 & -0.3 \\ 7.9 & 34.5 & -0.1\\ -0.3 & -0.1 & 28.6 \end{bmatrix}} \end{equation*}      
                              &
      \begin{equation*} \scriptsize{\begin{bmatrix} 37.8 & 8.0 & 0.1 \\ 8.0 & 34.2 & 0.0\\ 0.1 & 0.0 & 28.4 \end{bmatrix}} \end{equation*}   
                              &
      \begin{equation*} \scriptsize{\begin{bmatrix} 38.1 & 8.0 & 0.1 \\ 8.0 & 34.1 & -0.3\\ 0.1 & -0.3 & 28.2 \end{bmatrix}} \end{equation*}     
      \\
      \hline
      \begin{minipage}{.2\textwidth}
        \centering
        \vspace{0.2cm}
        Diffused cracking  \\
        Mid level of damage \\
        \vspace{0.2cm}
        \includegraphics[width=0.5\textwidth]{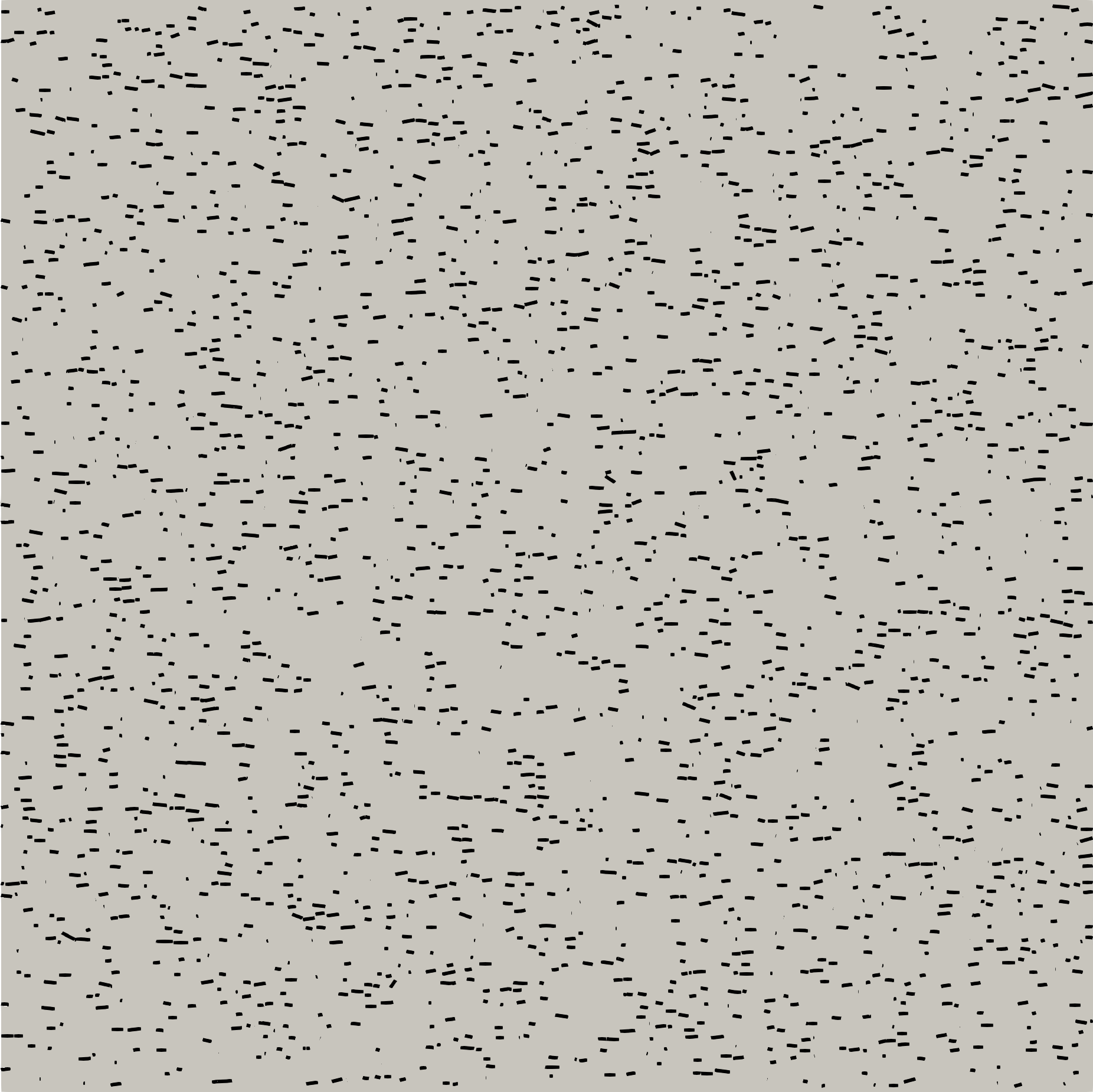}
        \vspace{0.2cm}
      \end{minipage}
                              &
      \begin{equation*} \scriptsize{\begin{bmatrix} 37.7 & 7.4 & 0.1 \\ 7.4 & 23.5 & -0.3\\ 0.1 & -0.3 & 24.6 \end{bmatrix}} \end{equation*}         
                              &
      \begin{equation*} \scriptsize{\begin{bmatrix} 37.8 & 7.3 & -0.3 \\ 7.3 & 23.6 & -0.1\\ -0.3 & -0.1 & 24.8 \end{bmatrix}} \end{equation*}       
                              &
      \begin{equation*} \scriptsize{\begin{bmatrix} 37.2 & 7.3 & 0.1 \\ 7.3 & 23.0 & -0.1\\ 0.1 & -0.1 & 24.6 \end{bmatrix}} \end{equation*} 
                              &
      \begin{equation*} \scriptsize{\begin{bmatrix} 37.6 & 7.3 & 0.1 \\ 7.3 & 23.0 & -0.1\\ 0.1 & -0.1 & 24.4 \end{bmatrix}} \end{equation*}  
      \\
      \hline
      \begin{minipage}{.2\textwidth}
        \centering
        \vspace{0.2cm}
        Diffused cracking  \\
        High level of damage \\
        \vspace{0.2cm}
        \includegraphics[width=0.5\textwidth]{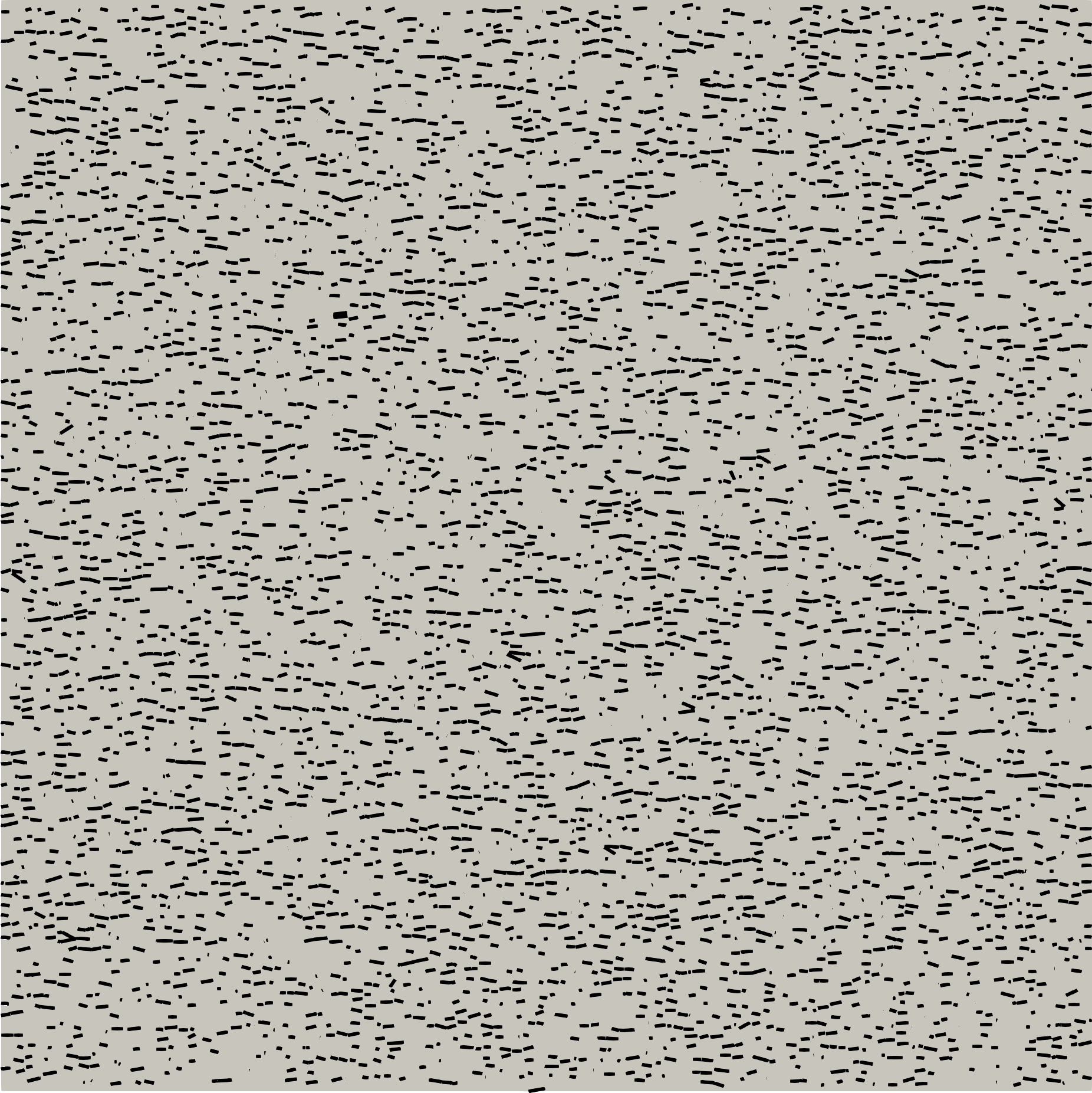}
        \vspace{0.2cm}
      \end{minipage}
                              &
      \begin{equation*} \scriptsize{\begin{bmatrix} 36.2 & 6.3 & 0.1 \\ 6.3 & 12.5 & 0.1\\ 0.1 & 0.1 & 18.6 \end{bmatrix}} \end{equation*}               
                              &
      \begin{center}
        not converged
      \end{center}
                              &
      \begin{equation*} \scriptsize{\begin{bmatrix} 35.6 & 5.8 & 0.1 \\ 5.8 & 11.7 & 0.1\\ 0.1 & 0.1 & 18.2 \end{bmatrix}} \end{equation*}     
                              &
      \begin{equation*} \scriptsize{\begin{bmatrix} 36.2 & 6.1 & 0.3 \\ 6.1 & 12.0 & 0.1\\ 0.3 & 0.1 & 18.2 \end{bmatrix}} \end{equation*}    
      \\
      \hline
    \end{tabular}
    \caption{Comparison of the measurement loadings used to extract the effective elasticity tensor $[\bC]$ (in GPa, Kelvin representation).}
    \label{tab:mes_loads}
  \end{center}
\end{table*}

Measurements of the effective elasticity tensors of cracked specimens are a more important feature of the present study. For this purpose, two cracking tests are carried out: a localized cracking test and a diffuse cracking test. In both cases, the orientation of the loads is such that the micro-cracks are globally perpendicular to the $y$ direction. For each type of cracking, we extract the crack patterns and the associated effective elasticity tensors for three levels of damage $D$ (computed as a relative loss of stiffness in the $y$ direction, \cite{LC1985}): a low level of damage ($D \approx 0.1$), a mid level of damage ($D \approx 0.4$) and a high level of damage ($D \approx 0.7$) as presented in table~\ref{tab:mes_loads}. The goal here is then to compare the four measurement loadings in order to select one of them for the rest of the study.

Regardless of the type or amount of cracking, the two strain loads (DEF\_WoC and DEF\_BT\_C) give similar results. This is expected and confirms that it is sufficient to deactivate the contact within the discrete model in order to obtain a measurement loading equivalent to a loading keeping the cracks open. Since the strain loading with additional bi-tension (DEF\_BT\_C) can cause convergence problems, it is important to be able to validate this equivalence with the elementary strain loading where the contact is deactivated (DEF\_WoC).

In the same way, we can observe that the two elementary stress loadings (SIG\_F\_WoC and SIG\_D\_WoC) give identical results, up to the measurement noise, whether forces or displacements are applied on the boundaries. Although this was not the case here, stress loading with applied forces may prove to be less robust because the blocking of the central particle, to prevent rigid body movements, can lead to convergence problems if this particle is in an area of significant cracking.

\begin{figure}[htp]
  \centering
  \includegraphics[align=c,width=0.8\linewidth]{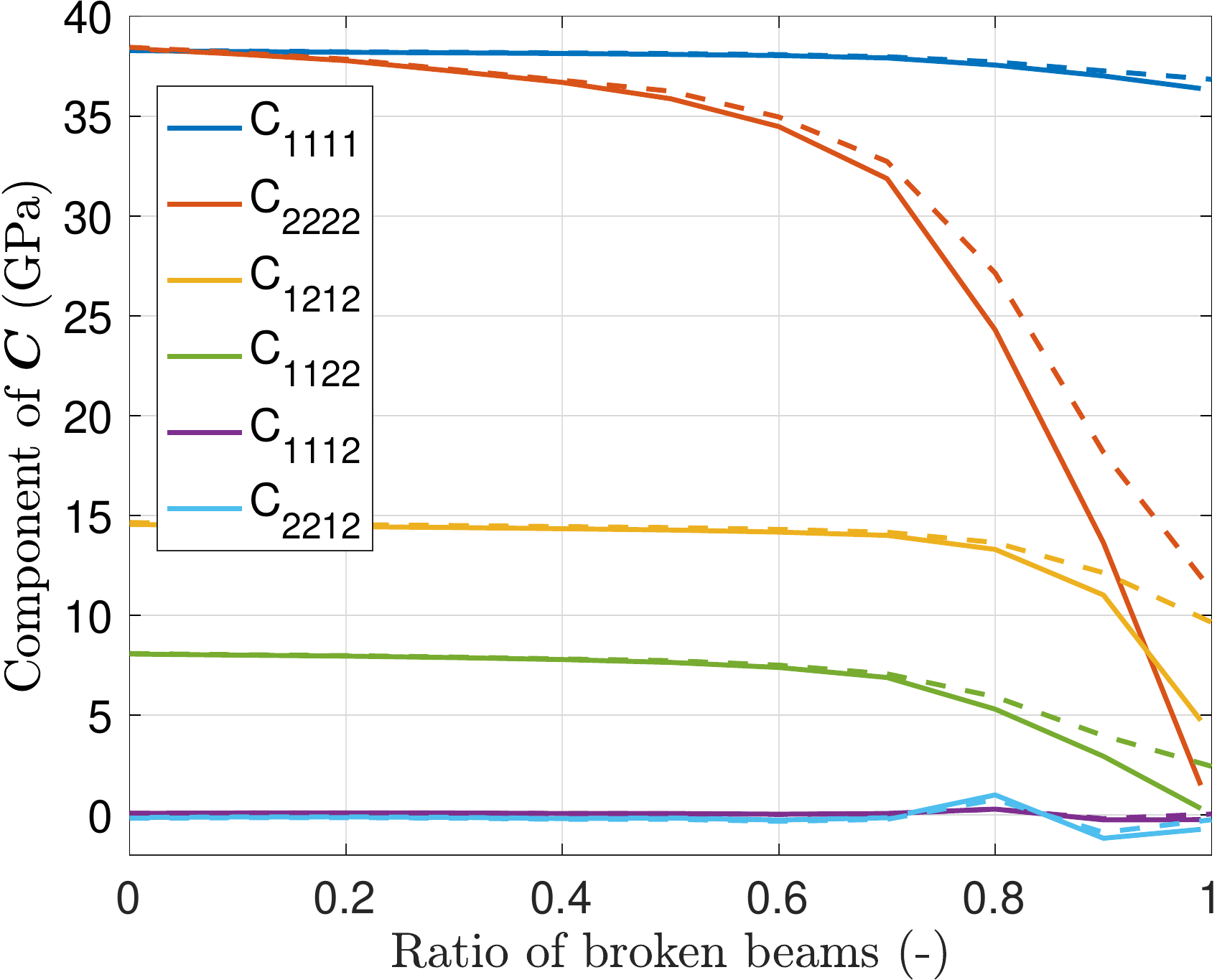}
  \caption{Evolution of the components of the effective elasticity tensor for the localized cracking test according to the elementary strain loading measurement (DEF\_WoC in dashed lines) and the elementary stress loading measurement with applied displacements (SIG\_D\_WoC in plain lines)}
  \label{fig:evol_telas_mes_sigu_def}
\end{figure}

Finally, strain and stress loadings give identical results in the case of diffuse cracking even for a significant level of damage. On the other hand, the results diverge for the case of localized cracking (see figure~\ref{fig:evol_telas_mes_sigu_def}). Indeed, if we pursue the simulation until the complete failure of the specimen in the localized case, we notice that the $C_{2222}$ modulus approaches 0 for the measurement loading in stress, which is not the case for the measurement loading in strain. Here, we encounter a limitation in the use of damage models for quasi-brittle materials such as concrete. Indeed, the damage is intended to represent diffuse cracking and is thus not appropriate for cases of localized cracking. However, damage models are commonly used up to high levels of degradation, where the cracks are almost necessarily localized in a quasi-brittle material. In order to correctly represent the loss of stiffness due to this localized cracking while keeping the simplicity of a kinematic loading, we decide to choose thereafter the elementary stress loading with applied displacements without contact (SIG\_D\_WoC).

\section{Deviation from isotropy of the initial tensor}
\label{sec:isotropy}

\subsection{Geometrical isotropy}

\begin{figure*}[htb]
  \begin{subfigure}{.33\textwidth}
    \centering
    \includegraphics[width=.8\linewidth]{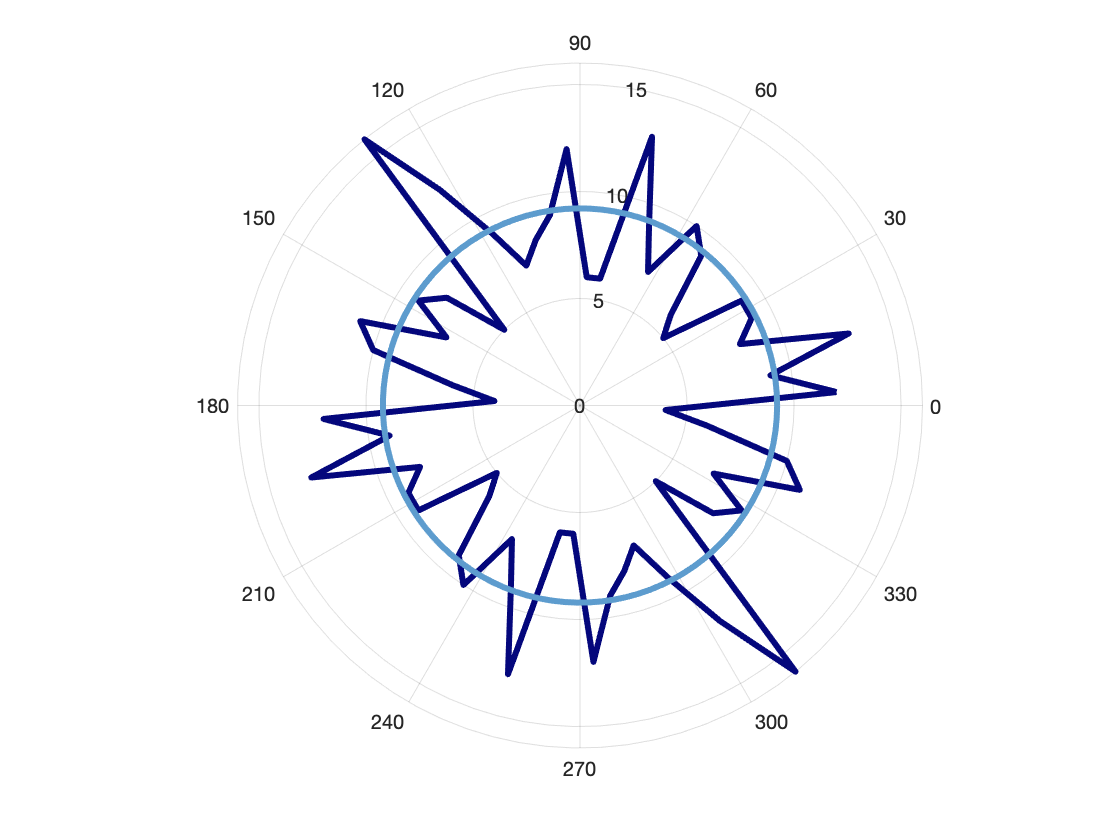}

    \begin{equation*} [\bC] = \left[ \begin{matrix}  43.13 &  8.42 &  -0.81\\    8.42  & 42.65  &  0.48\\  -0.81 &   0.48  & 30.62 \end{matrix}    \right]   \end{equation*} 

    \begin{equation*} [\bC_{iso}] = \left[ \begin{matrix} 41.92 & 9.39 & 0.00\\ 9.39 & 41.92 & 0.00\\ 0.00 & 0.00 & 32.54 \end{matrix}     \right] \end{equation*} 

    $\kappa = 25.65$ GPa, $\mu = 16.27$ Gpa\\
    $E=38.82$ GPa, $\nu=0.22$

    \caption{$10 \times 10$}
    \label{fig:iso_orientation_10}
  \end{subfigure}
  \begin{subfigure}{.33\textwidth}
    \centering
    \includegraphics[width=.8\linewidth]{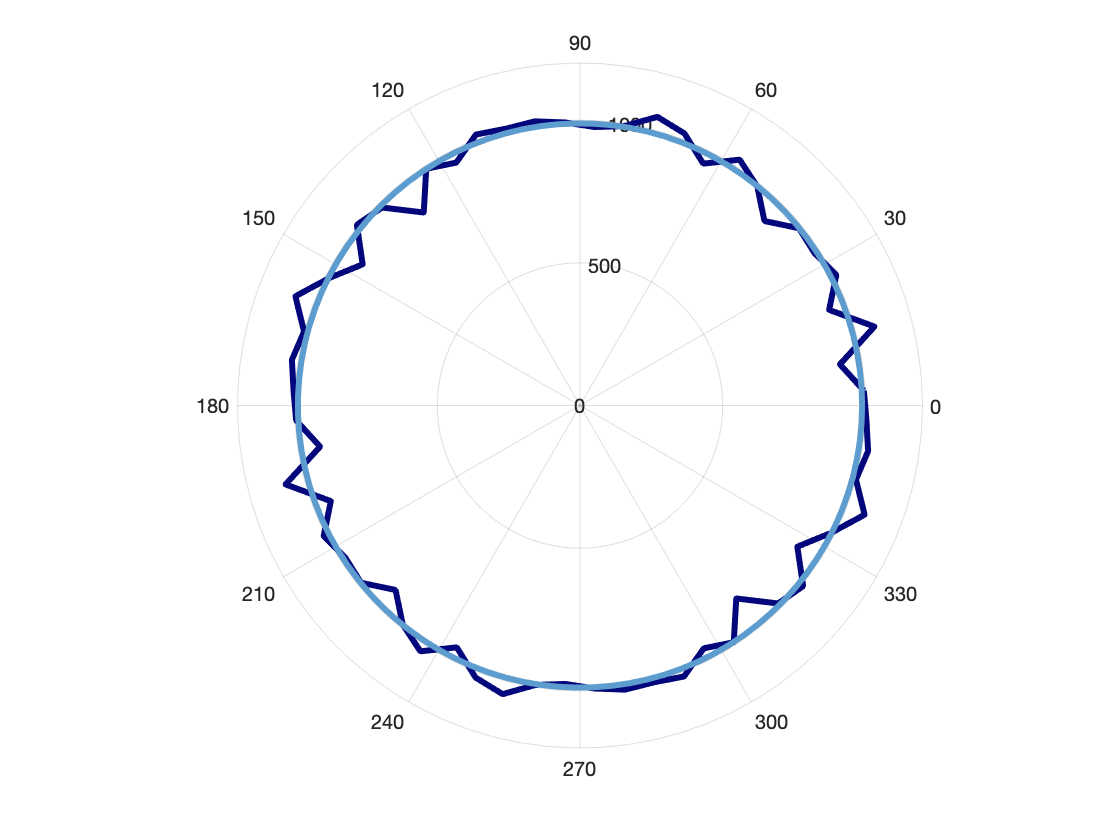}

    \begin{equation*} [\bC] = \left[ \begin{matrix}  38.72 &    7.90 &   -0.24\\    7.90 &   38.44 &  -0.10\\  -0.24 &  -0.10 &  29.00 \end{matrix}    \right] \end{equation*} 

    \begin{equation*} [\bC_{iso}] = \left[ \begin{matrix} 38.16 & 8.32 & 0.00\\ 8.32 & 38.16 & 0.00\\ 0.00 & 0.00 & 29.84 \end{matrix}    \right] \end{equation*} 

    $\kappa = 23.24$ GPa, $\mu = 14.92$ Gpa\\
    $E=36.35$ GPa, $\nu=0.22$

    \caption{$100 \times 100$}
    \label{fig:iso_orientation_100}
  \end{subfigure}
  \begin{subfigure}{.33\textwidth}
    \centering
    \includegraphics[width=.8\linewidth]{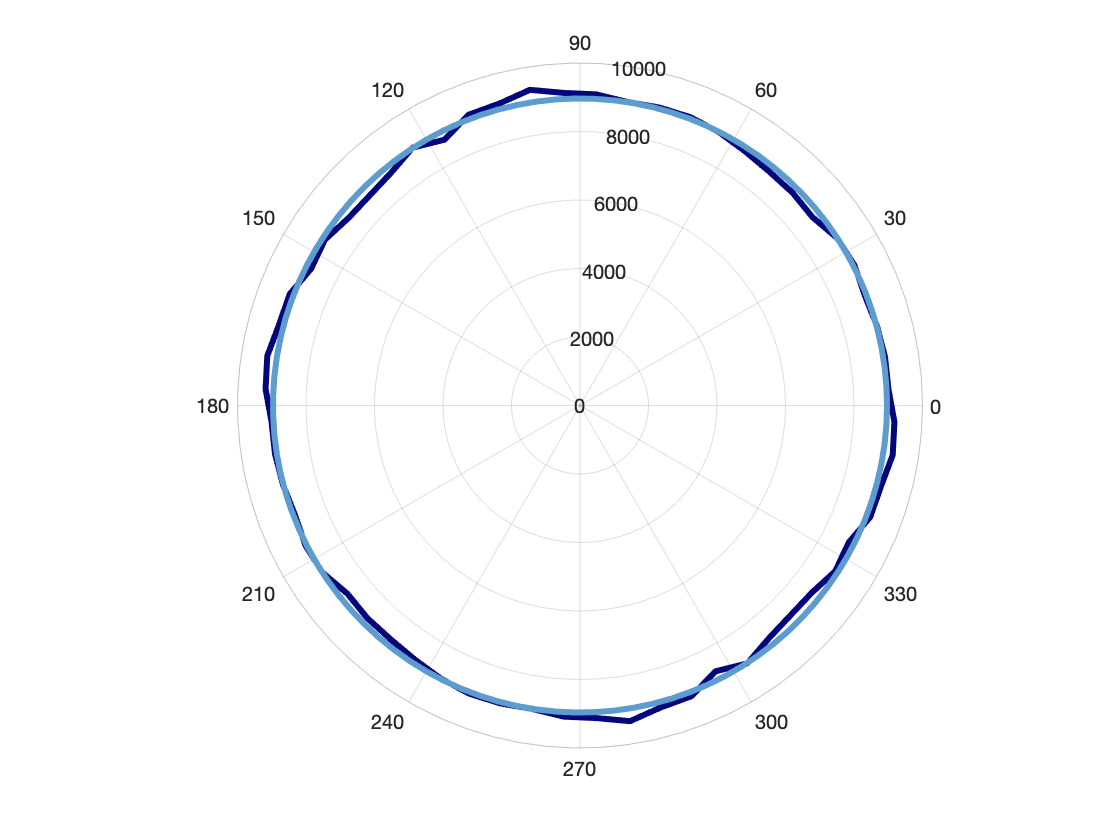}

    \begin{equation*} [\bC] = \left[ \begin{matrix}  38.06  &  8.09  &  0.00\\    8.09 &  38.07 &  -0.03\\    0.00 &  -0.03  & 29.00 \end{matrix}     \right] \end{equation*} 

    \begin{equation*} [\bC_{iso}] = \left[ \begin{matrix}    37.82 & 8.34 & 0.00\\    8.34 & 37.82 & 0.00\\    0.00 & 0.00 & 29.48 \end{matrix}    \right] \end{equation*} 

    $\kappa = 23.08$ GPa, $\mu = 14.74$ Gpa\\
    $E=35.98$ GPa, $\nu=0.22$

    \caption{$300 \times 300$}
    \label{fig:iso_orientation_300}
  \end{subfigure}
  \caption{Polar histograms of the beams orientation for different mesh densities, elasticity tensors and associated isotropic tensors (in GPa, Kelvin representation).}
  \label{fig:iso_orientation}
\end{figure*}

The isotropy of discrete media is generally studied from a geometric point of view \citep{andre2012discrete}. Indeed, the geometrical isotropy of the mesh is considered as a good criterion to approach the mechanical isotropy of the material if the elasticity parameters are homogeneous over the domain. This geometric isotropy can be visualized simply by plotting the polar histogram associated with the orientation of the beams. The figure~\ref{fig:iso_orientation} presents those polar histograms for three meshes of different densities. One can see that the polar histogram tends towards a circle as the density increases, which shows geometrical isotropy convergence.

\subsection{Mechanical isotropy}

Through the harmonic decomposition of the elasticity tensors of each material samples, it is possible to quantify their deviation from isotropy. The evaluation of the relative distance of the elasticity tensors to the isotropic class thus provides a criterion for quantifying the validity of the mechanical isotropy hypothesis.

\begin{figure}[htp]
  \centering
  \includegraphics[width=0.4\textwidth]{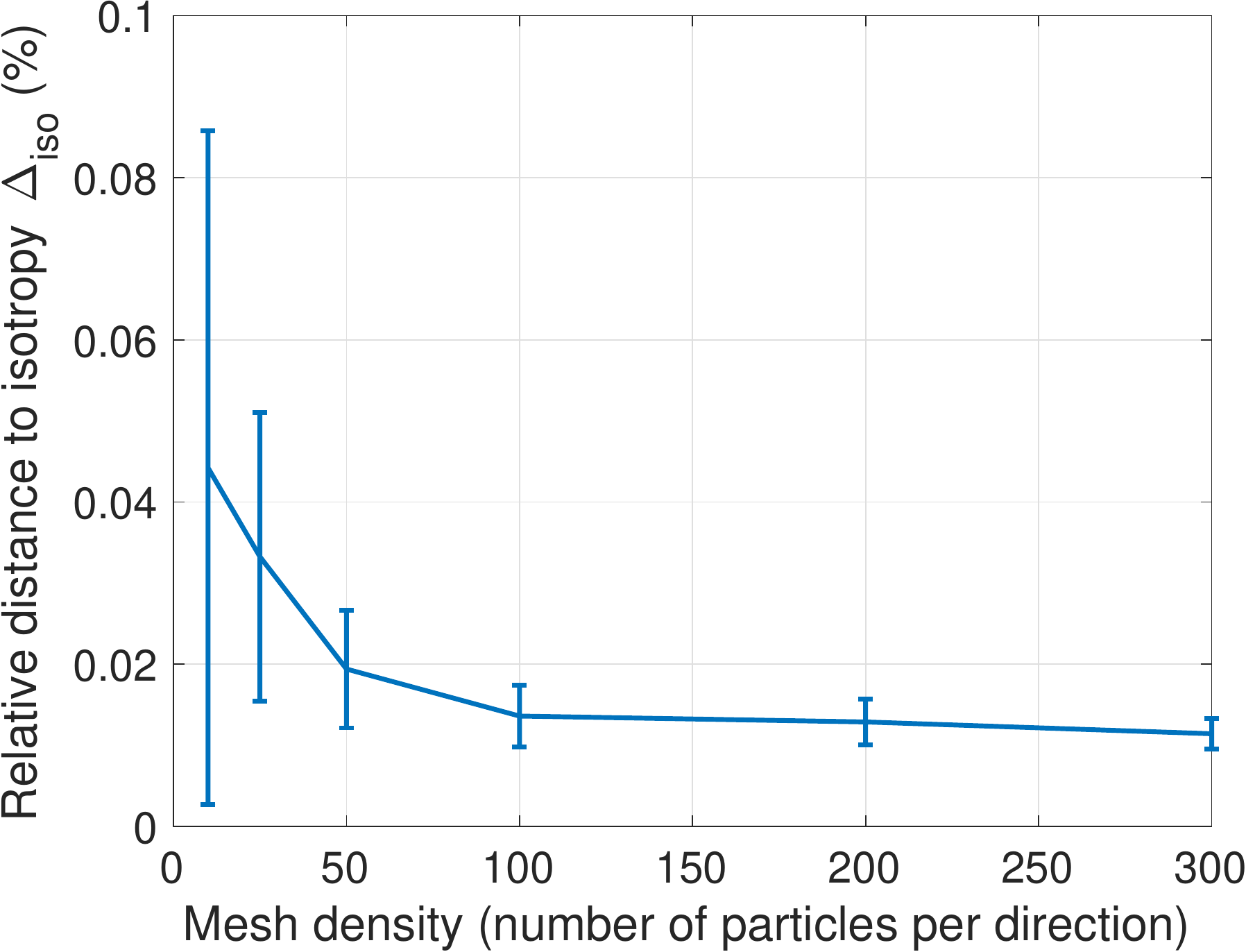}
  \caption{Evolution of the relative distance to isotropy \eqref{eq:DeltaIso} with the mesh density.}
  \label{fig:distance_isotropy_poisson_grid}
\end{figure}

Figure~\ref{fig:distance_isotropy_poisson_grid} presents the evolution with the mesh size of the relative distance to isotropy $\Delta_{iso}$ defined as :

\begin{equation}\label{eq:DeltaIso}
  \Delta_{iso} = \frac{\| \bC- \bC_{iso}\|}{\| \bC\|}
\end{equation}
with $\bC_{iso}= 2 \mu \bJ + \kappa \Idd \otimes \Idd$ the isotropic projection of $\bC$. The elasticity parameters $\mu$ and $\kappa$ are determined by Eq. \eqref{eq:HarmDecompComponents}.

For each mesh density, several simulations have been performed (from 200 simulations for the $10 \times 10$ particles mesh to 50 simulations for a $300 \times 300$ particles mesh). As expected, the assumption of an initial isotropic medium is correct if the number of particles is sufficient. In Figure~\ref{fig:iso_orientation}, the elasticity tensors obtained for three simulations are given along with the associated isotropic tensors and elasticity coefficients.

Given the above results, it seems safe to adopt a mesh density of $100 \times 100$ particles to ensure the mechanical isotropy of the initial, uncracked, state.

\section{Multiaxial analyses of effective elasticity tensors up to high level of damage}
\label{sec:telas_analysis}

In this part, a 20cm$\times$20cm specimen representative of mortar is subjected to various loads. As mentioned before, the simulation is done with $100 \times 100$ particles, which results in an average beam size $\overline{l_b}$ of 2 mm.

The parameters of the discrete model, given in the table~\ref{tab:num_param_exp}, are chosen so as to reproduce macroscopically the behaviour of a mortar of Young's modulus $E = 36.35$ GPa and Poisson ratio $\nu = 0.22$ (corresponding to $\kappa = 23.24$ GPa and $\mu = 14.92$ GPa), of tensile strength $f_t = 3$ MPa and of compressive strength $f_c = 40$ Mpa.

\begin{table}[ht]
  \begin{center}
    \begin{tabular}[c]{|c|c|c|}
      \hline
      \textbf{Symbol}           & \textbf{Values} & \textbf{Unit} \\
      \hline
      \hline
      \\[-1em]
      $\overline{l_b}$          & 2               & mm            \\
      \hline
      \hline
      \\[-1em]
      $E_b$                     & 46              & GPa           \\
      \hline
      \\[-1em]
      $\alpha$                  & 0.83            & -             \\
      \hline
      \hline
      \\[-1em]
      $\lambda_{\epsilon_{cr}}$ & 2.4 $10^{-4}$   & -             \\
      \hline
      \\[-1em]
      $\lambda_{\theta_{cr}}$   & 3.3 $10^{-3}$   & -             \\
      \hline
      \\[-1em]
      $k$                       & 2.8             & -             \\
      \hline
      \hline
      \\[-1em]
      $\tan \phi$               & 0.7             & -             \\
      \hline
    \end{tabular}
    \caption{Numerical parameters for the beam-particle model.}
    \label{tab:num_param_exp}
  \end{center}
\end{table}

\subsection{Studied loadings}

\begin{figure*}[htp]
  \centering
  \includegraphics[align=c,width=0.2\linewidth]{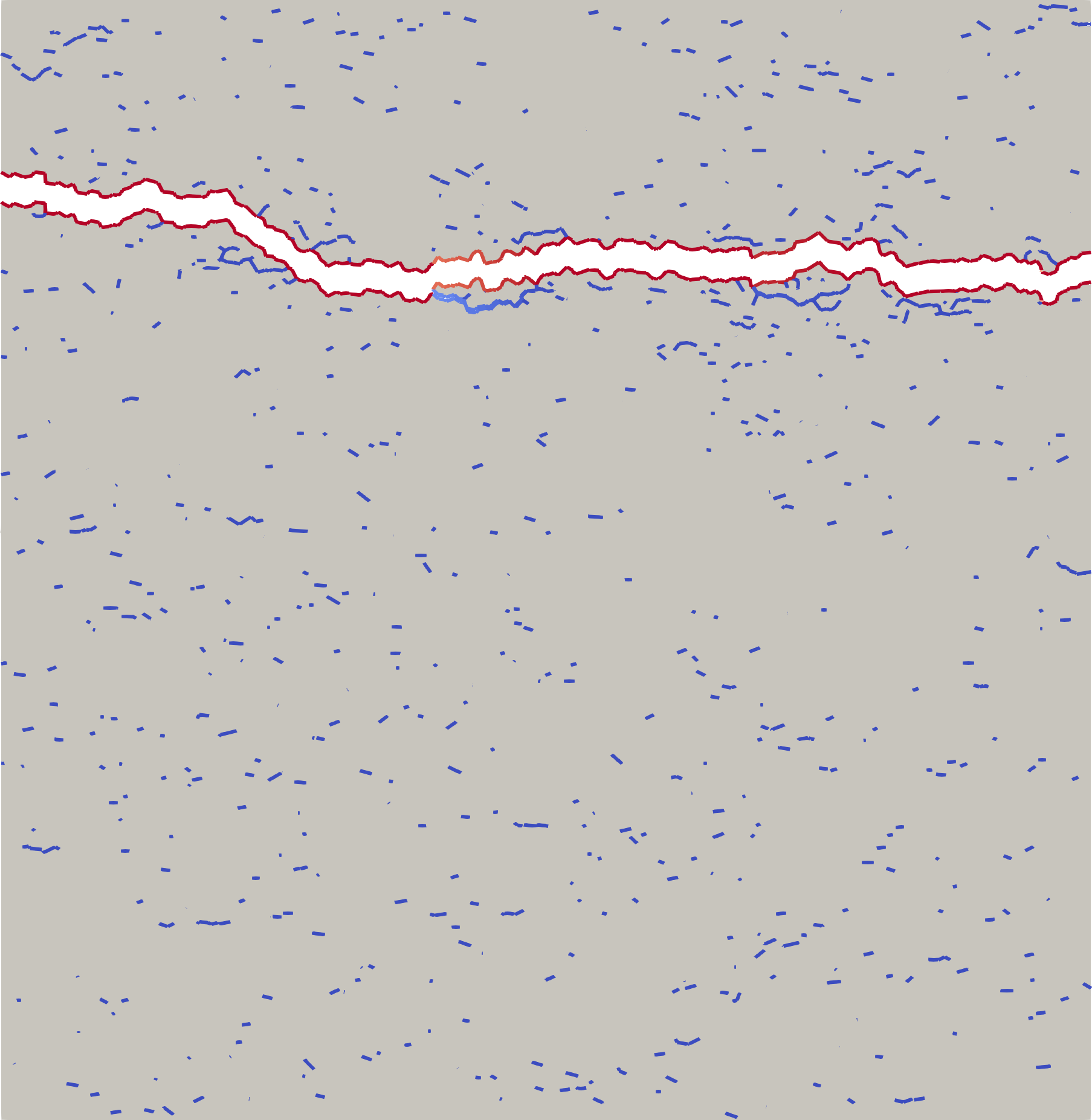}
  \hspace{1cm}
  \includegraphics[align=c,width=0.35\linewidth]{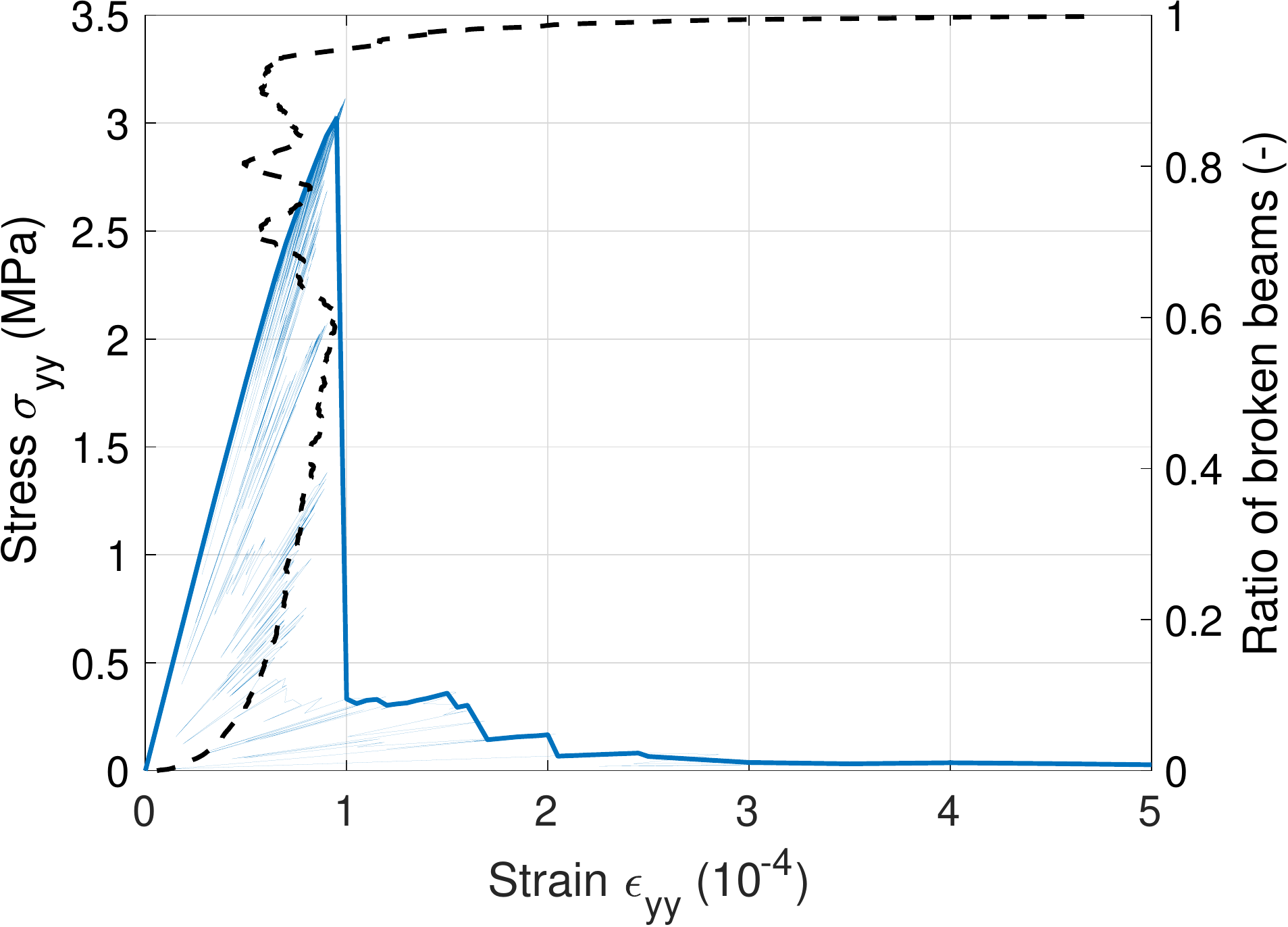}
  \caption{Tensile loading: final crack pattern, macroscopic responses (complete and smoothed) and evolution of the ratio of broken beams (dashed line).}
  \label{fig:tension}
\end{figure*}

\begin{figure*}[htp]
  \centering
  \includegraphics[align=c,width=0.2\linewidth]{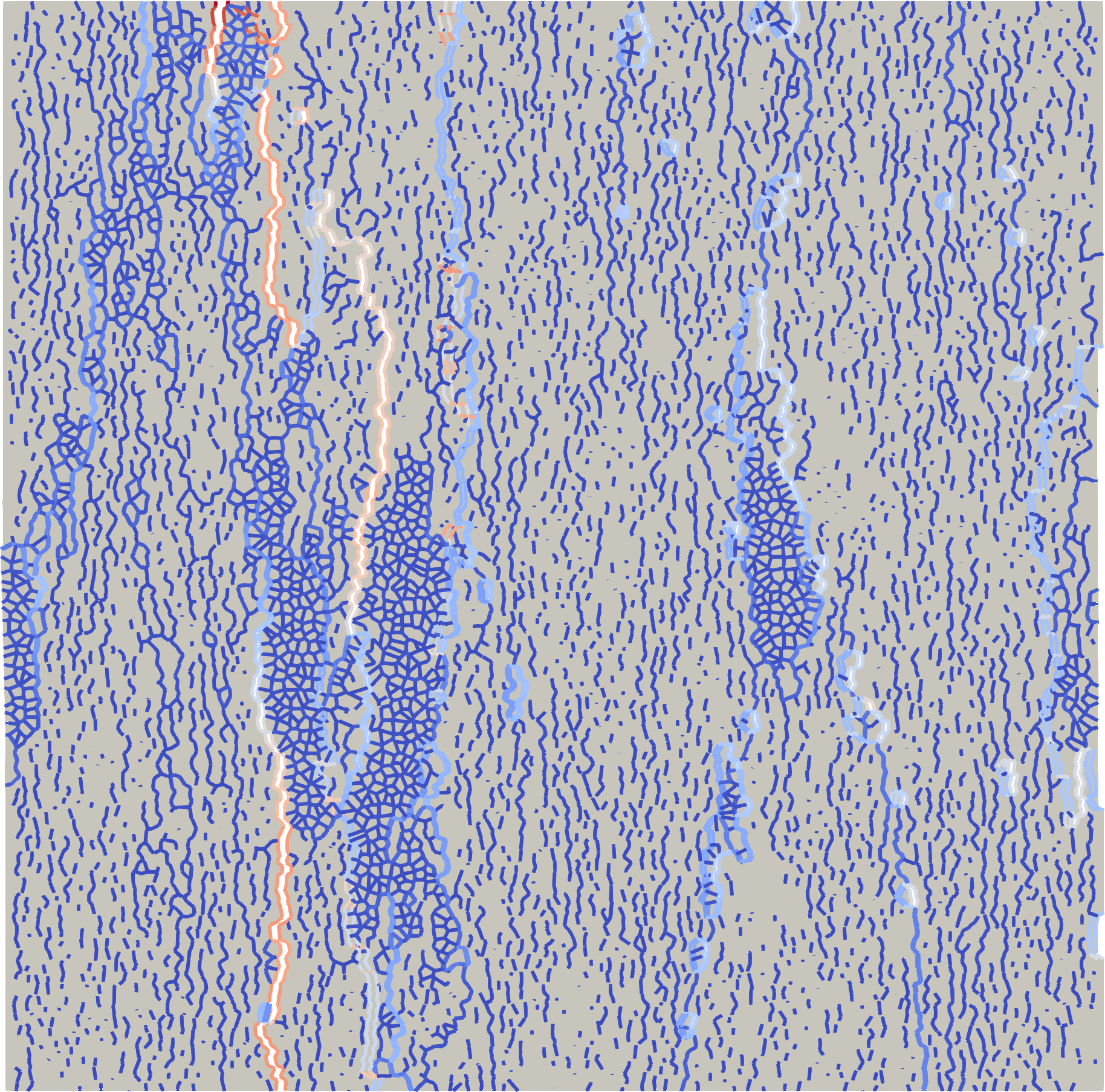}
  \hspace{1cm}
  \includegraphics[align=c,width=0.35\linewidth]{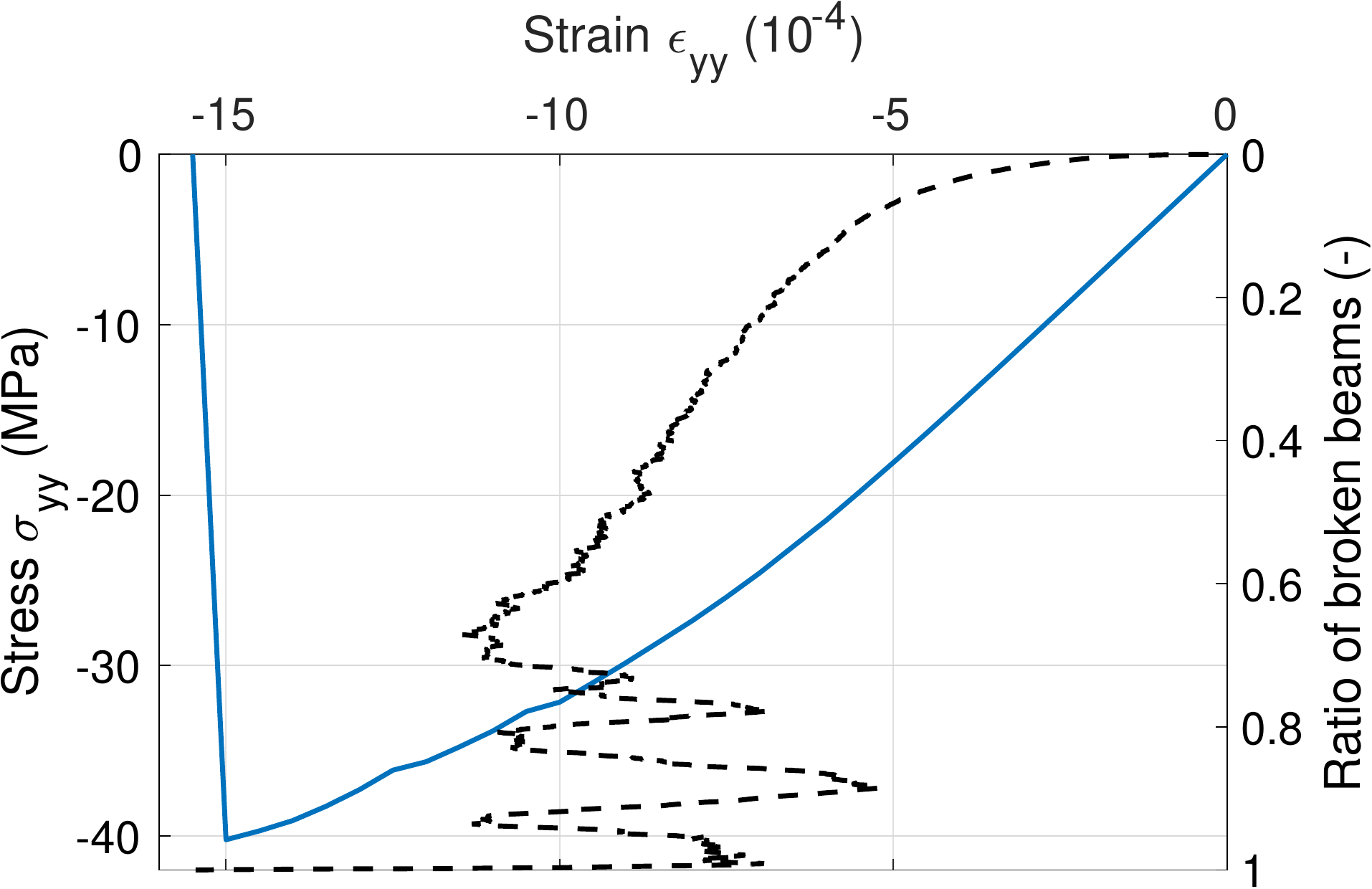}
  \caption{Compressive loading: final crack pattern, macroscopic response and evolution of the ratio of broken beams (dashed line).}
  \label{fig:compression}
\end{figure*}

\begin{figure*}[htp]
  \centering
  \includegraphics[align=c,width=0.2\linewidth]{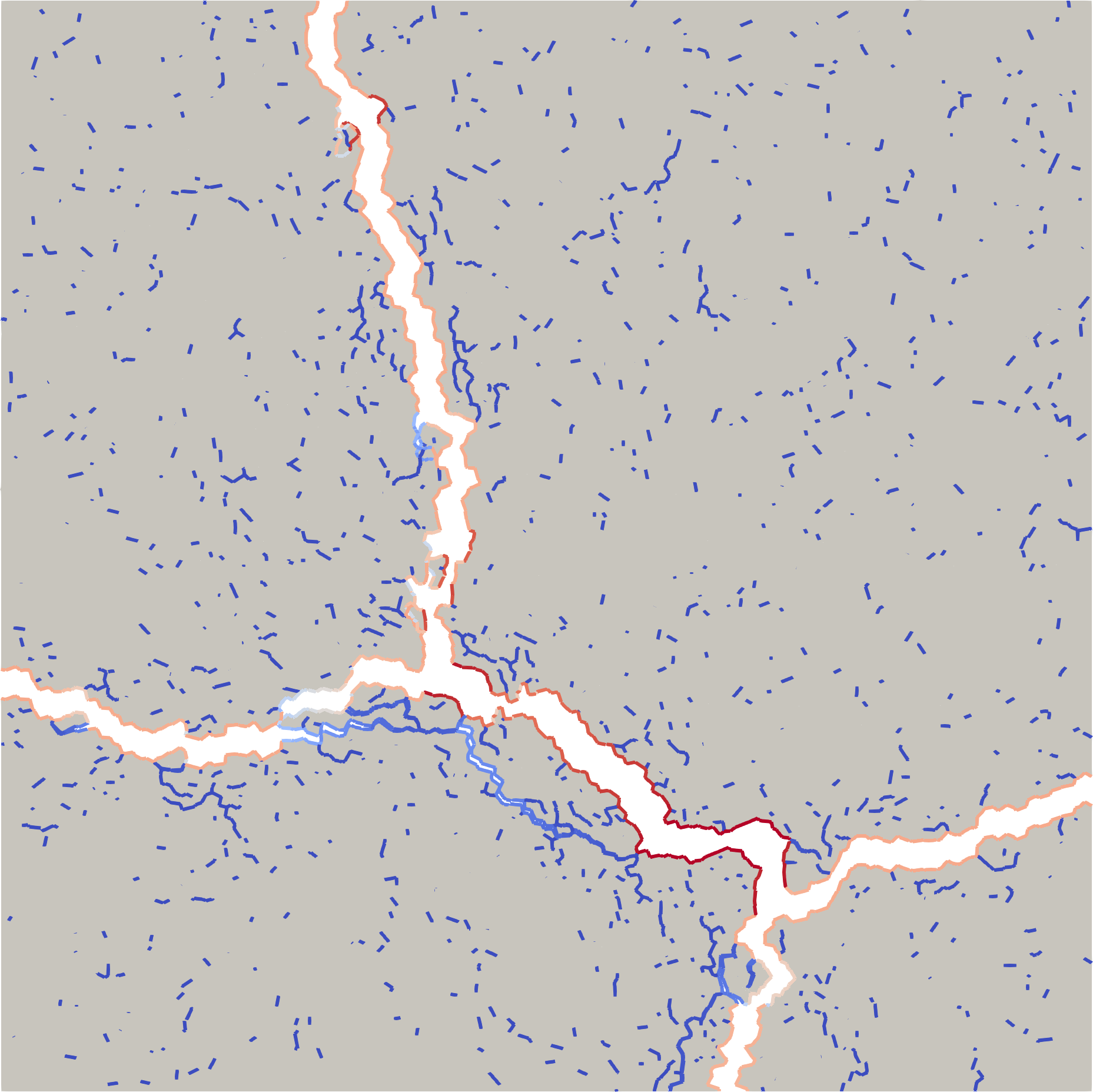}
  \hspace{1cm}
  \includegraphics[align=c,width=0.35\linewidth]{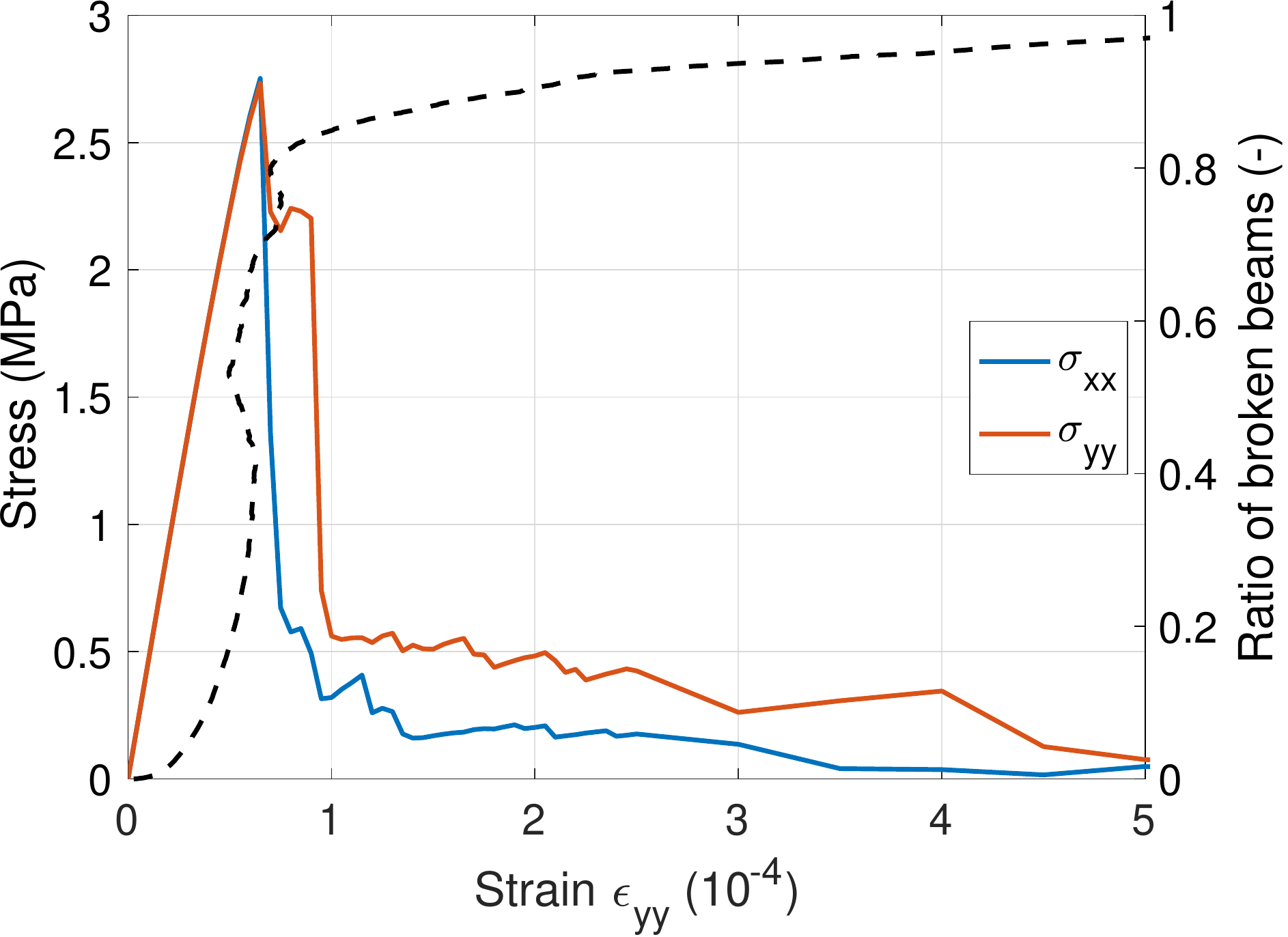}
  \caption{Bi-Tensile loading: final crack pattern, macroscopic response and evolution of the ratio of broken beams (dashed line).}
  \label{fig:bi-tension}
\end{figure*}

\begin{figure*}[htp]
  \centering
  \includegraphics[align=c,width=0.2\linewidth]{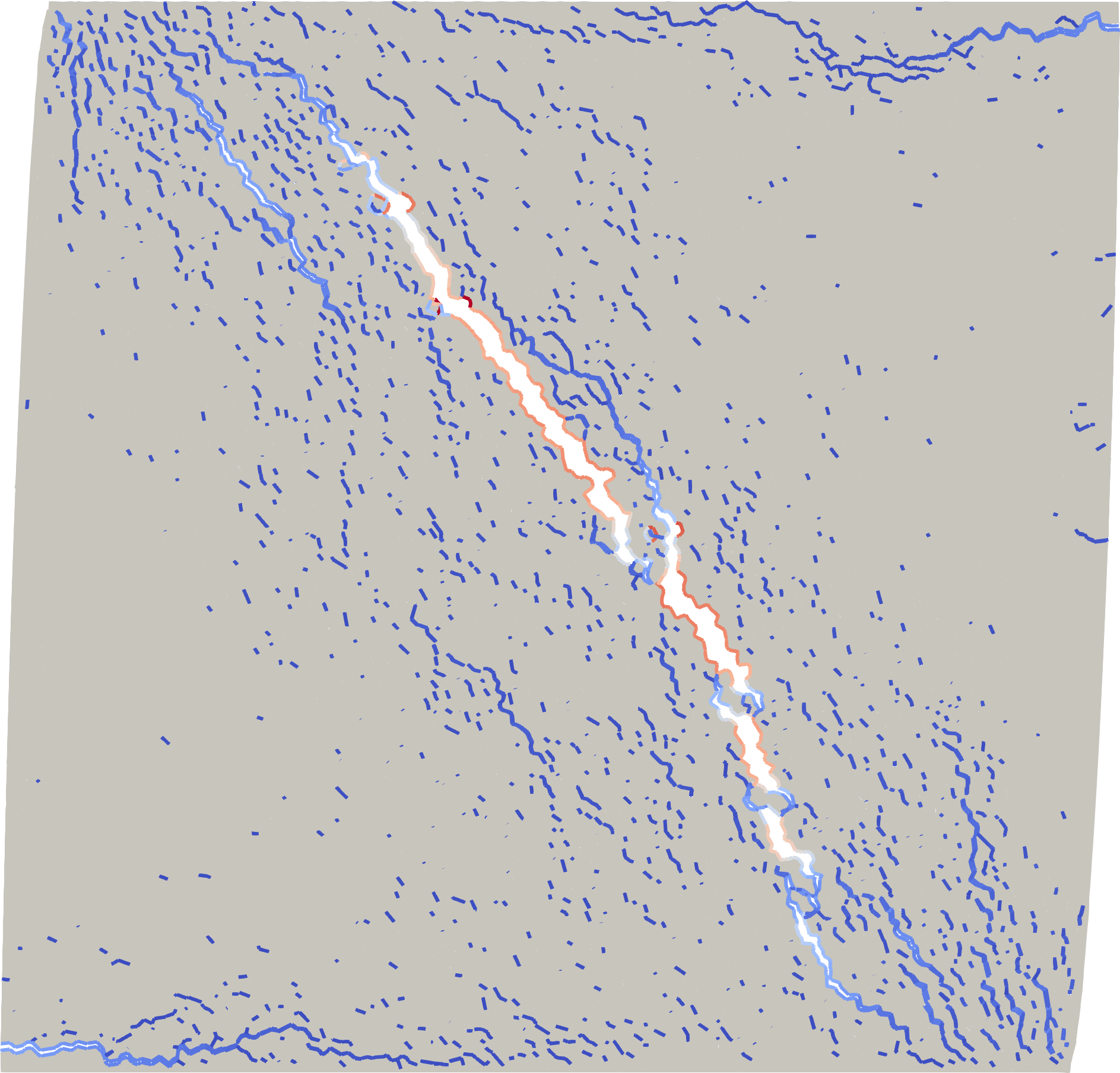}
  \hspace{1cm}
  \includegraphics[align=c,width=0.35\linewidth]{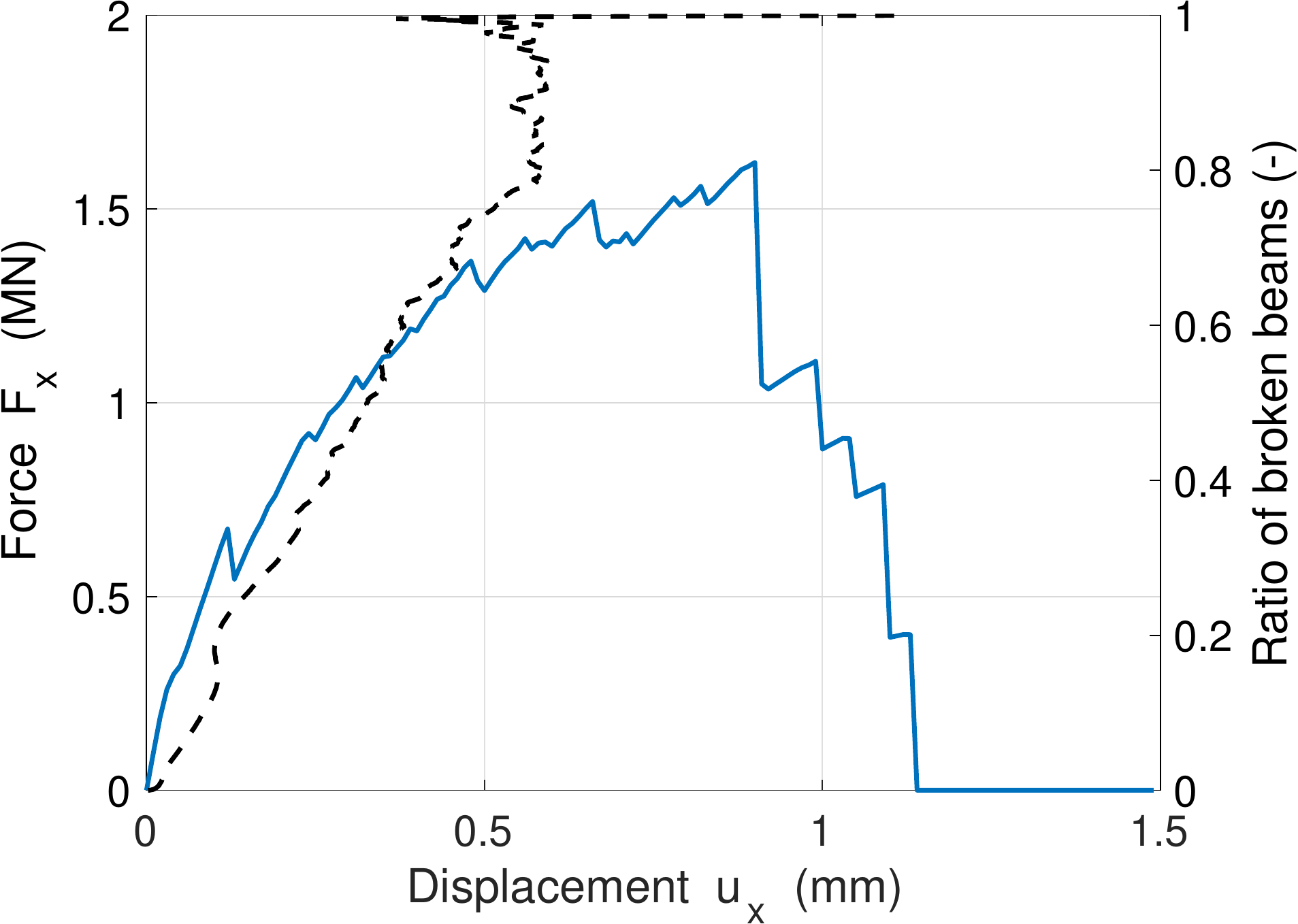}
  \caption{Simple shear loading: final crack pattern, macroscopic response and evolution of the ratio of broken beams (dashed line).}
  \label{fig:simple_shear}
\end{figure*}

\begin{figure*}[htp]
  \centering
  \includegraphics[align=c,width=0.2\linewidth]{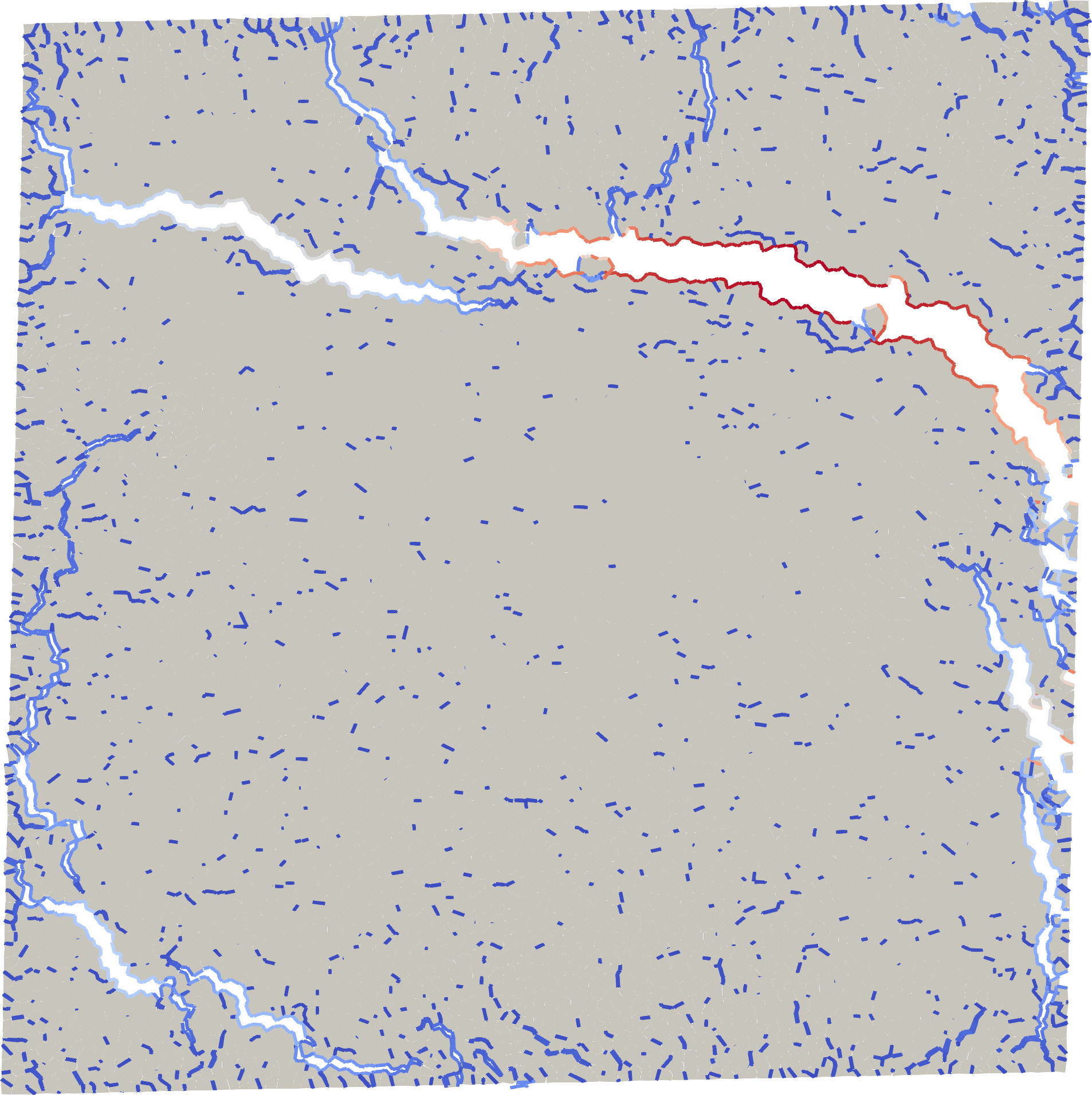}
  \hspace{1cm}
  \includegraphics[align=c,width=0.35\linewidth]{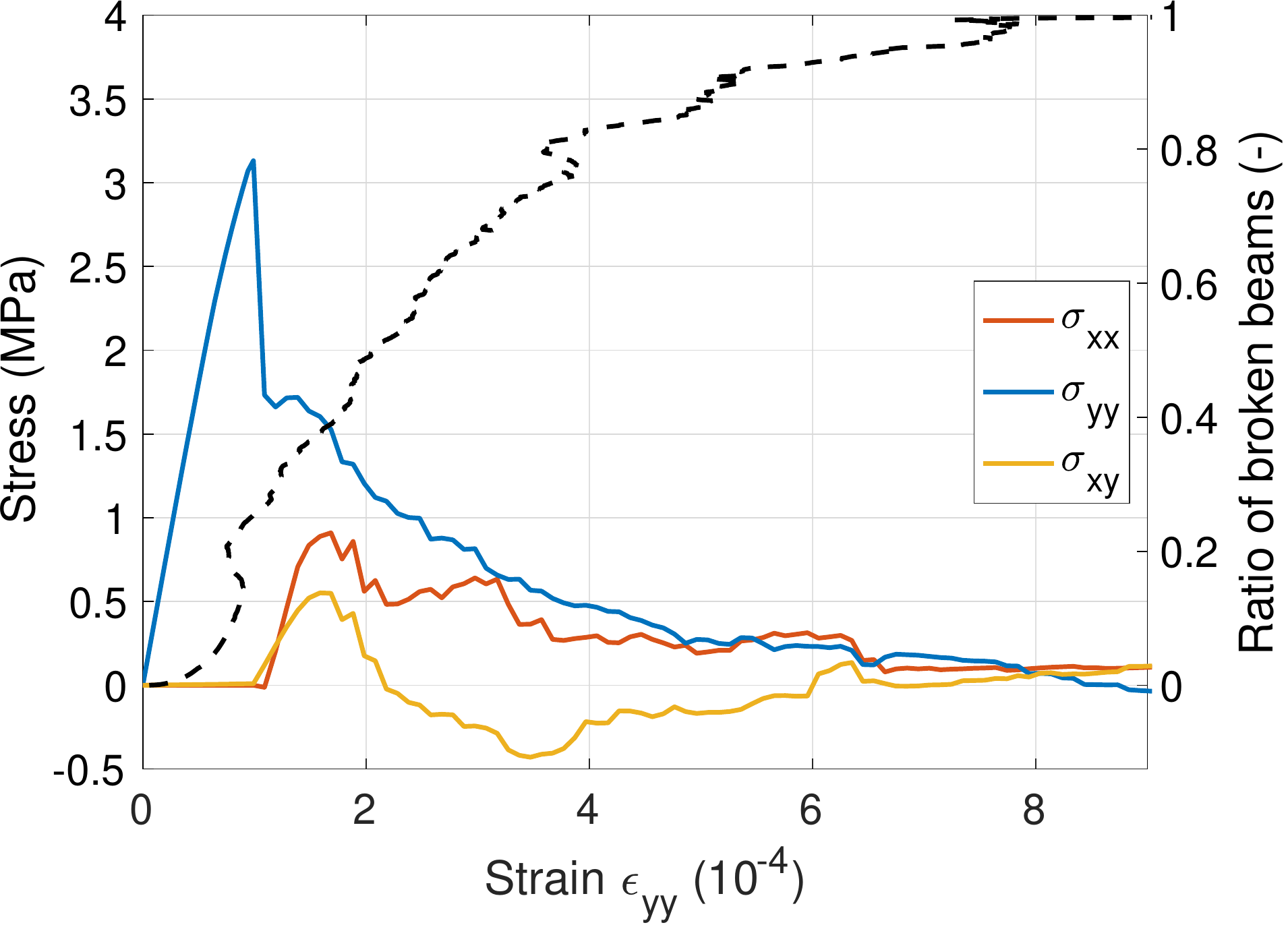}
  \caption{Willam loading: final crack pattern, macroscopic response and evolution of the ratio of broken beams (dashed line).}
  \label{fig:willam}
\end{figure*}

The loadings are chosen so as to obtain different cracking paths:
\begin{itemize}
  \item Tension along the $y$-axis (see figure~\ref{fig:tension});
  \item Compression along the $y$-axis with unrestrained boundary conditions along the $x$-axis (see figure~\ref{fig:compression});
  \item Bi-tension along the $x$-axis and $y$-axis (see figure~\ref{fig:bi-tension});
  \item Simple shear (see figure~\ref{fig:simple_shear});
  \item Willam loading \citep{willam1989fundamental} consisting on a simple tension along the $y$-axis pursued up to the tensile stress followed by a combination of bi-tension and shear with an increase of the strain components $\epsilon_{xx}$, $\epsilon_{yy}$ and $\epsilon_{xy}$ in the proportions 1.5/1.0/0.5 (see figure~\ref{fig:willam}). This numerical test was proposed to induce a rotation of the principal stress/strain directions leading to a misalignment with the material axes of orthotropy associated with the initial crack direction.
\end{itemize}

For each of them, we first observe a development of diffuse micro-cracks and then a localization of these micro-cracks leading to the propagation of the macro-crack.

In order to follow the breakage of the specimen, the evolution of the ratio of broken beams is plotted in comparison with the macroscopic response. This ratio is initially worth $0$, when the specimen is intact, and reaches $1$ at the end of loading when the specimen can no longer sustain any effort. The total number of broken beams is usually far from the number of beams in the specimen.

We can notice snap-back phenomena in the evolution curves of the ratio of broken beams. These snap-backs are related to avalanche breakage phenomena \citep{rinaldi2007statistical}, classical for lattice simulations of quasi-brittle materials, and are visible in the complete macroscopic responses (see figure~\ref{fig:tension}). These macroscopic responses are smoothed on the figures~\ref{fig:tension} to~\ref{fig:willam}  to get closer to the type of data observed experimentally. However, the extraction of the effective elasticity tensors is not impacted by this post-treatment of the curves.

These loadings allow us to quantify the impact on the effective (damaged) elasticity tensor of the presence of micro-cracks in different directions, which may nucleate or rotate.

\subsection{Orthotropy of the effective stiffness tensors}

\begin{figure}[htp]
  \centering
  \includegraphics[align=c,width=0.8\linewidth]{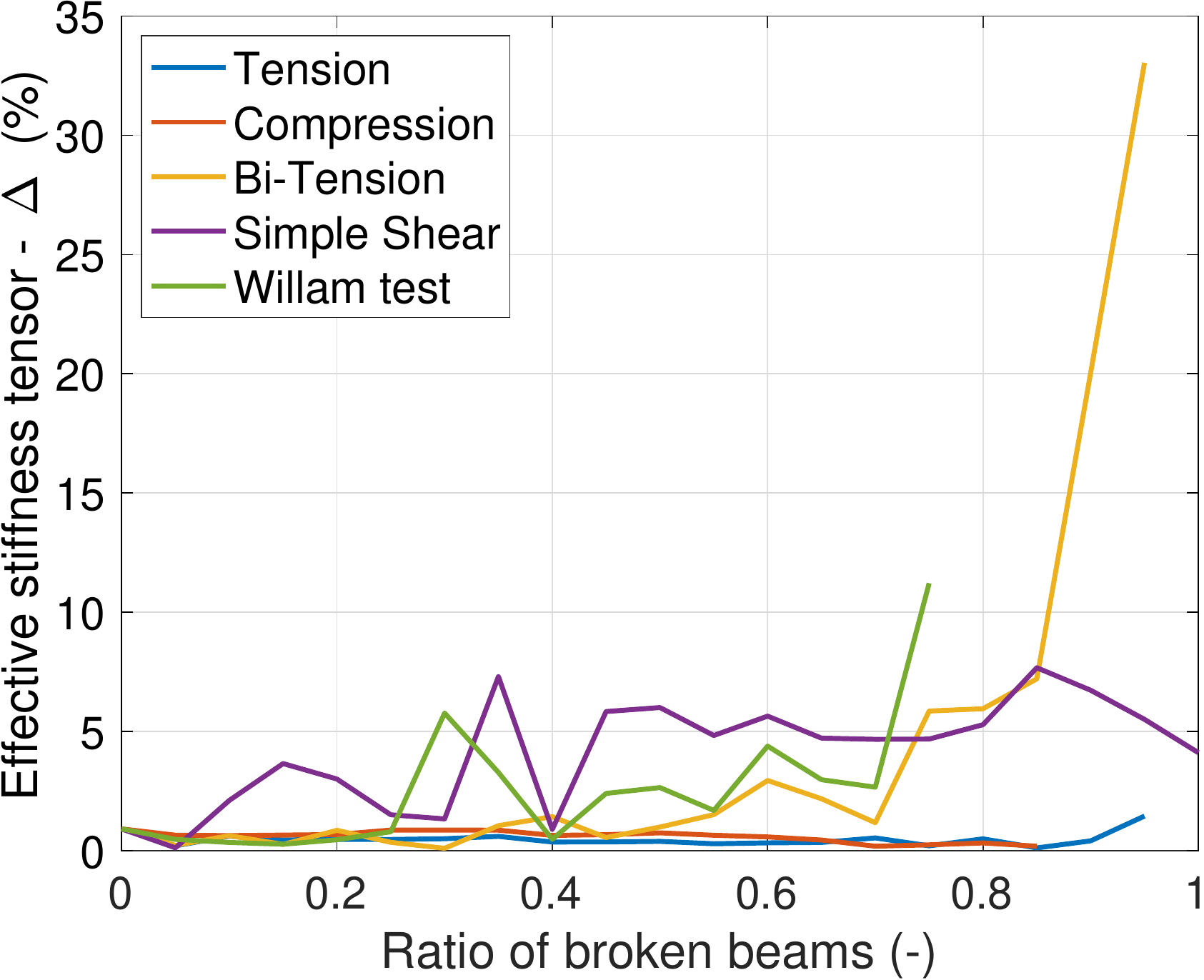}
  \caption{Upper-bound of distance to orthotropy for the effective stiffness tensors}
  \label{fig:stiff_orth}
\end{figure}

\begin{figure}[htp]
  \centering
  \includegraphics[align=c,width=0.8\linewidth]{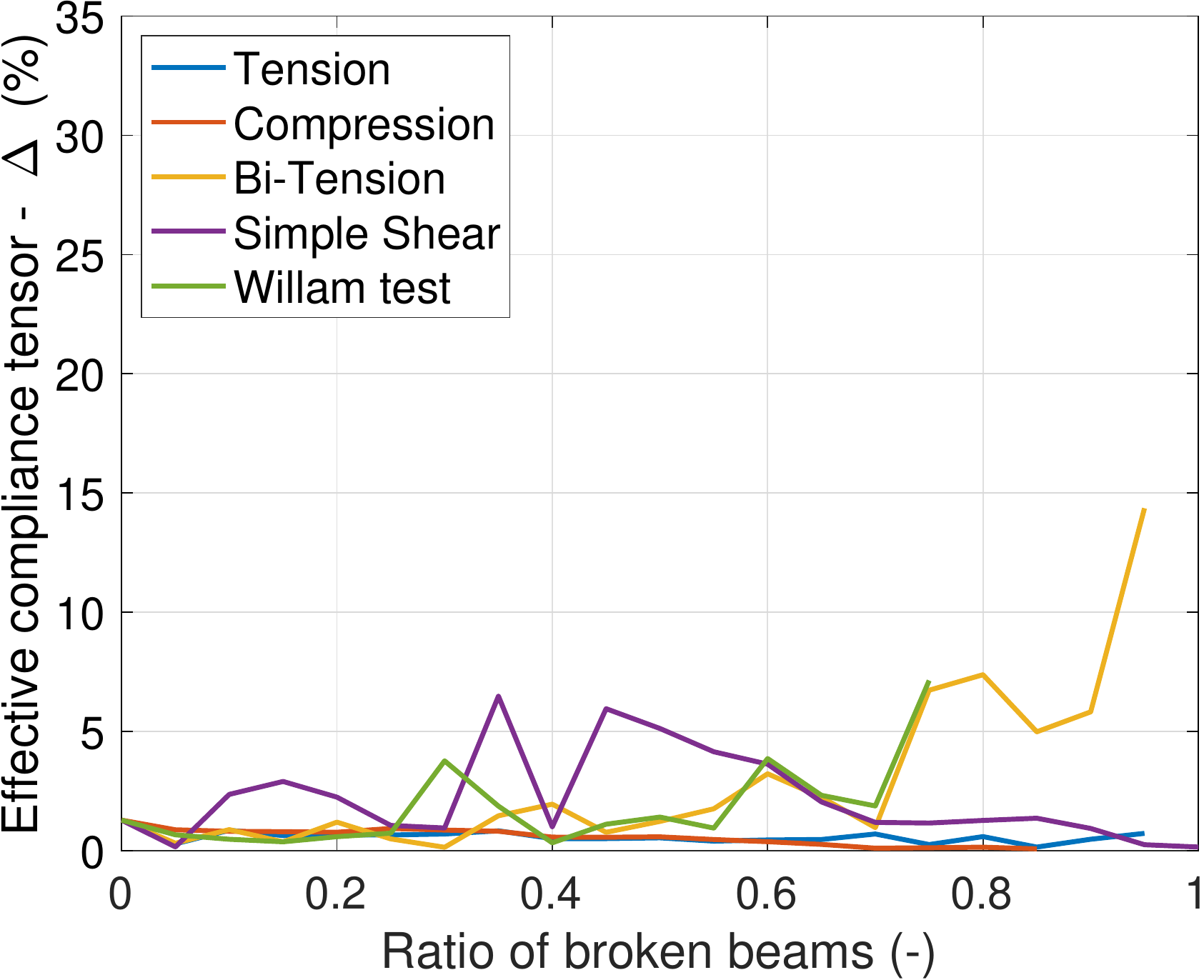}
  \caption{Upper-bound of distance to orthotropy for the effective compliance tensors}
  \label{fig:comp_orth}
\end{figure}

For the previously defined loadings, the stiffness and compliance tensors were extracted at different ratio of broken beams. The evolution of $\Delta$, which is an upper bound of the distance to the orthotropic class for an elasticity tensor as defined in corollary~\ref{cor:Delta}, is plotted in the figures~\ref{fig:stiff_orth} and~\ref{fig:comp_orth}. It should be noted that it is not always possible to extract the elasticity tensors for very high levels of damage.

Even at high levels of damage, the elasticity tensors remain close to the orthotropic symmetry class. This is even truer in the case of the compliance tensors. It is interesting to note that in the case of Willam loading, which results in a non-orthotropic cracking pattern, the deviations from orthotropy remain small. The largest deviation observed corresponds to the case of the stiffness tensor for bi-tension loading. This may be related to the coalescence of the main two orthogonal cracks.

One should note that the orthotropy of the effective elasticity tensor $\tilde \bC=\tilde \bS^{-1}$ is a sufficient condition to define an orthotropic damage tensor (see next), since the initial elasticity tensor $\bC=\bS^{-1}$ is isotropic.

\subsection{Comparison with micromechanics of cracked solids}

In the bi-dimensional micromechanical approach summarized by \cite{kachanov1992effective}, the material is considered to be initially isotropic with an initial (undamaged) compliance tensor,
\begin{equation}
  \centering
  \bS= \frac{1}{2 \mu} \bJ + \frac{1}{4 \kappa} \Idd \otimes \Idd .
\end{equation}

Damage is introduced using several networks of cracks with no interactions between each other. A family of cracks $p$ is characterized by its orientation $\vec n^{(p)}$, which is the normal to all the cracks in this family, a length $2 l^{(p)}$ and a density of micro-cracks $\omega^{(p)}={\pi l^{(p)}}/{A}$ with $A$ the area of the RAE.

A crack density tensor is then introduced for the network of all the families of cracks:
\begin{equation}
  \bomega = \sum_{p} \omega^{(p)} \vec n^{(p)} \otimes \vec n^{(p)}
\end{equation}

Gibbs free enthalpy density of the cracked solid writes, in case of open cracks \citep{kachanov1992effective}:
\begin{equation}
  \begin{split}
    \rho\psi^\star &= \frac{1}{2} \bsigma : \tilde \bS : \bsigma \\
    &= \frac{1}{2} \bsigma : \bS : \bsigma +\frac{1}{E} \tr\left(\bsigma\cdot \bomega \cdot \bsigma\right) \\
    &=\frac{1}{2} \bsigma : \bS : \bsigma + \frac{1}{2E} ~\bsigma: \left(\Idd  \otimesbar \bomega + \bomega  \otimesbar \Idd  \right) :\bsigma
  \end{split}
  \label{eq:enthalpy_kachanov}
\end{equation}
with $\bS$ the initial (isotropic) compliance tensor ($E$ being the Young modulus of the isotropic uncracked solid) and where $\tilde \bS$ is the effective compliance tensor.

We recall here the harmonic decomposition of the compliance tensor (see remark~\ref{rem1}):

\begin{equation}\label{eq:comp_harmonic}
  \begin{aligned}
    \tilde{\bS} & =  \tilde\bS_{iso}+  \frac{1}{2} \left( \Idd \otimes \bd'(\tilde{\bS}) +\bd'(\tilde{\bS}) \otimes \Idd\right)+ \bH(\tilde{\bS})
    \\
                & = \tilde \bS_{iso}+ \frac{1}{2} \left( \Idd \otimesbar \bd'(\tilde{\bS}) +\bd'(\tilde{\bS}) \otimesbar \Idd\right)+ \bH(\tilde{\bS})
  \end{aligned}
\end{equation}
with $ \bd(\tilde{\bS})= \tr_{12} \tilde{\bS}$ (in the same way $\bv(\tilde{\bS})= \tr_{13} \tilde{\bS}$) and where (see Eq. \eqref{eq:HarmDecompComponents})
\begin{equation*}
  \tilde\bS_{iso}= \frac{1}{2 \tilde \mu} \bJ + \frac{1}{4 \tilde \kappa} \Idd \otimes \Idd,
  \quad
  \begin{cases}
    \frac{1}{2\tilde \mu}= \frac{1}{4}(2\tr \bv(\tilde \bS)- \tr \bd(\tilde \bS)),
    \\
    \frac{1}{4\tilde \kappa} = \frac{1}{4} \tr \bd(\tilde \bS).
  \end{cases}
\end{equation*}
By comparing equations \eqref{eq:enthalpy_kachanov} and \eqref{eq:comp_harmonic}, it can be concluded that in the case of Kachanov micro-cracking theory (representing networks of non interacting open micro-cracks), one has $\tilde \bS_{iso}=\bS+\frac{1}{2E}\tr \bomega\, \Idd \otimesbar \Idd$ (with
$\Idd \otimesbar \Idd=\bI=\bJ+\frac{1}{2} \Idd \otimes \Idd$), $\bd'(\bS)=2 \bomega'/E$ and the harmonic part of the compliance tensor $\bH(\tilde{\bS})$ vanishes.

\begin{figure}[htp]
  \centering
  \includegraphics[align=c,width=0.8\linewidth]{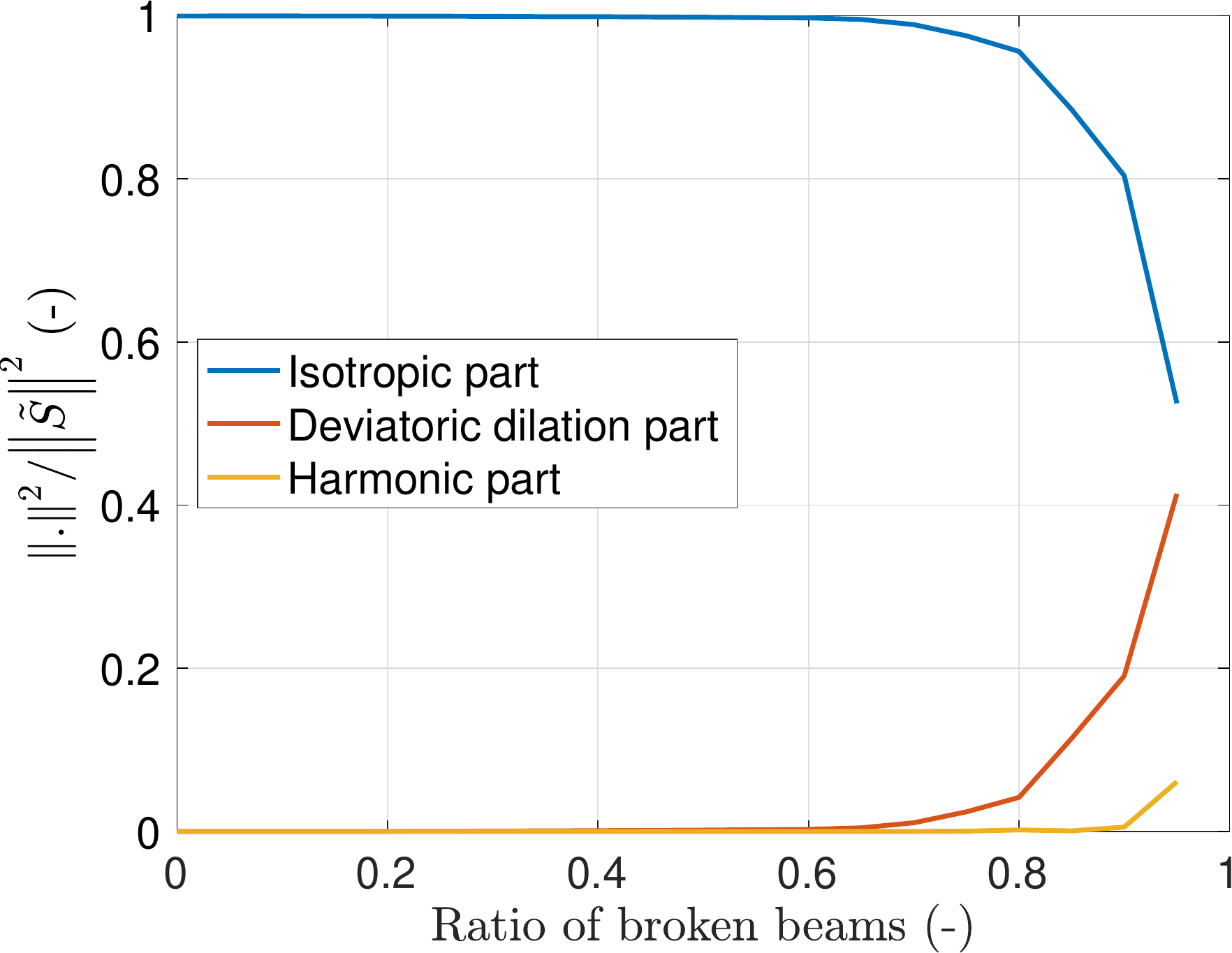}
  \caption{Tensile loading: Evolution of the relative parts of the effective compliance tensor with the ratio of broken beams.}
  \label{fig:cont_comp_tension}
\end{figure}

\begin{figure}[htp]
  \centering
  \includegraphics[align=c,width=0.8\linewidth]{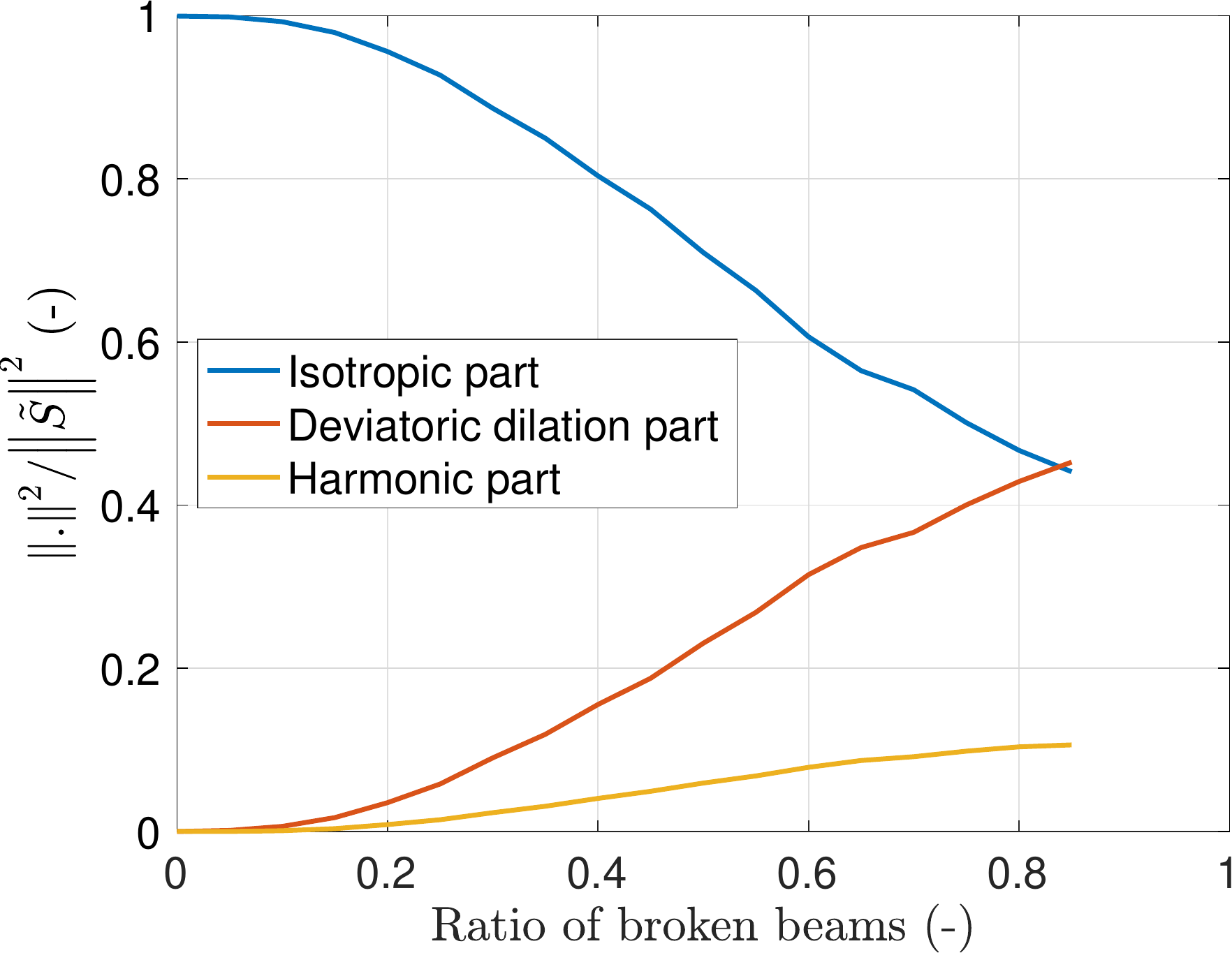}
  \caption{Compressive loading: Evolution of the relative parts of the effective compliance tensor with the ratio of broken beams.}
  \label{fig:cont_comp_compression}
\end{figure}

\begin{figure}[htp]
  \centering
  \includegraphics[align=c,width=0.8\linewidth]{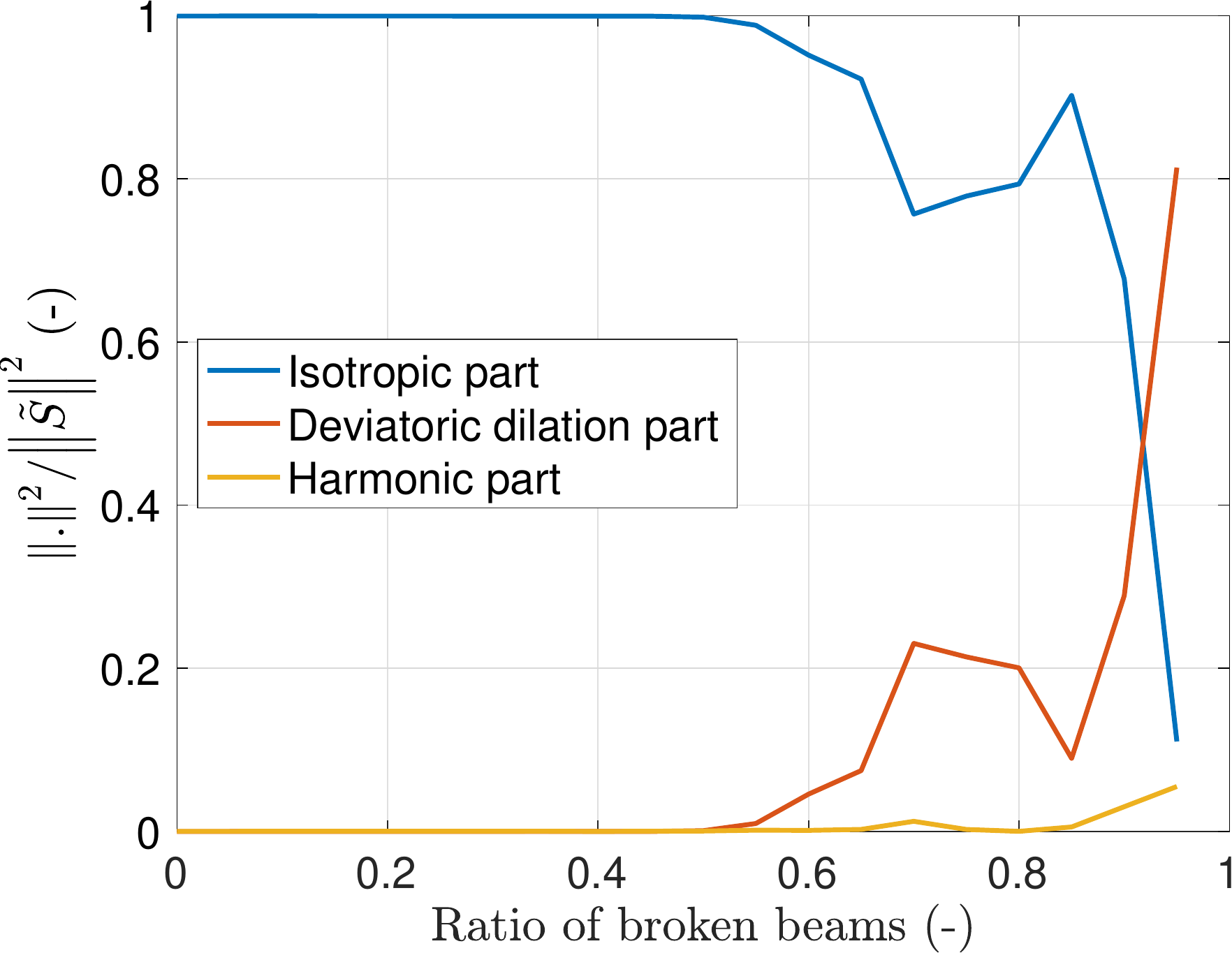}
  \caption{Bi-Tensile loading: Evolution of the relative parts of the effective compliance tensor with the ratio of broken beams.}
  \label{fig:cont_comp_bitension}
\end{figure}

\begin{figure}[htp]
  \centering
  \includegraphics[align=c,width=0.8\linewidth]{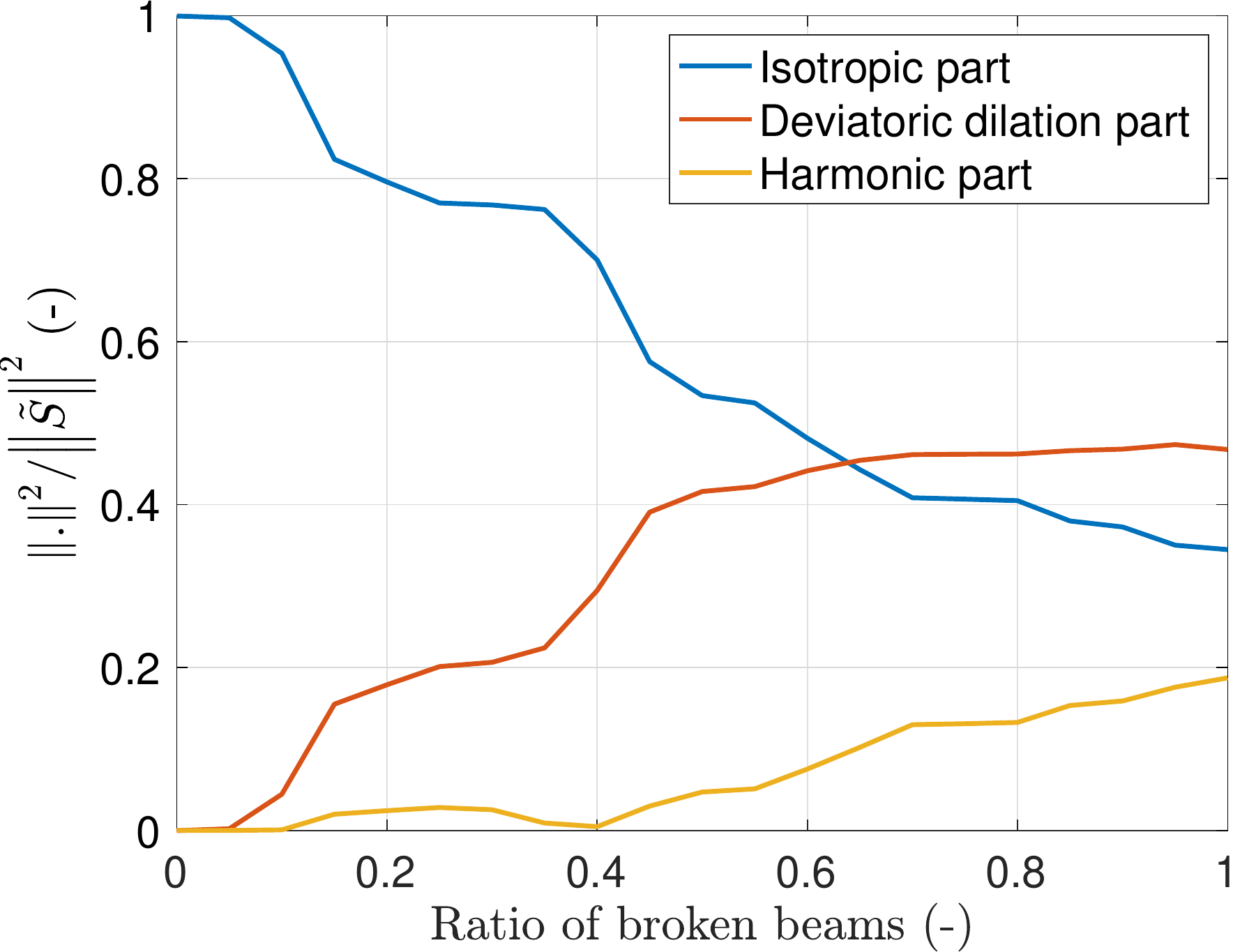}
  \caption{Simple shear loading: Evolution of the relative parts of the effective compliance tensor with the ratio of broken beams.}
  \label{fig:cont_comp_shear}
\end{figure}

\begin{figure}[htp]
  \centering
  \includegraphics[align=c,width=0.8\linewidth]{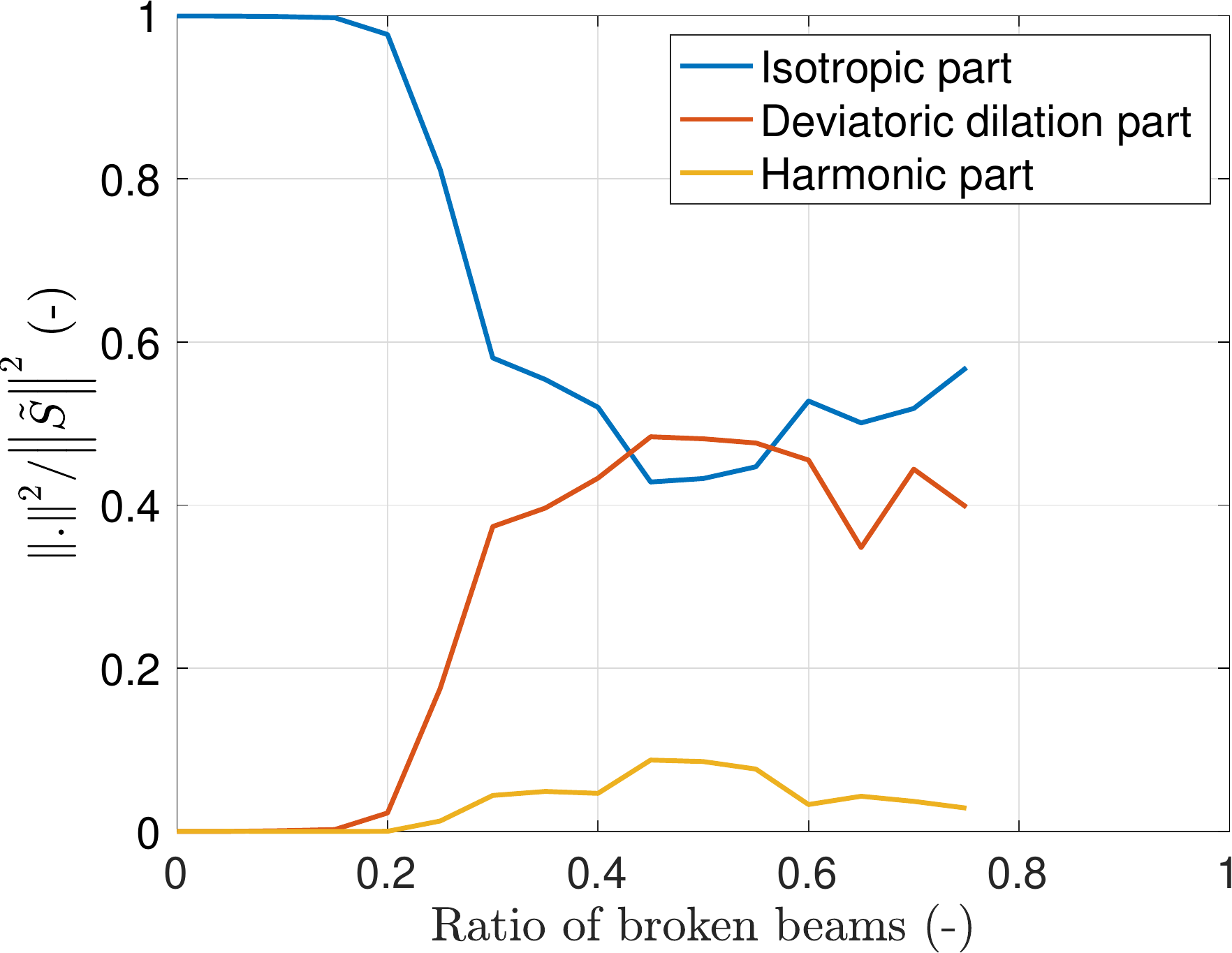}
  \caption{Willam loading: Evolution of the relative parts of the effective compliance tensor with the ratio of broken beams.}
  \label{fig:cont_comp_willam}
\end{figure}

In order to test the validity of the property $\bH(\tilde{\bS})=0$, eventually when cracks interaction takes place, we plot in figures~\ref{fig:cont_comp_tension} 
to~\ref{fig:cont_comp_willam} the evolution of the three parts of the effective compliance tensor $\tilde{\bS}$ (computed thanks to the beam-particle method):
\begin{itemize}
  \item the isotropic part $\tilde \bS_{iso}$,

  \item the deviatoric dilatation part $\frac{1}{2} ( \Idd \otimes \bd'(\tilde{\bS}) +\bd'(\tilde{\bS}) \otimes \Idd)$,

  \item and the harmonic part $\bH(\tilde{\bS})$,
\end{itemize}
according to the ratio of broken beams.
One should remember that the deviatoric dilatation part is always orthotropic while the harmonic part can have the square symmetry or be zero.

With these curves, we can distinguish four stages of cracking:
\begin{itemize}
  \item[$i)$] Initially, the medium is non-cracked and therefore isotropic, which implies that the parts related to $\bd'$ and $\bH$ are null.

  \item[$ii)$] At the beginning of loading, only the weakest beams break without influence of the loading direction. The orientation of the cracks therefore remains isotropic and here again the parts linked to $\bd'$ and $\bH$ are zero.

  \item[$iii)$] After a certain level, the beams failure will be related to the direction of loading and the microcracks begin to orient themselves while remaining diffuse. The loss of isotropy implies that the $\bd'$-related part becomes non-zero. On the other hand, the harmonic part remains null. This is a case of crack-induced orthotropy.

  \item[$iv)$] Eventually, the cracks will start to interact or coalesce and the harmonic part becomes non-zero. So we deviate from the framework of Kachanov's theory. Depending on the symmetry class of the harmonic tensor, it can be induced orthotropy or no-symmetry induced anisotropy.
\end{itemize}

\begin{figure*}[hp]
  \centering
  \begin{subfigure}{.18\textwidth}
    \centering
    \includegraphics[width=.9\linewidth]{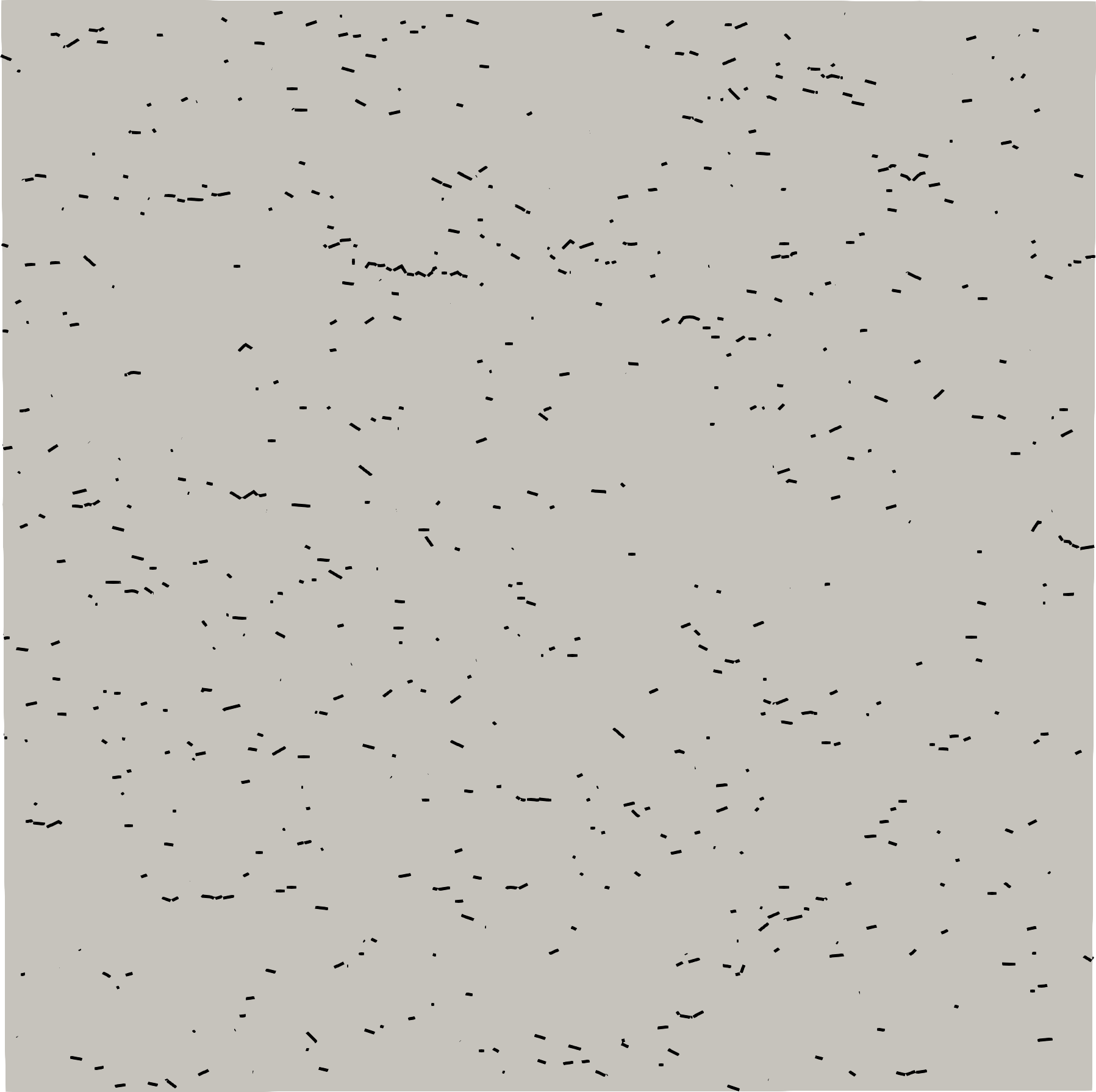}
    \caption{60 \%}
    \label{fig:traction_60}
  \end{subfigure}
  \begin{subfigure}{.18\textwidth}
    \centering
    \includegraphics[width=.9\linewidth]{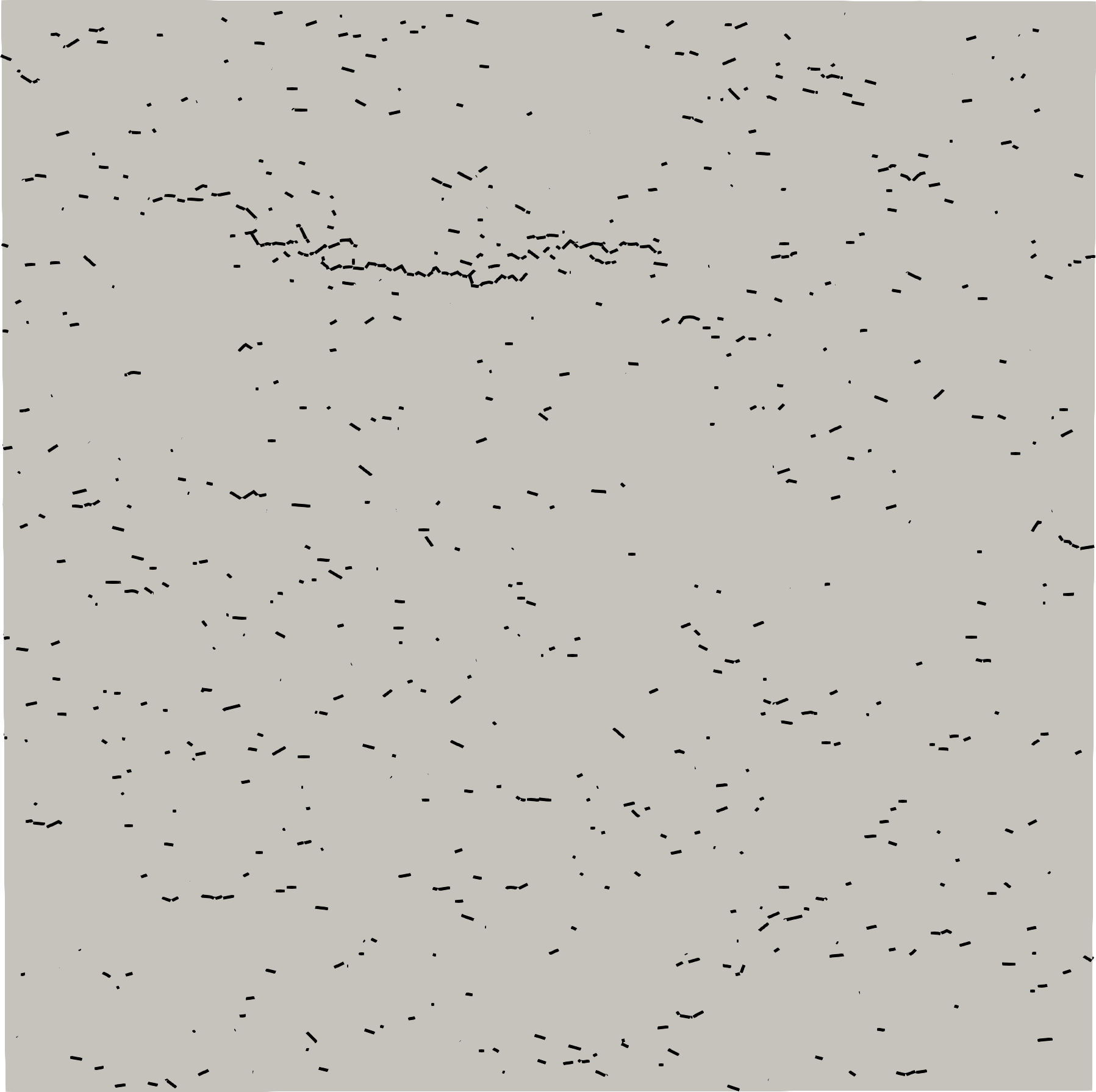}
    \caption{70 \%}
    \label{fig:traction_70}
  \end{subfigure}
  \begin{subfigure}{.18\textwidth}
    \centering
    \includegraphics[width=.9\linewidth]{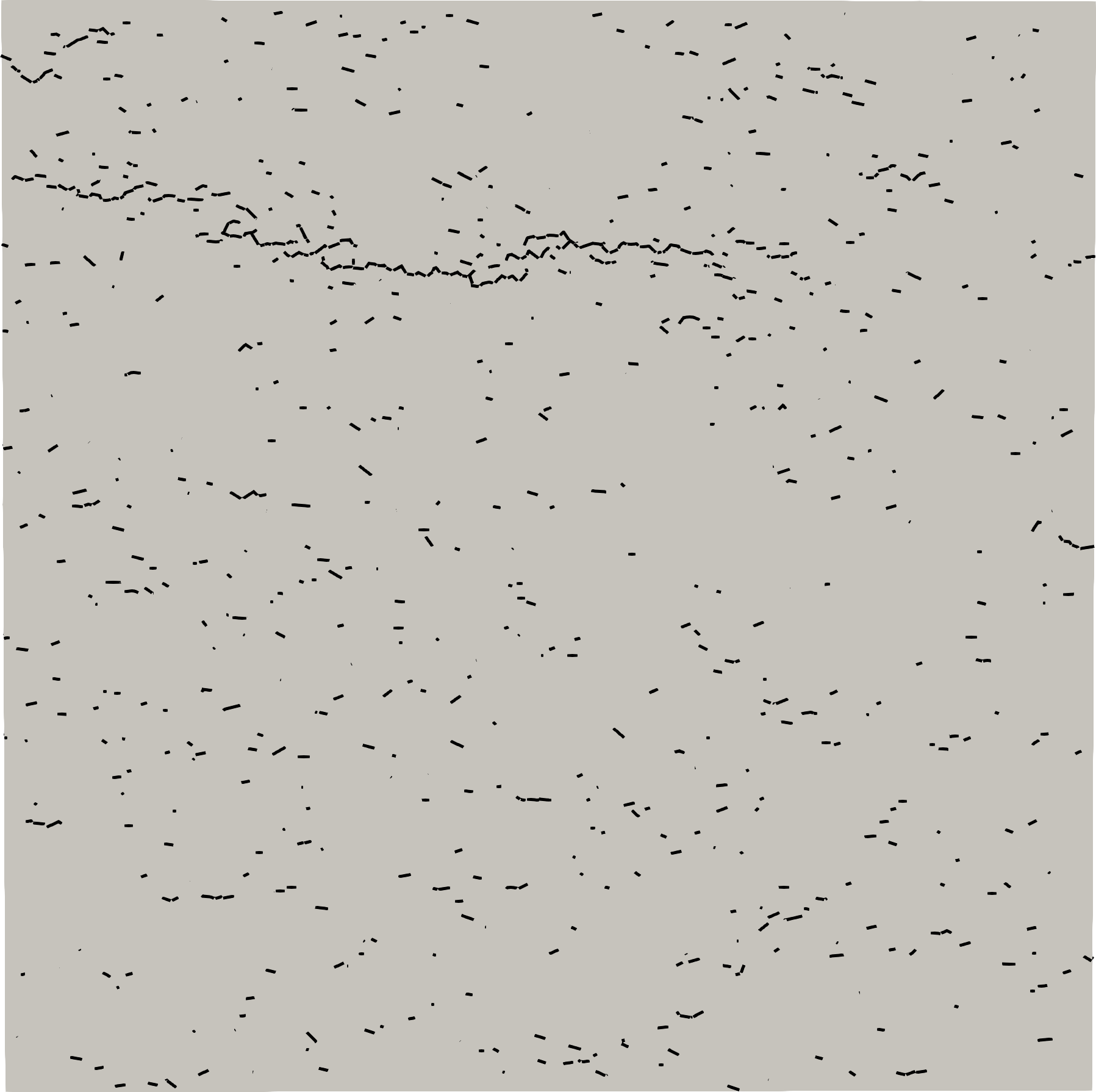}
    \caption{80 \%}
    \label{fig:traction_80}
  \end{subfigure}
  \begin{subfigure}{.18\textwidth}
    \centering
    \includegraphics[width=.9\linewidth]{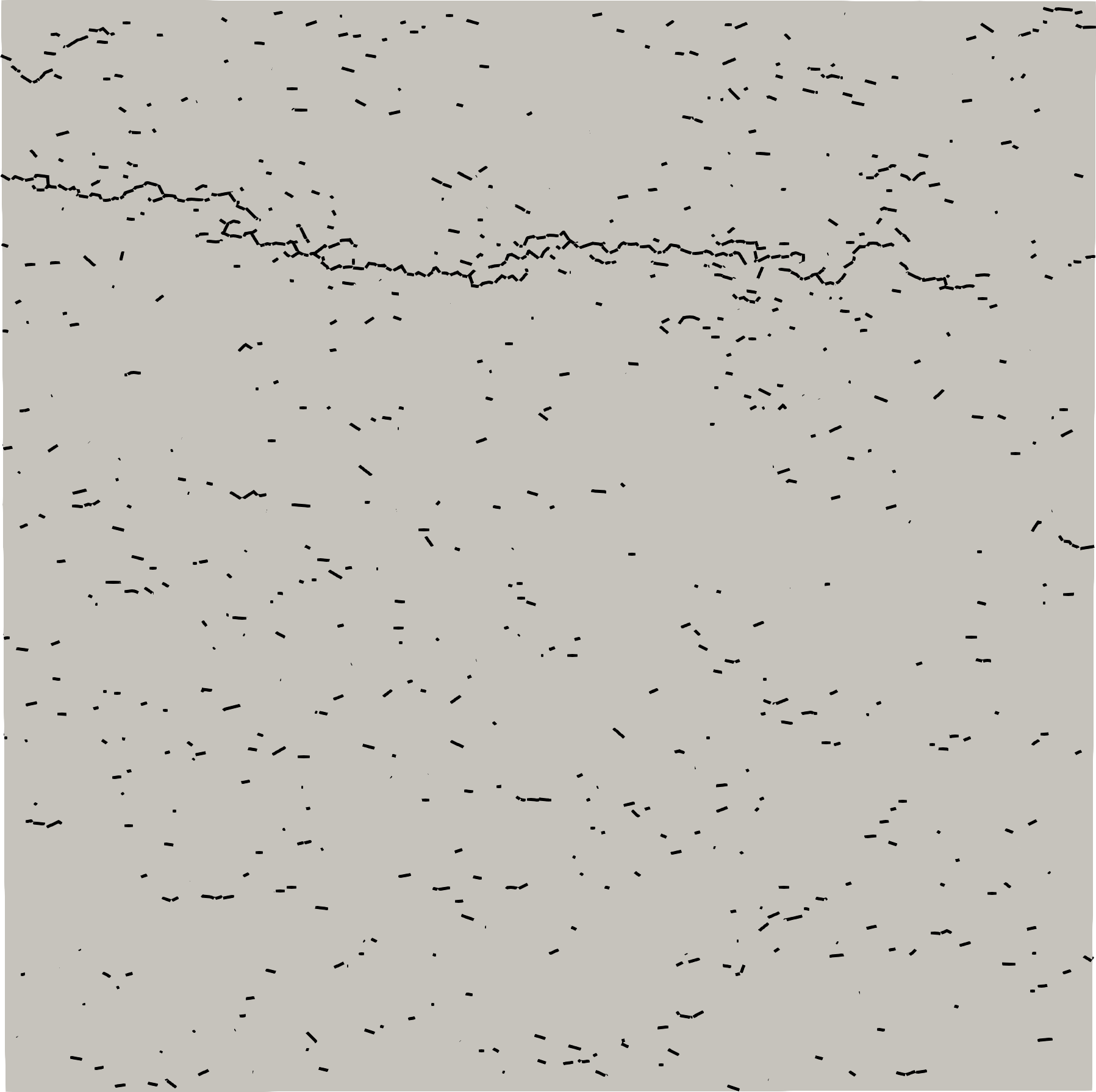}
    \caption{90 \%}
    \label{fig:traction_90}
  \end{subfigure}
  \begin{subfigure}{.18\textwidth}
    \centering
    \includegraphics[width=.9\linewidth]{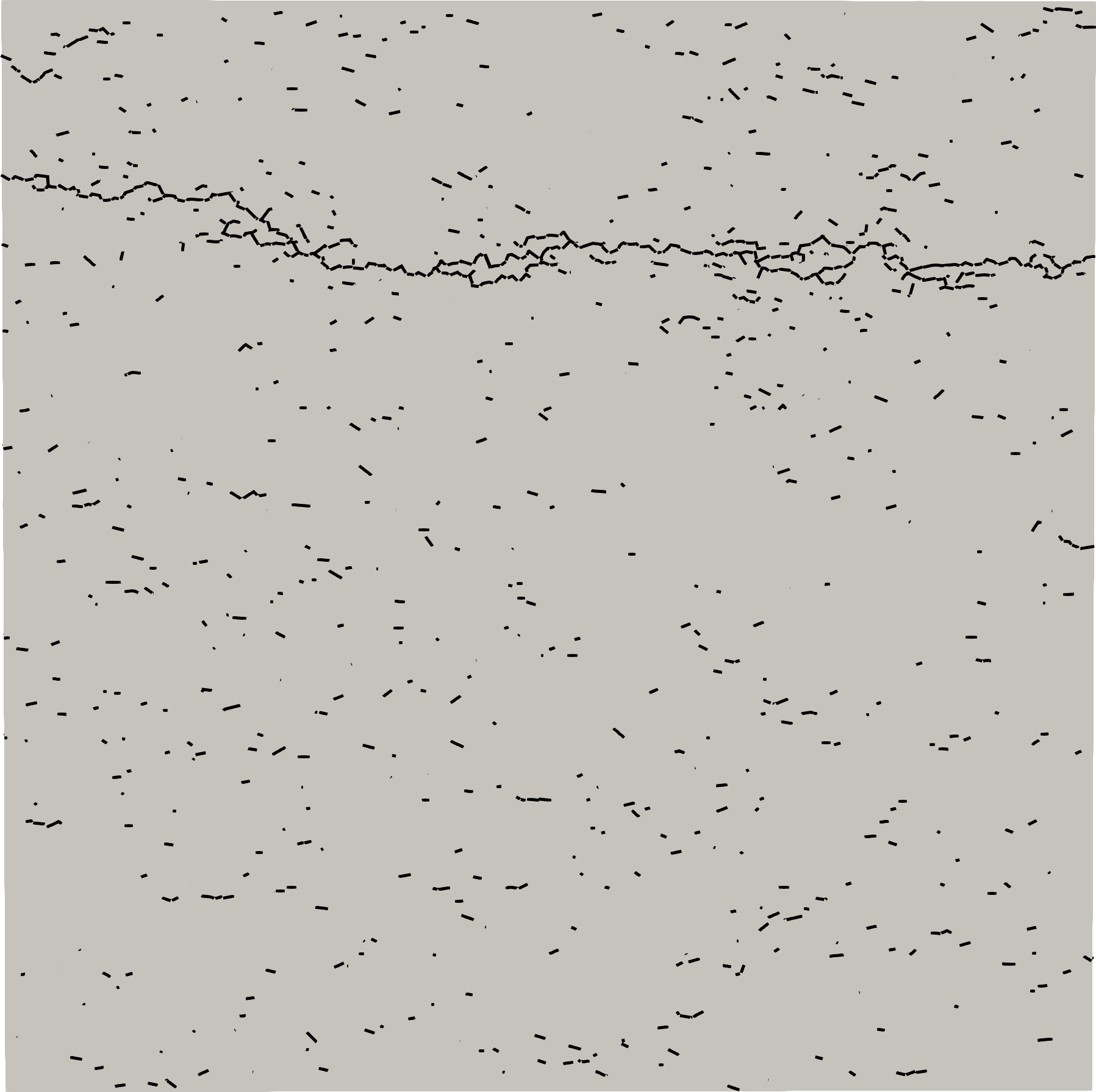}
    \caption{100 \%}
    \label{fig:traction_100}
  \end{subfigure}
  \caption{Tensile loading: evolution of the crack pattern with the ratio of broken beams.}
  \label{fig:traction_pattern}
\end{figure*}

\begin{figure*}[hp]
  \centering
  \begin{subfigure}{.18\textwidth}
    \centering
    \includegraphics[width=.9\linewidth]{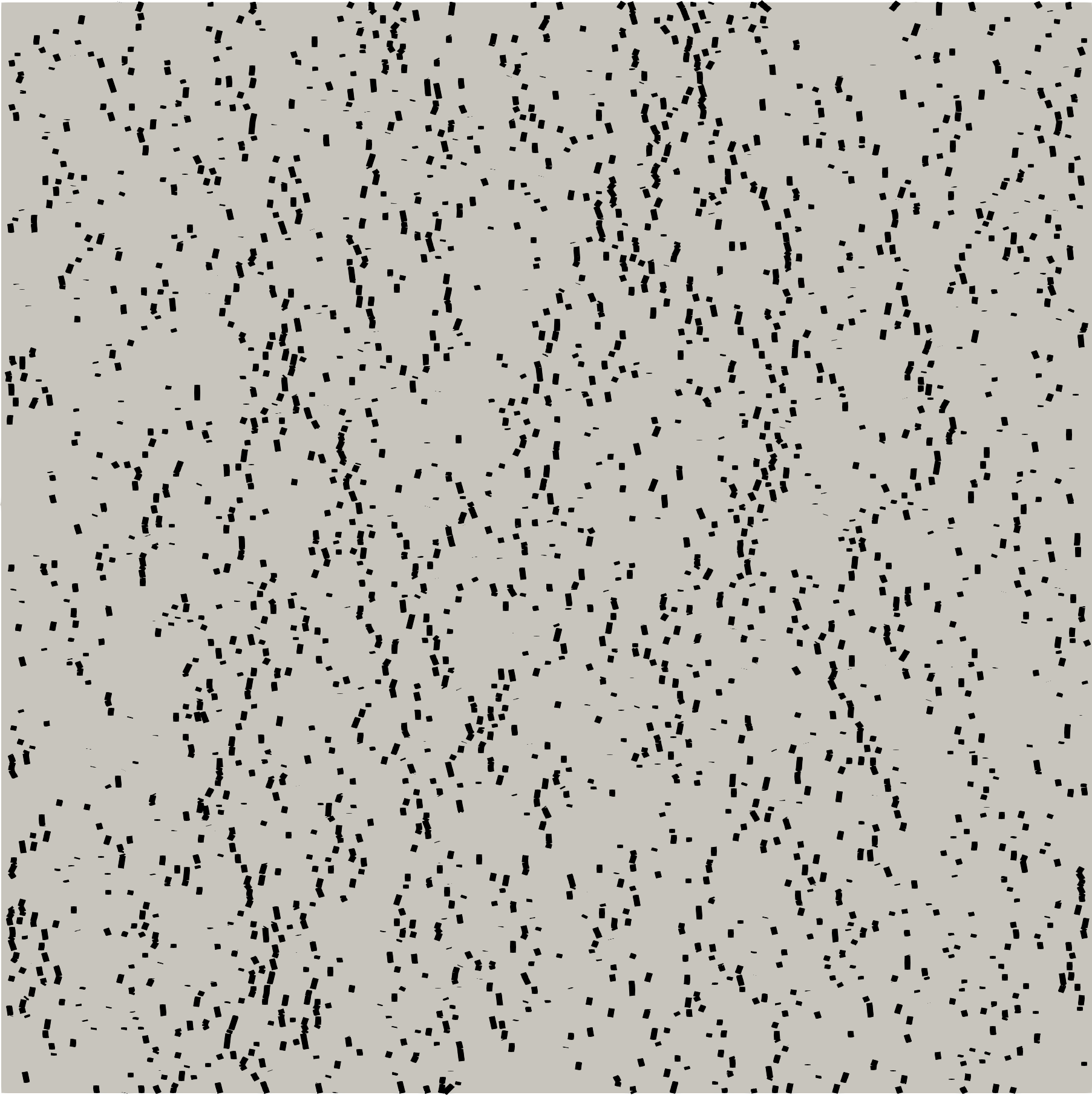}
    \caption{20 \%}
    \label{fig:compression_20}
  \end{subfigure}
  \begin{subfigure}{.18\textwidth}
    \centering
    \includegraphics[width=.9\linewidth]{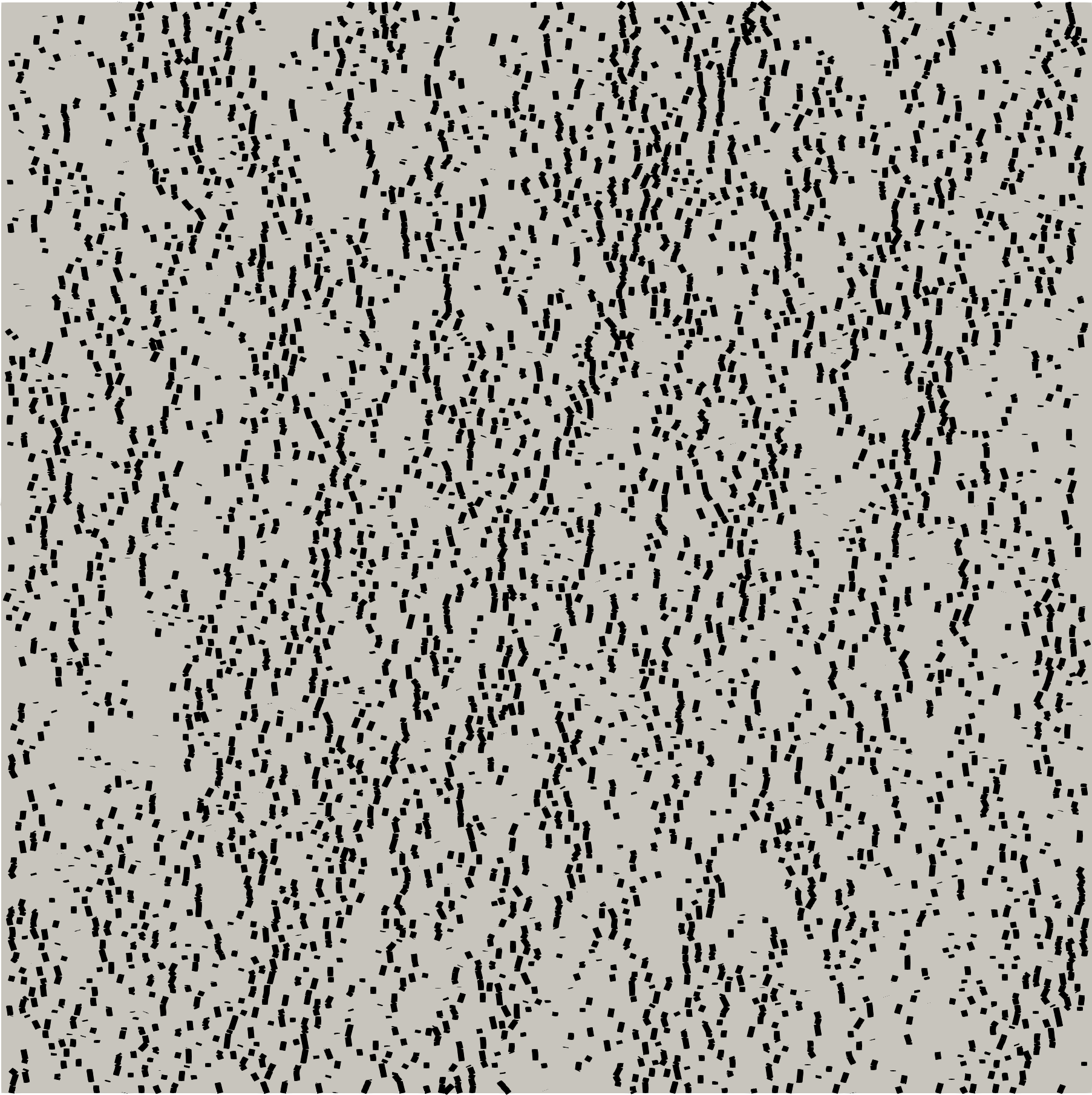}
    \caption{40 \%}
    \label{fig:compression_40}
  \end{subfigure}
  \begin{subfigure}{.18\textwidth}
    \centering
    \includegraphics[width=.9\linewidth]{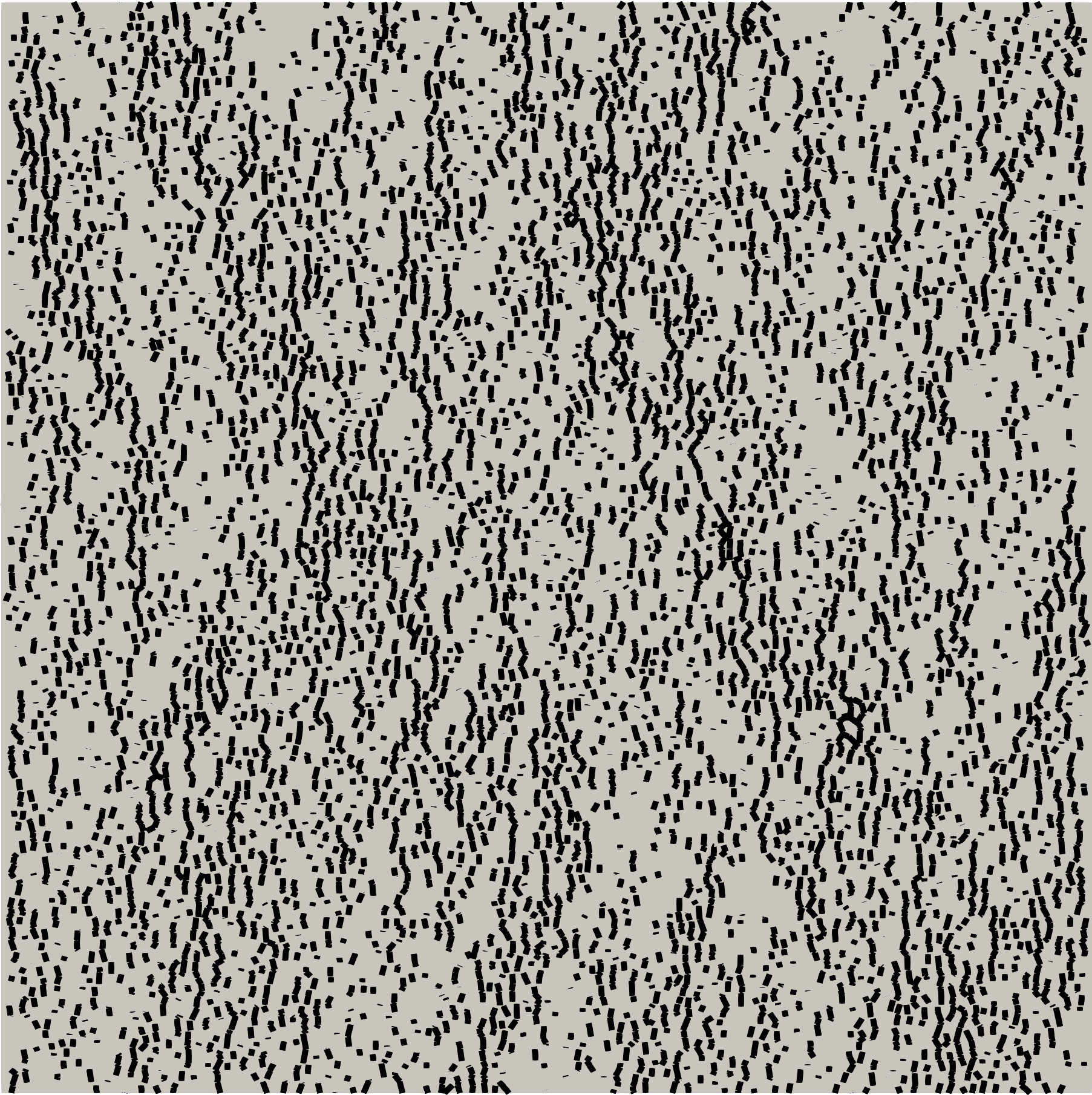}
    \caption{60 \%}
    \label{fig:compression_60}
  \end{subfigure}
  \begin{subfigure}{.18\textwidth}
    \centering
    \includegraphics[width=.9\linewidth]{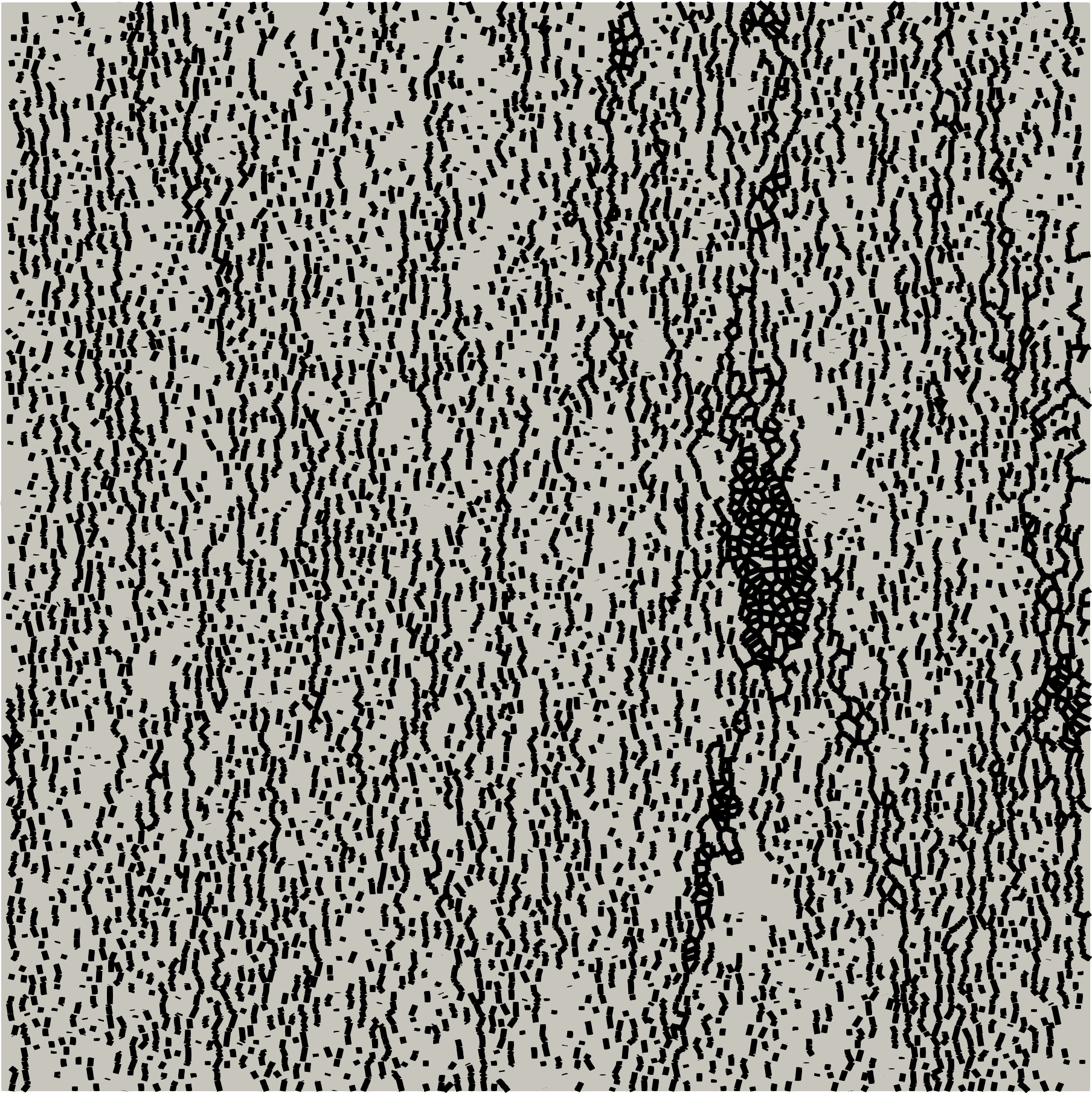}
    \caption{80 \%}
    \label{fig:compression_80}
  \end{subfigure}
  \caption{Compressive loading: evolution of the crack pattern with the ratio of broken beams.}
  \label{fig:compression_pattern}
\end{figure*}

\begin{figure*}[hp]
  \centering
  \begin{subfigure}{.18\textwidth}
    \centering
    \includegraphics[width=.9\linewidth]{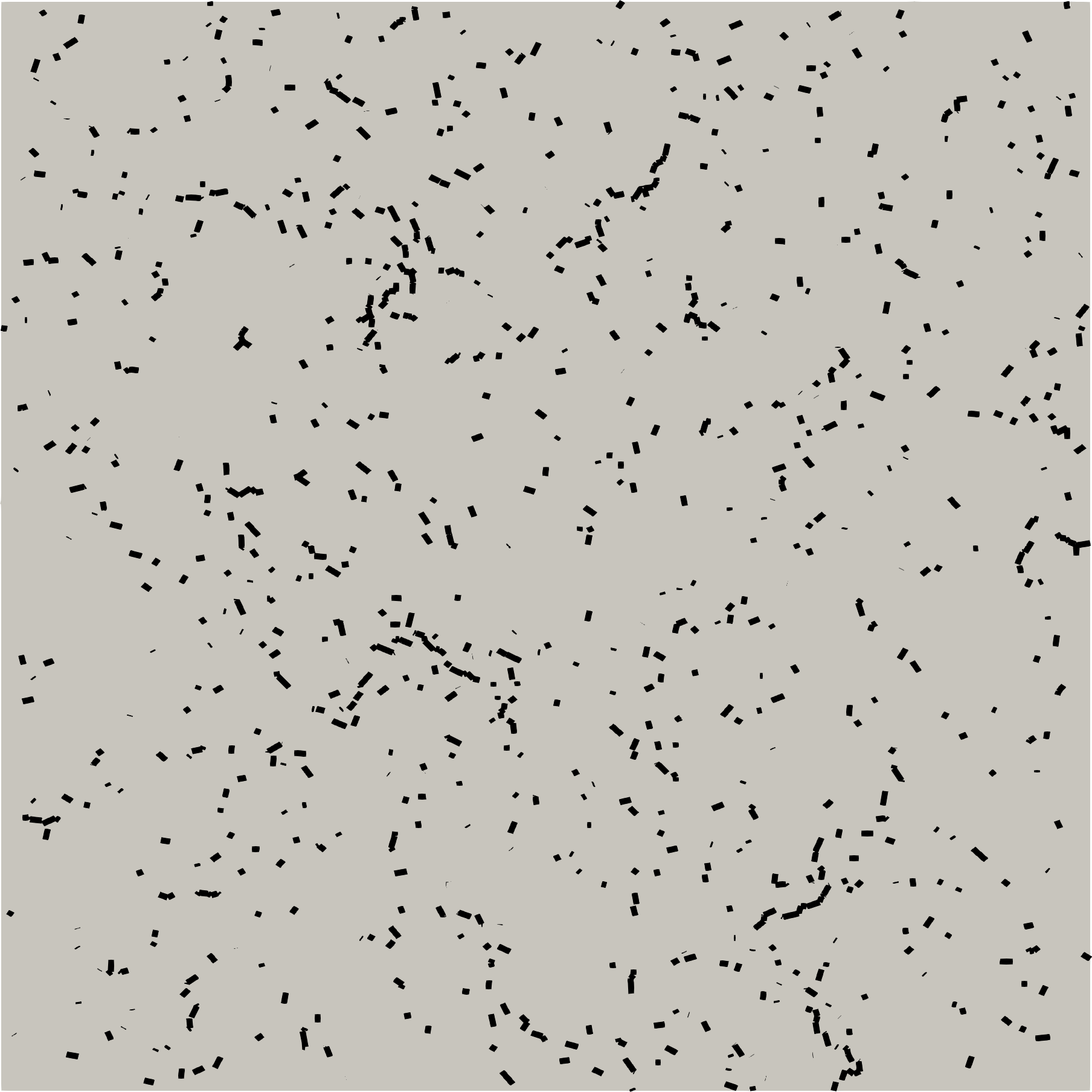}
    \caption{45 \%}
    \label{fig:bitraction_45}
  \end{subfigure}
  \begin{subfigure}{.18\textwidth}
    \centering
    \includegraphics[width=.9\linewidth]{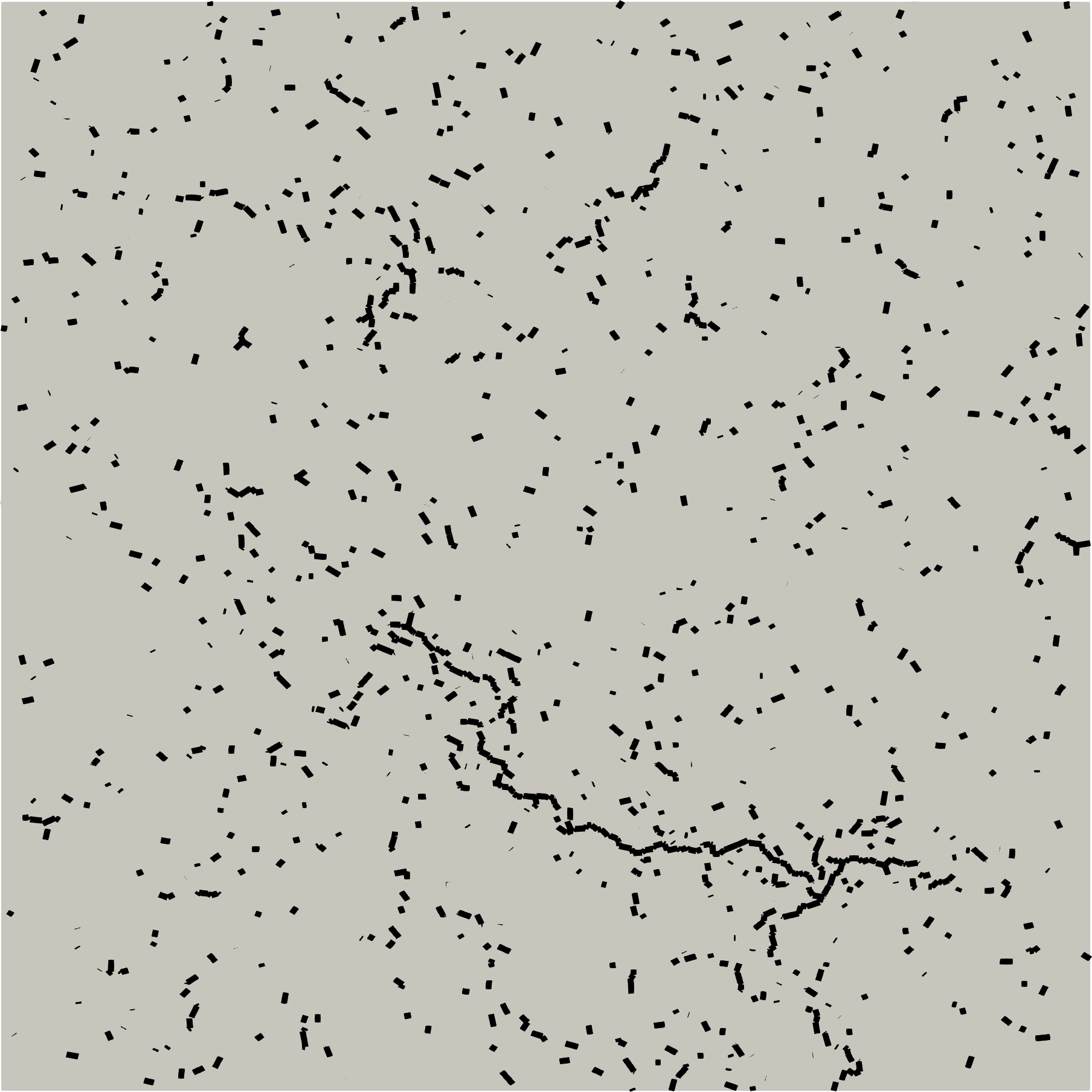}
    \caption{55 \%}
    \label{fig:bitraction_55}
  \end{subfigure}
  \begin{subfigure}{.18\textwidth}
    \centering
    \includegraphics[width=.9\linewidth]{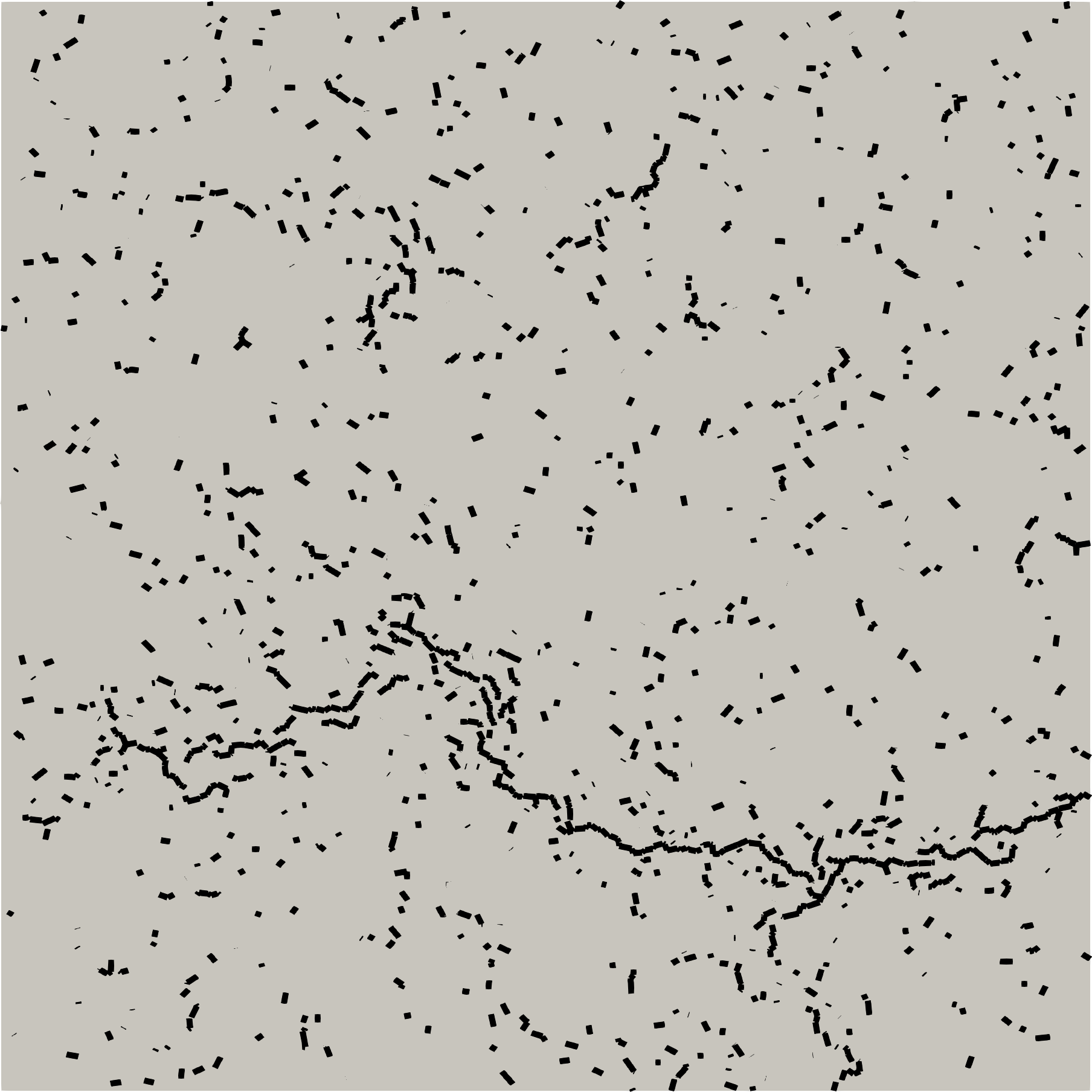}
    \caption{65 \%}
    \label{fig:bitraction_65}
  \end{subfigure}
  \begin{subfigure}{.18\textwidth}
    \centering
    \includegraphics[width=.9\linewidth]{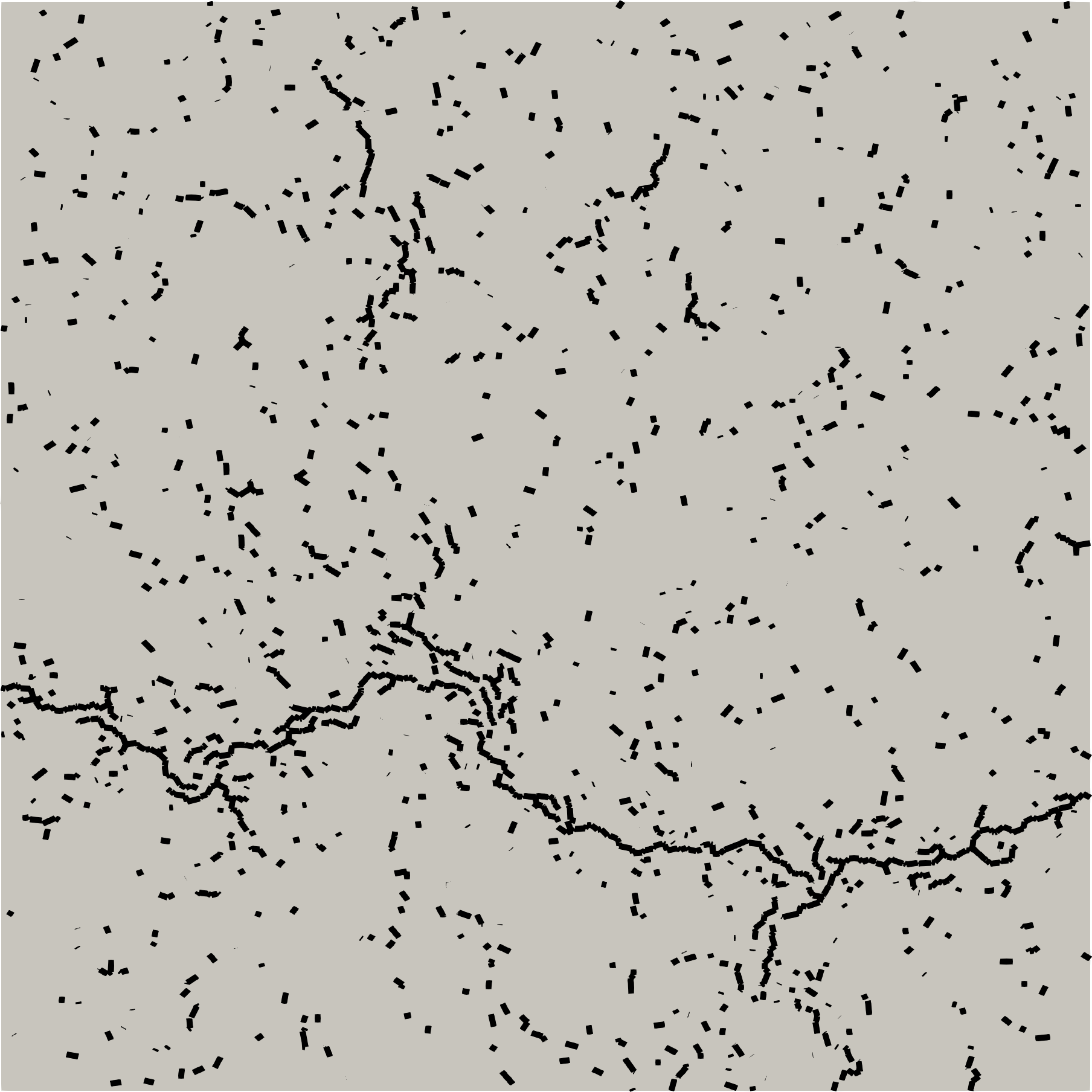}
    \caption{75 \%}
    \label{fig:bitraction_75}
  \end{subfigure}
  \begin{subfigure}{.18\textwidth}
    \centering
    \includegraphics[width=.9\linewidth]{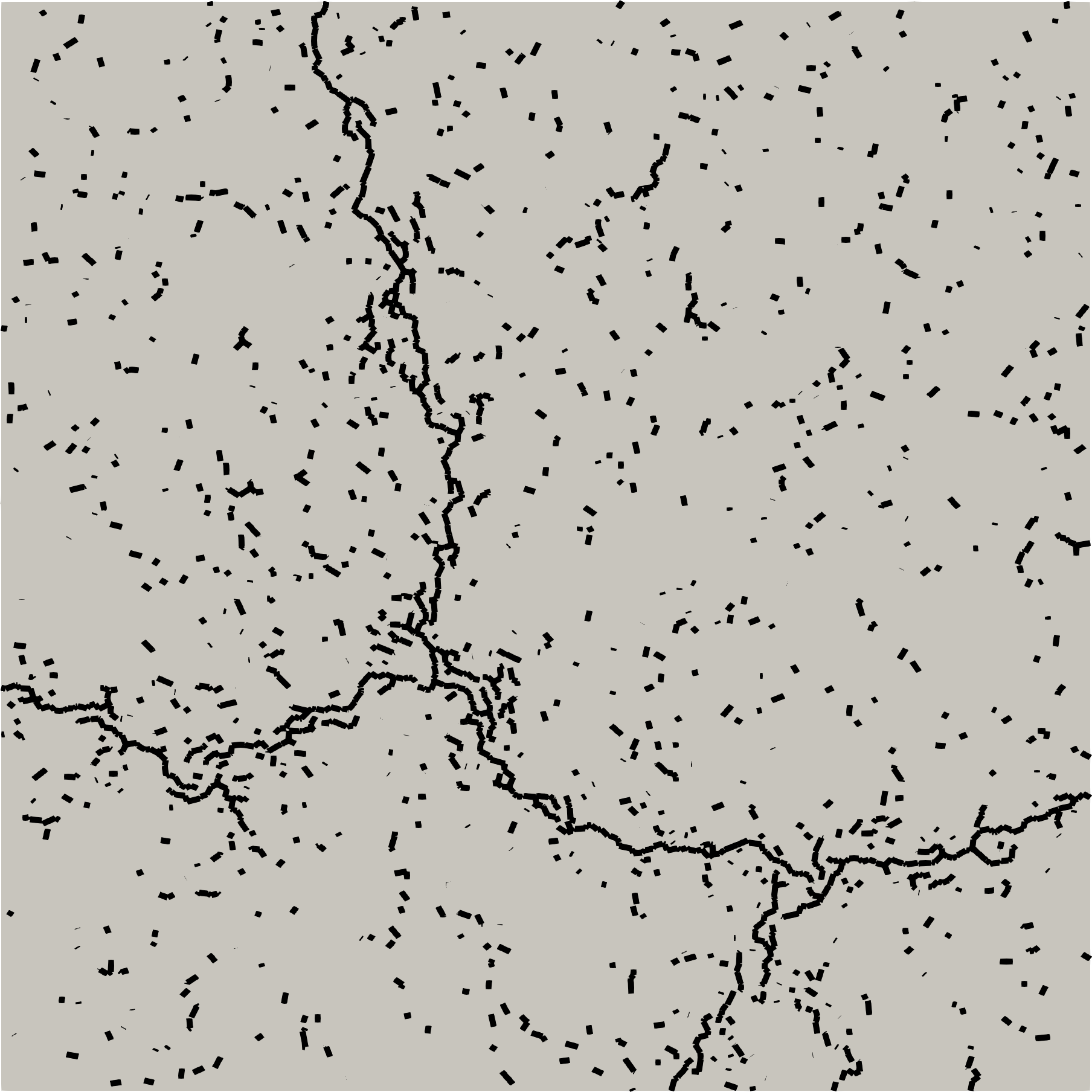}
    \caption{85 \%}
    \label{fig:bitraction_85}
  \end{subfigure}
  \caption{Bi-Tensile loading: evolution of the crack pattern with the ratio of broken beams.}
  \label{fig:bitraction_pattern}
\end{figure*}

\begin{figure*}[hp]
  \centering
  \begin{subfigure}{.18\textwidth}
    \centering
    \includegraphics[width=.9\linewidth]{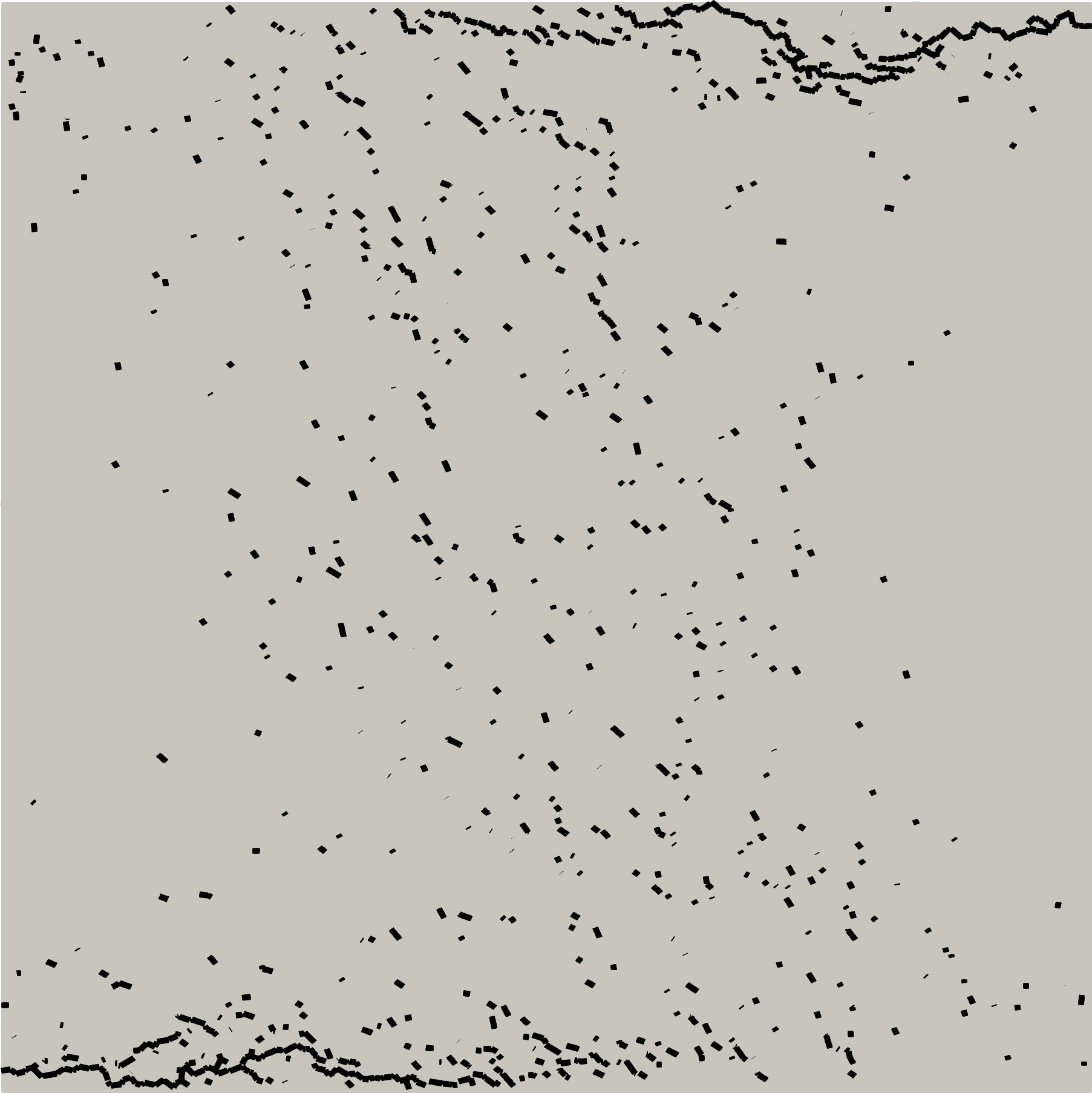}
    \caption{20 \%}
    \label{fig:shear_20}
  \end{subfigure}
  \begin{subfigure}{.18\textwidth}
    \centering
    \includegraphics[width=.9\linewidth]{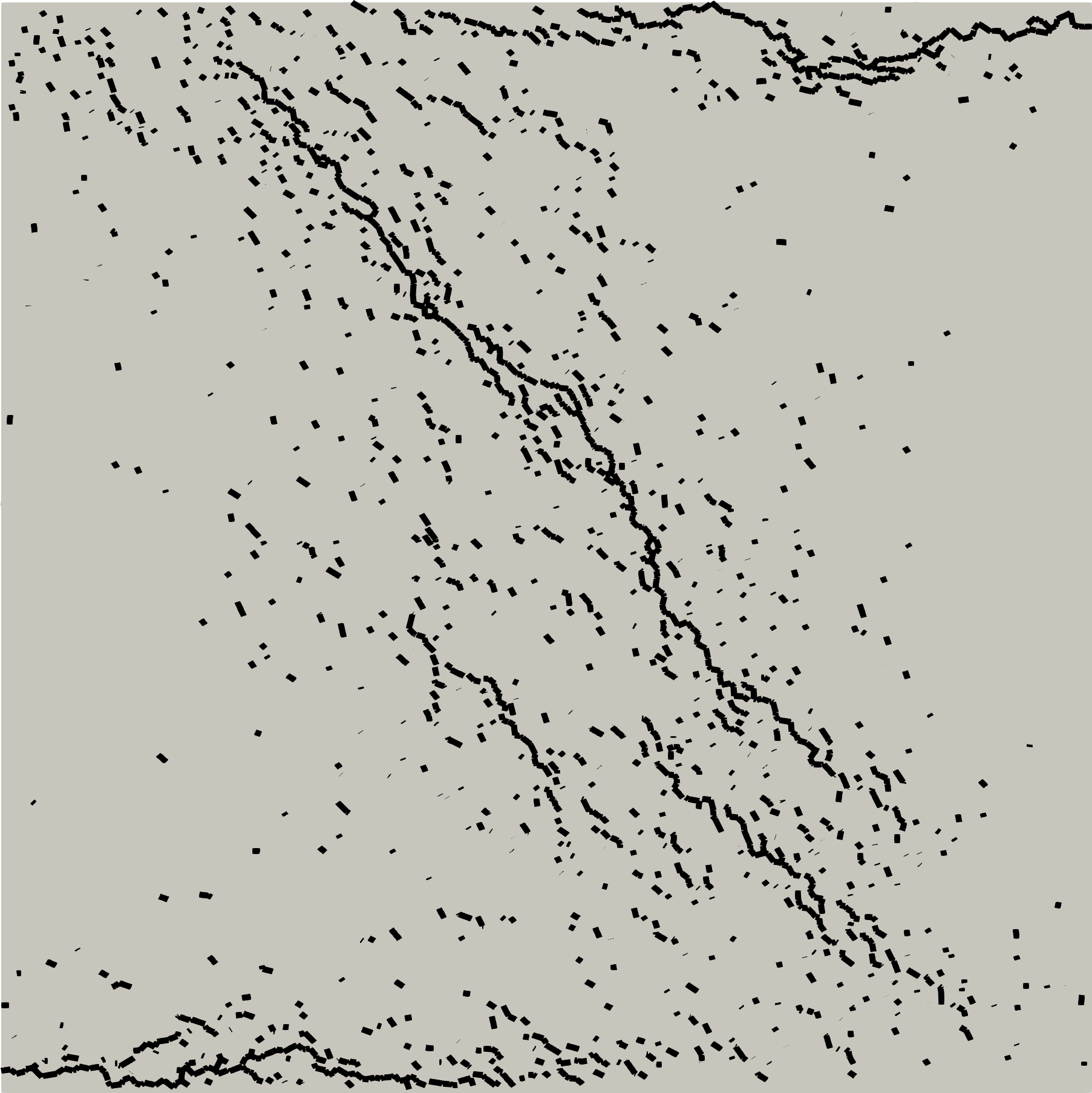}
    \caption{40 \%}
    \label{fig:shear_40}
  \end{subfigure}
  \begin{subfigure}{.18\textwidth}
    \centering
    \includegraphics[width=.9\linewidth]{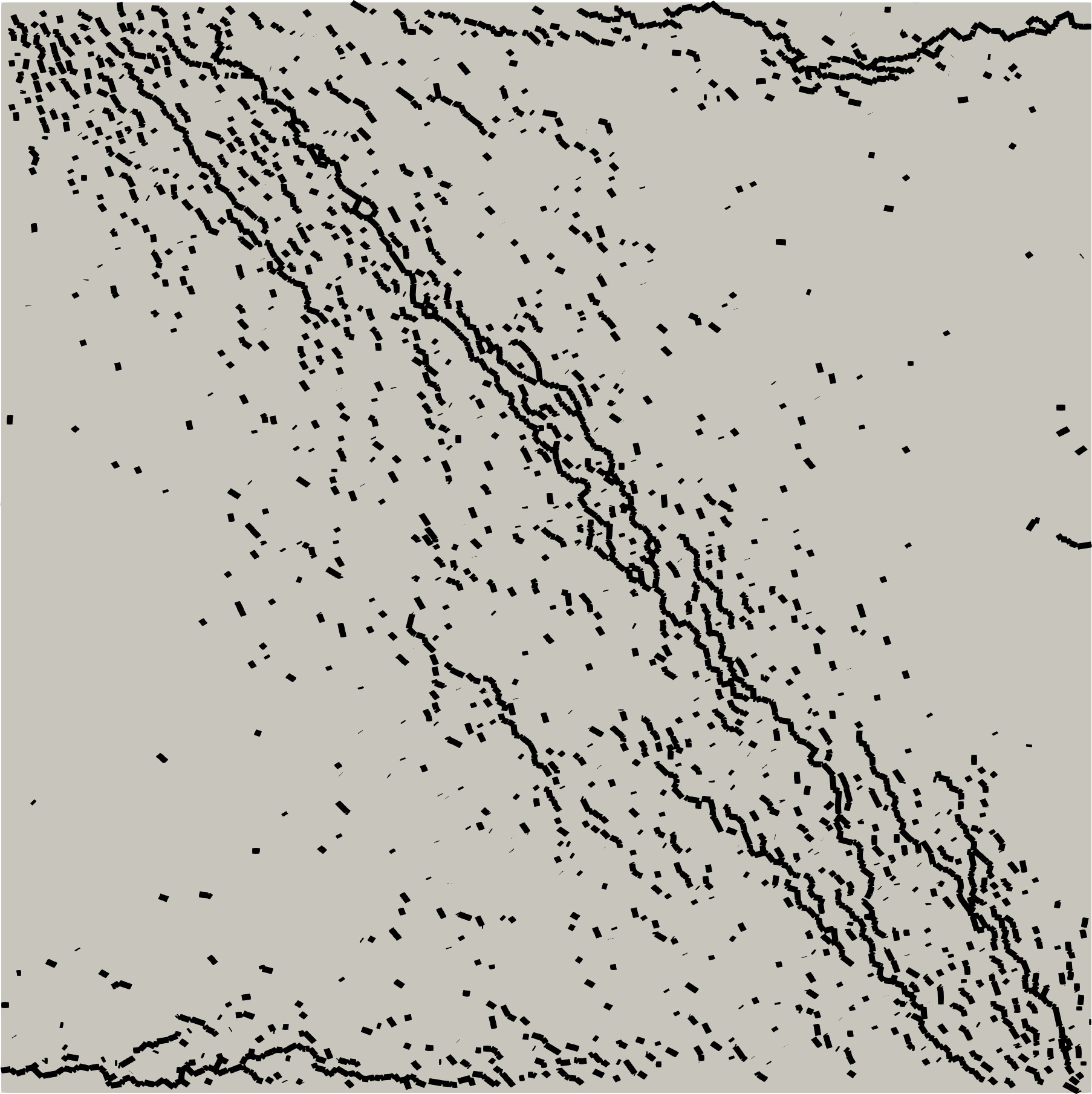}
    \caption{60 \%}
    \label{fig:shear_60}
  \end{subfigure}
  \begin{subfigure}{.18\textwidth}
    \centering
    \includegraphics[width=.9\linewidth]{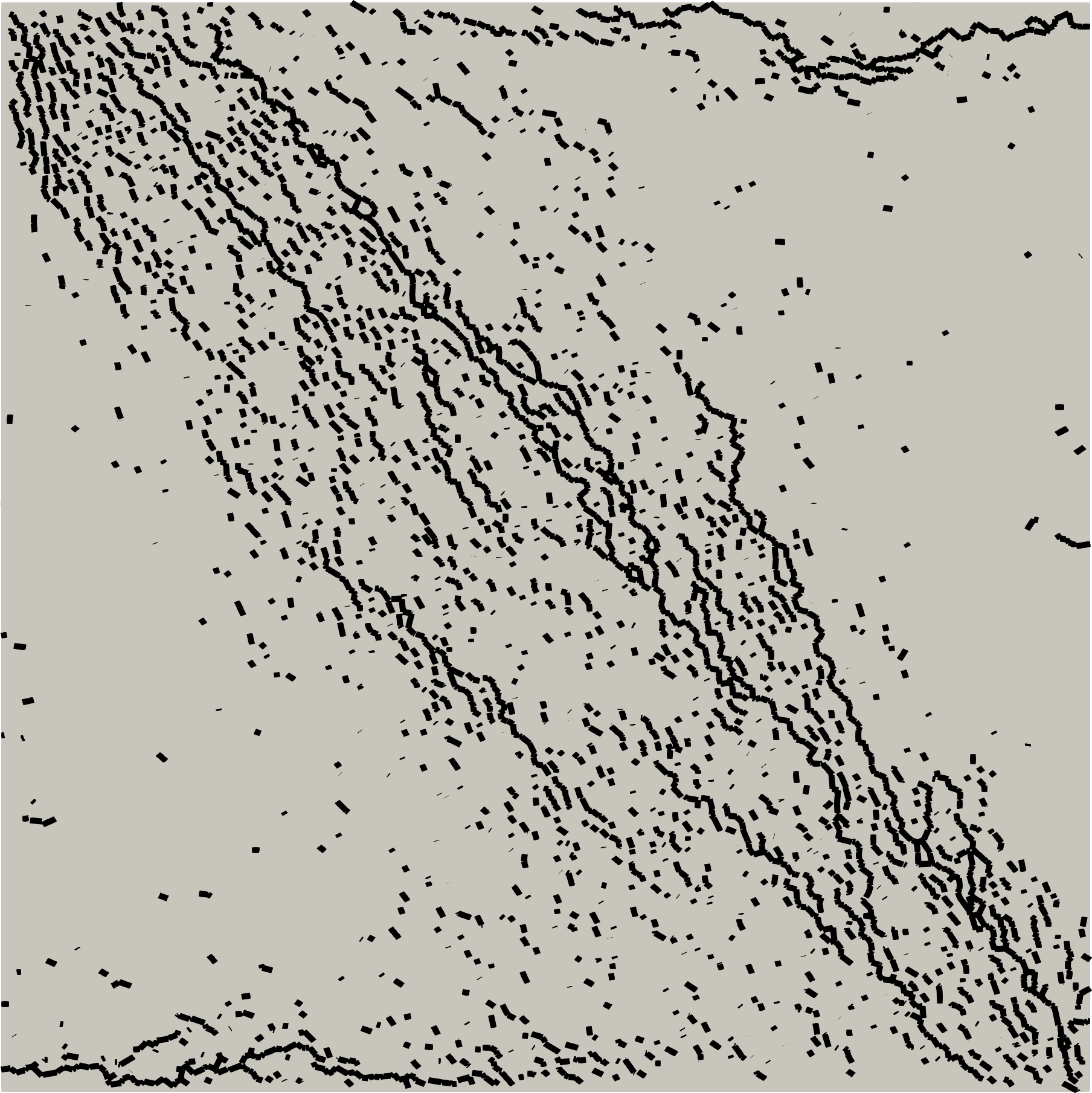}
    \caption{80 \%}
    \label{fig:shear_80}
  \end{subfigure}
  \caption{Simple shear loading: evolution of the crack pattern with the ratio of broken beams.}
  \label{fig:shear_pattern}
\end{figure*}

\begin{figure*}[hp]
  \centering
  \begin{subfigure}{.18\textwidth}
    \centering
    \includegraphics[width=.9\linewidth]{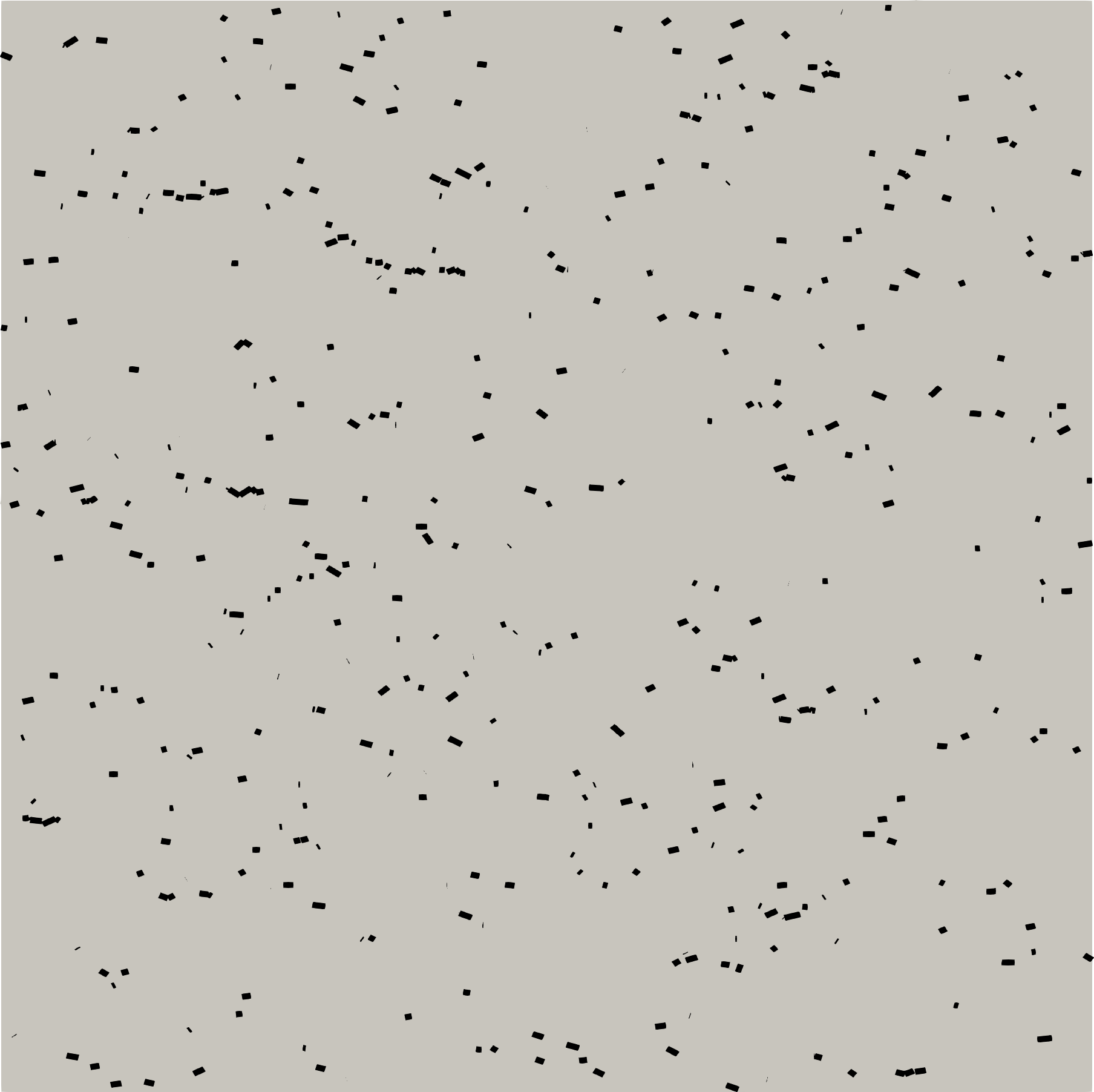}
    \caption{10 \%}
    \label{fig:willam_10}
  \end{subfigure}
  \begin{subfigure}{.18\textwidth}
    \centering
    \includegraphics[width=.9\linewidth]{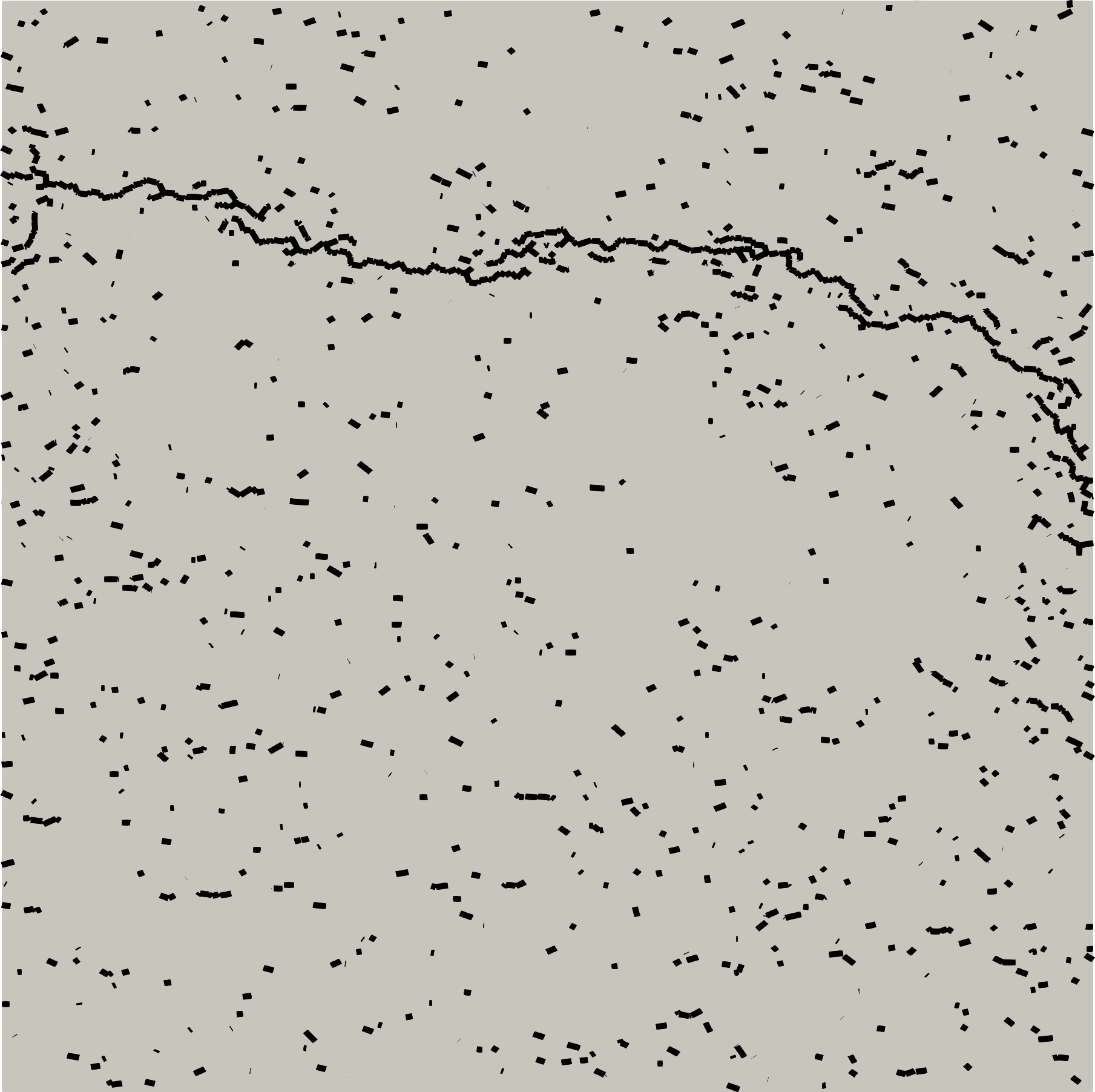}
    \caption{30 \%}
    \label{fig:willam_30}
  \end{subfigure}
  \begin{subfigure}{.18\textwidth}
    \centering
    \includegraphics[width=.9\linewidth]{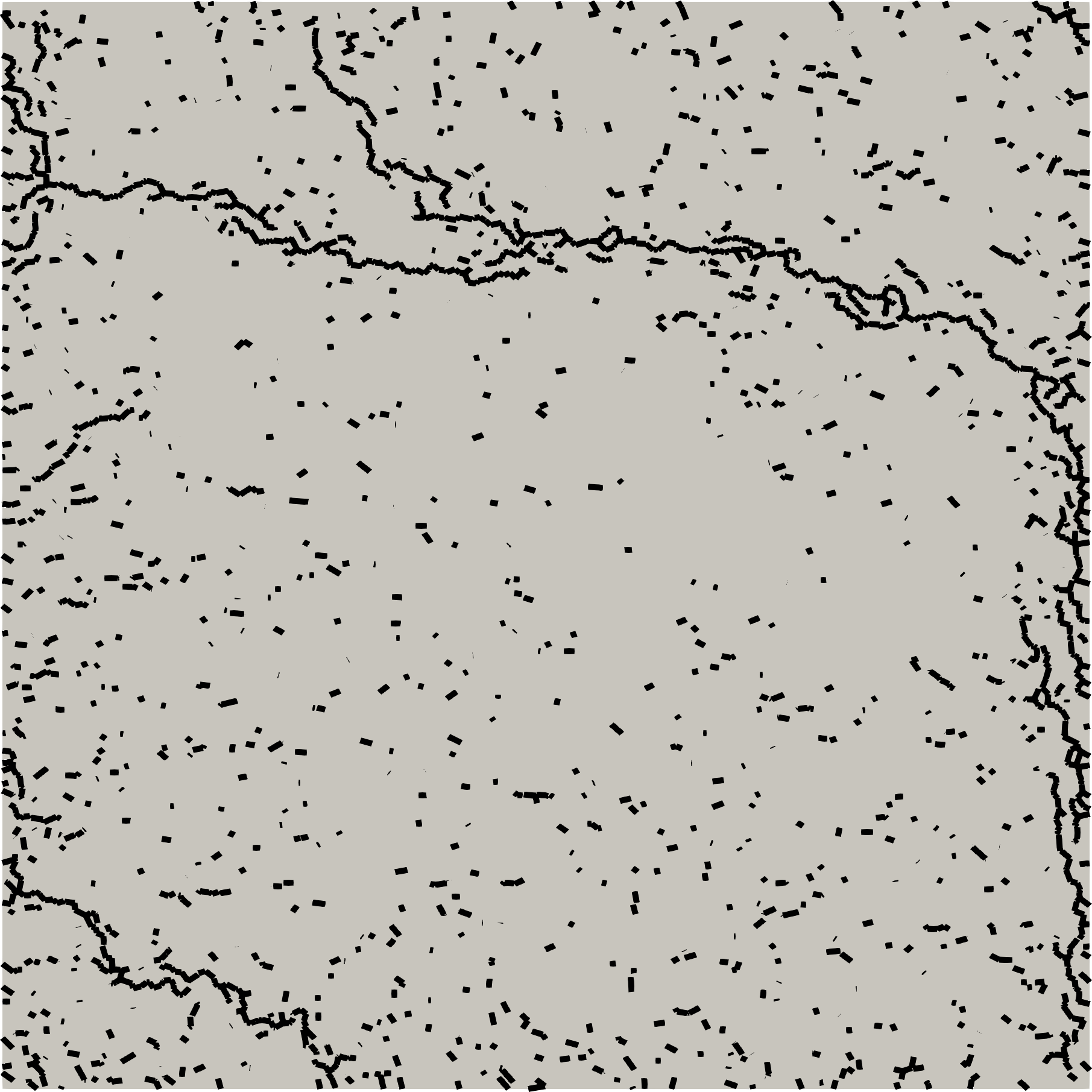}
    \caption{50 \%}
    \label{fig:willam_50}
  \end{subfigure}
  \begin{subfigure}{.18\textwidth}
    \centering
    \includegraphics[width=.9\linewidth]{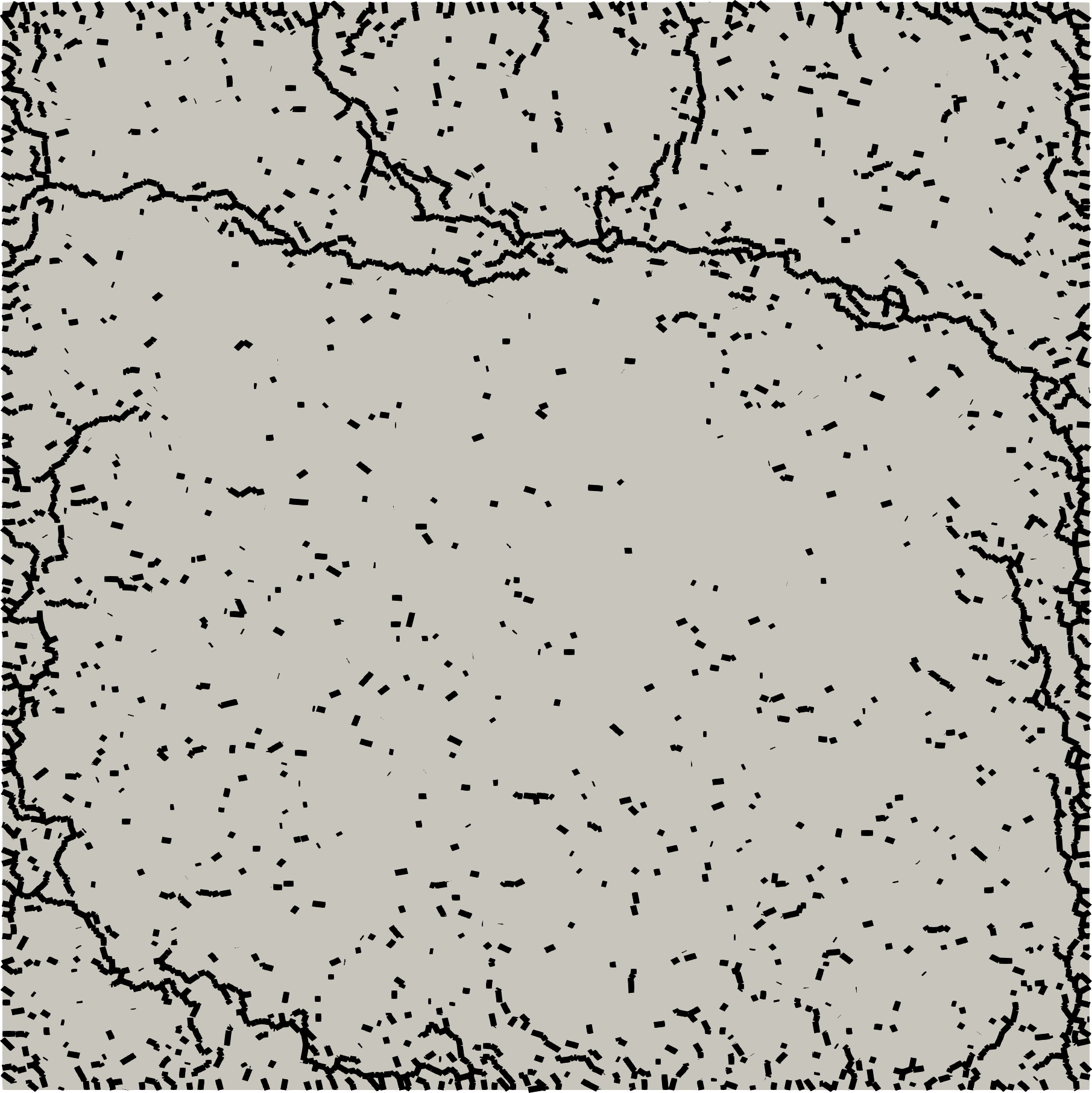}
    \caption{70 \%}
    \label{fig:willam_70}
  \end{subfigure}
  \caption{Willam loading: evolution of the crack pattern with the ratio of broken beams.}
  \label{fig:willam_pattern}
\end{figure*}

Those stages can be observed on the computed cracking patterns of figures~\ref{fig:traction_pattern}
to~\ref{fig:willam_pattern}.

We note that the results are very dependent on the loading studied. However, we can draw a general conclusion from the curves presented above: the effective compliance tensor is close to remain orthotropic even at very high levels of cracking. Indeed, the harmonic part of the effective compliance tensor remains negligible in most loading cases.
This last observation is consistent with Kachanov dilute (non interacting) micro-cracking theory, for which the harmonic part of the compliance tensor remains zero for any network of diffuse and open micro-cracks. Furthermore, it is most often satisfied here even in cases of localizing and interacting cracks.

When this harmonic part is not negligible, as in the cases of compression (see figure~\ref{fig:cont_comp_compression}) or shear (see figure~\ref{fig:cont_comp_shear}), the effective compliance tensor does not necessarily becomes anisotropic without symmetry. Indeed, as it is observed in figure~\ref{fig:comp_orth}, $\tilde \bS$  remains close to belong to the orthotropic symmetry class ($\Delta < 15 \%$), and this up to quite high values of damage.

\section{Extraction of damage tensors}
\label{sec:damage}

A natural definition of the damage variable from the effective elasticity tensor leads to a fourth-order tensor \citep{Cha1979,LO1980}. However, the use of a symmetric second-order tensor is common due to the simplicity of its interpretation \citep{CS1982,Mur1988}.

The definition of a second-order damage variable is justified here since we have checked that the elasticity tensors, and thus the fourth-order damage tensors, remain close to the orthotropic symmetric class -- the symmetry class of a generic second-order tensor-- even at high levels of damage.
Nonetheless, the definition of such a damage variable is not straightforward. After choosing between a definition based on the compliance tensor or the stiffness tensor, one must select the degradations rendered by this damage variable, which can be related either to the bulk modulus, to the shear modulus or possibly to a combination of both.

To validate the definition of a symmetric second-order tensor as damage variable, it must be checked that it is positive definite, that its eigenvalue are bounded by one and that its time derivative is positive definite, the later property being related to the second principle of thermodynamics \citep{Des2006}. One should note that those properties are verified by the fourth-order tensor per definition. This is on the other hand the modelling choices that will ensure, or not, those properties for the second-order tensor.

\subsection{From compliance tensor}

Following M. Kachanov, a natural definition for an anisotropic damage variable is related to the compliance tensor. More precisely, we propose here to define it from the difference between the invert of the effective bulk modulus, very sensitivitive to damage, and the invert of the initial bulk modulus,
\begin{equation*}
  \frac{1}{\tilde\kappa}-\frac{1}{\kappa}
  =\tr\bd(\tilde \bS- \bS)
  =\tr\tr_{12}\left( \tilde \bS- \bS\right).
\end{equation*}

We first define a dimensionless symmetric second-order tensor,
\begin{equation*}
  \bOmega:=\kappa\left(\bd(\tilde\bS)- \bd(\bS)\right)
  =\kappa \tr_{12}\left( \tilde\bS- \bS\right),
\end{equation*}
whose trace is $ \kappa/\tilde\kappa-1\geq 0$, and of eigenvalues expected to be positive but unbounded ($0\leq \Omega_i<\infty$). A dimensionless symmetric second-order damage variable is obtained as
\begin{equation*}
  \widehat \bD:= \bOmega \left(\Idd +\bOmega\right)^{-1}=\left(\Idd +\bOmega\right)^{-1}\bOmega .
\end{equation*}
One should note that the eigenvalues of $\widehat \bD$ remain positive and bounded by 1 provided those of $\bOmega$ are positive.

\begin{figure}[htp]
  \centering
  \includegraphics[align=c,width=0.8\linewidth]{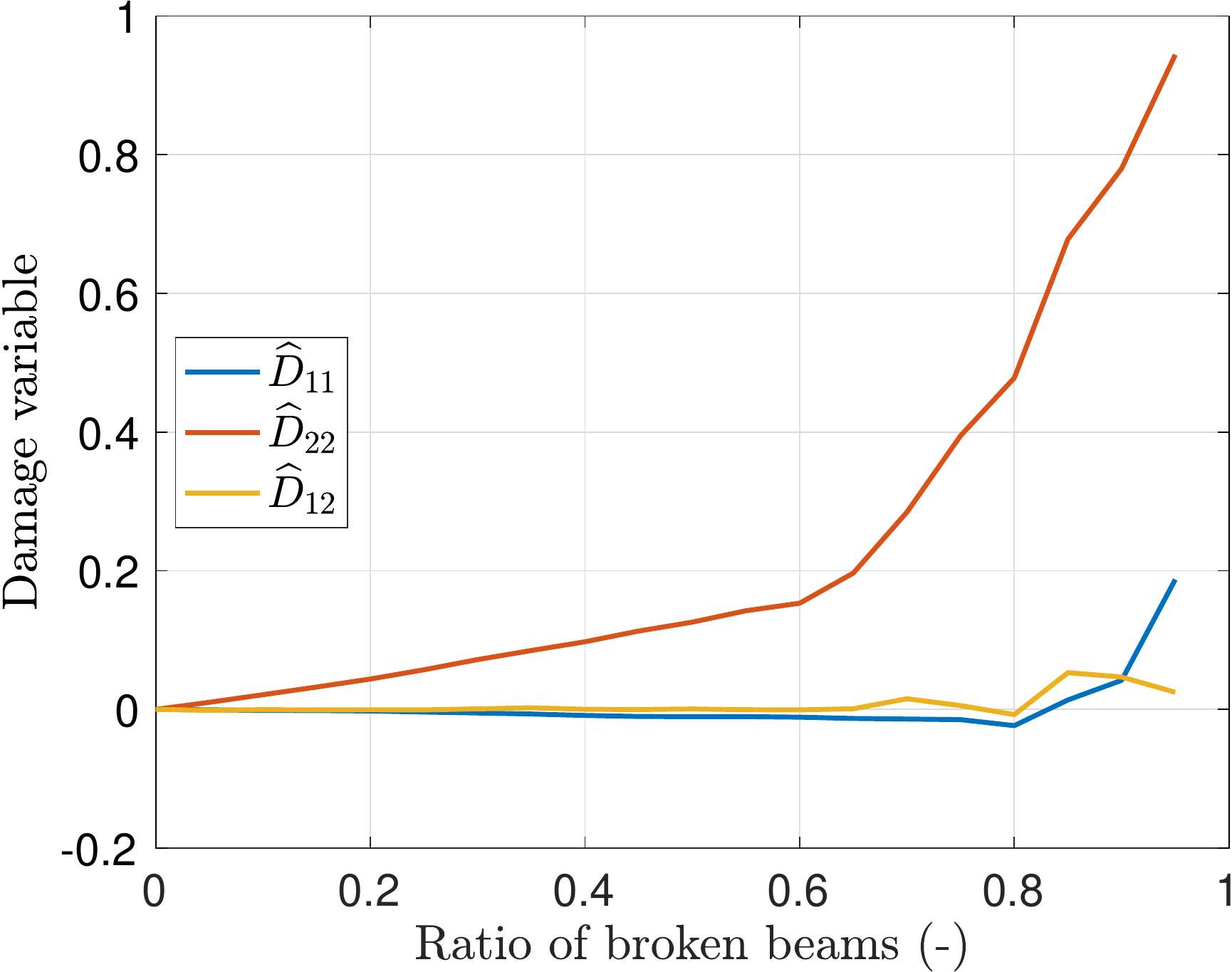}
  \caption{Tensile loading: Evolution of the components of the damage tensor $\widehat \bD$ with the ratio of broken beams.}
  \label{fig:endo_souplesse_traction_Dhat}
\end{figure}

\begin{figure}[htp]
  \centering
  \includegraphics[align=c,width=0.8\linewidth]{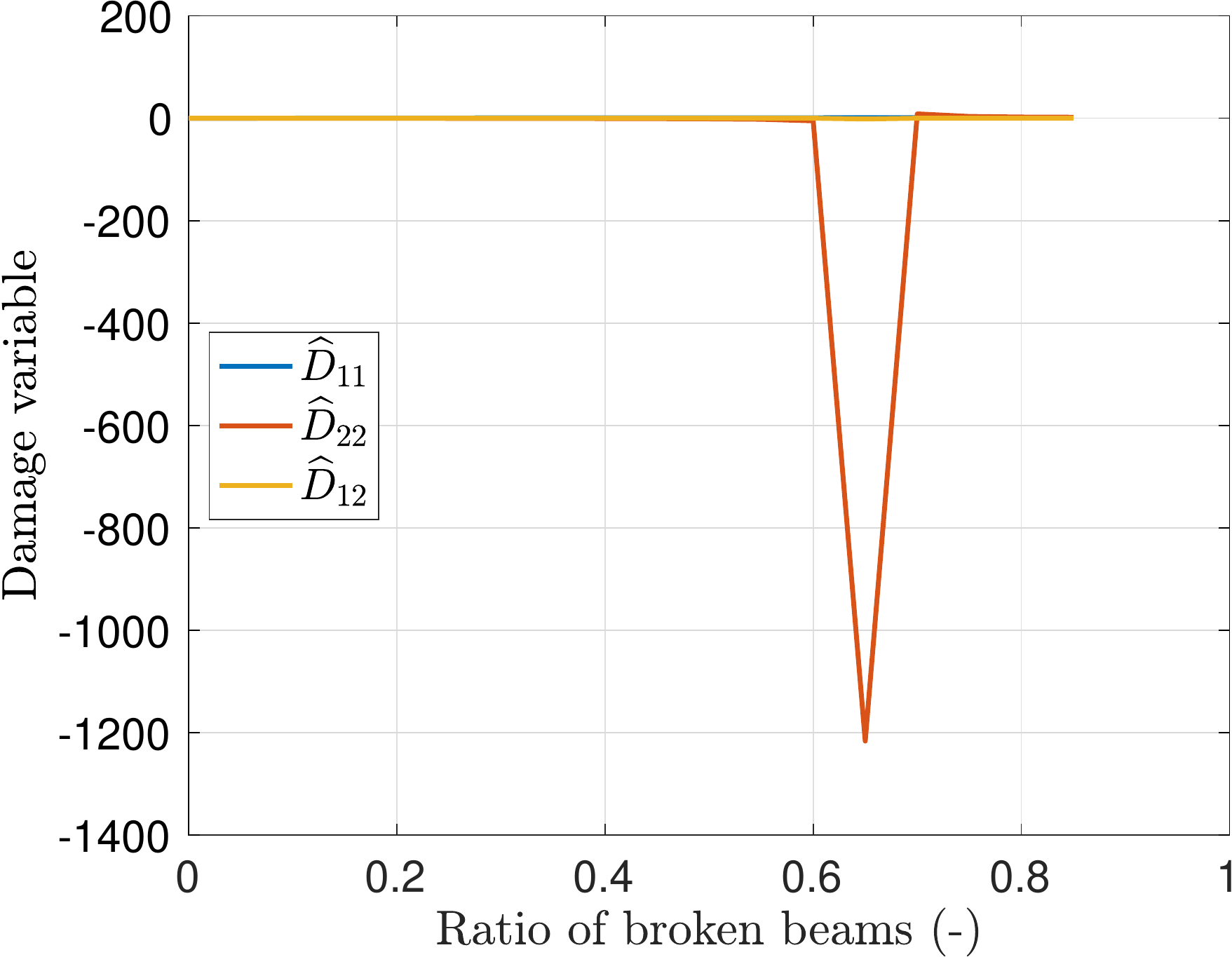}
  \caption{Compressive loading: Evolution of the components of the damage tensor $\widehat \bD$ with the ratio of broken beams.}
  \label{fig:endo_souplesse_compression_Dhat}
\end{figure}

\begin{figure}[htp]
  \centering
  \includegraphics[align=c,width=0.8\linewidth]{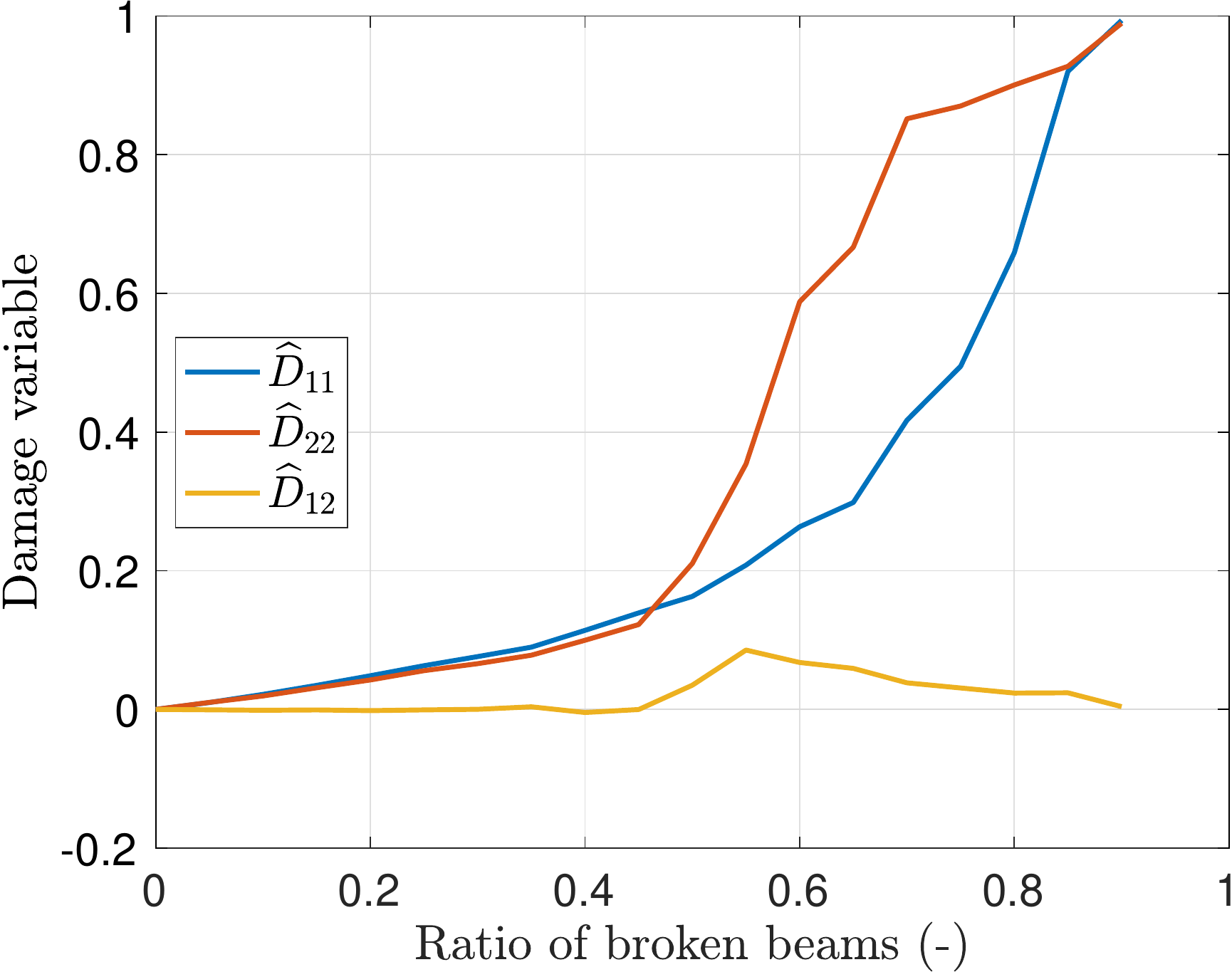}
  \caption{Bi-Tensile loading: Evolution of the components of the damage tensor $\widehat \bD$ with the ratio of broken beams.}
  \label{fig:endo_souplesse_bitraction_Dhat}
\end{figure}

\begin{figure}[htp]
  \centering
  \includegraphics[align=c,width=0.8\linewidth]{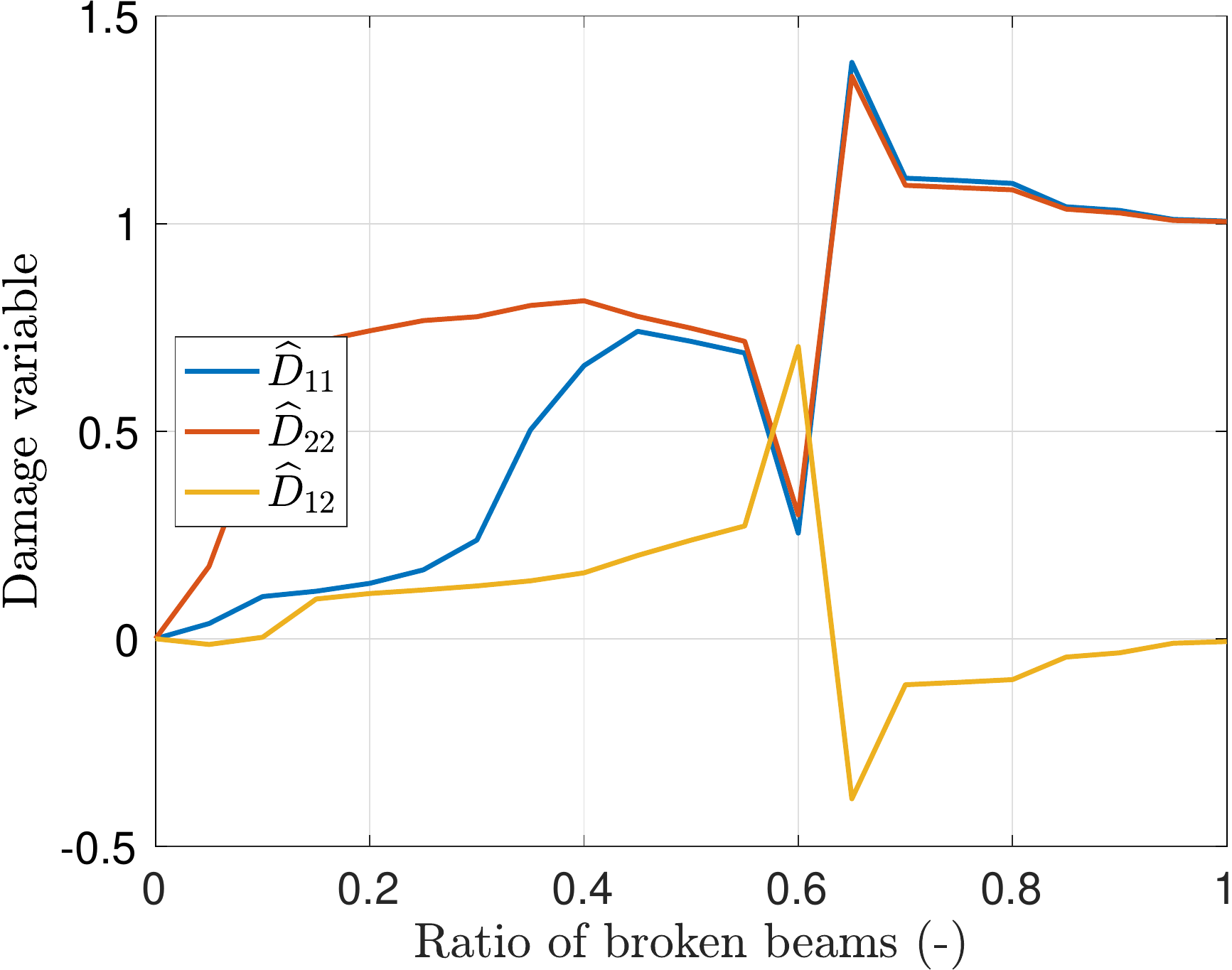}
  \caption{Simple shear loading: Evolution of the components of the damage tensor $\widehat \bD$ with the ratio of broken beams.}
  \label{fig:endo_souplesse_cisaillement_Dhat}
\end{figure}

\begin{figure}[htp]
  \centering
  \includegraphics[align=c,width=0.8\linewidth]{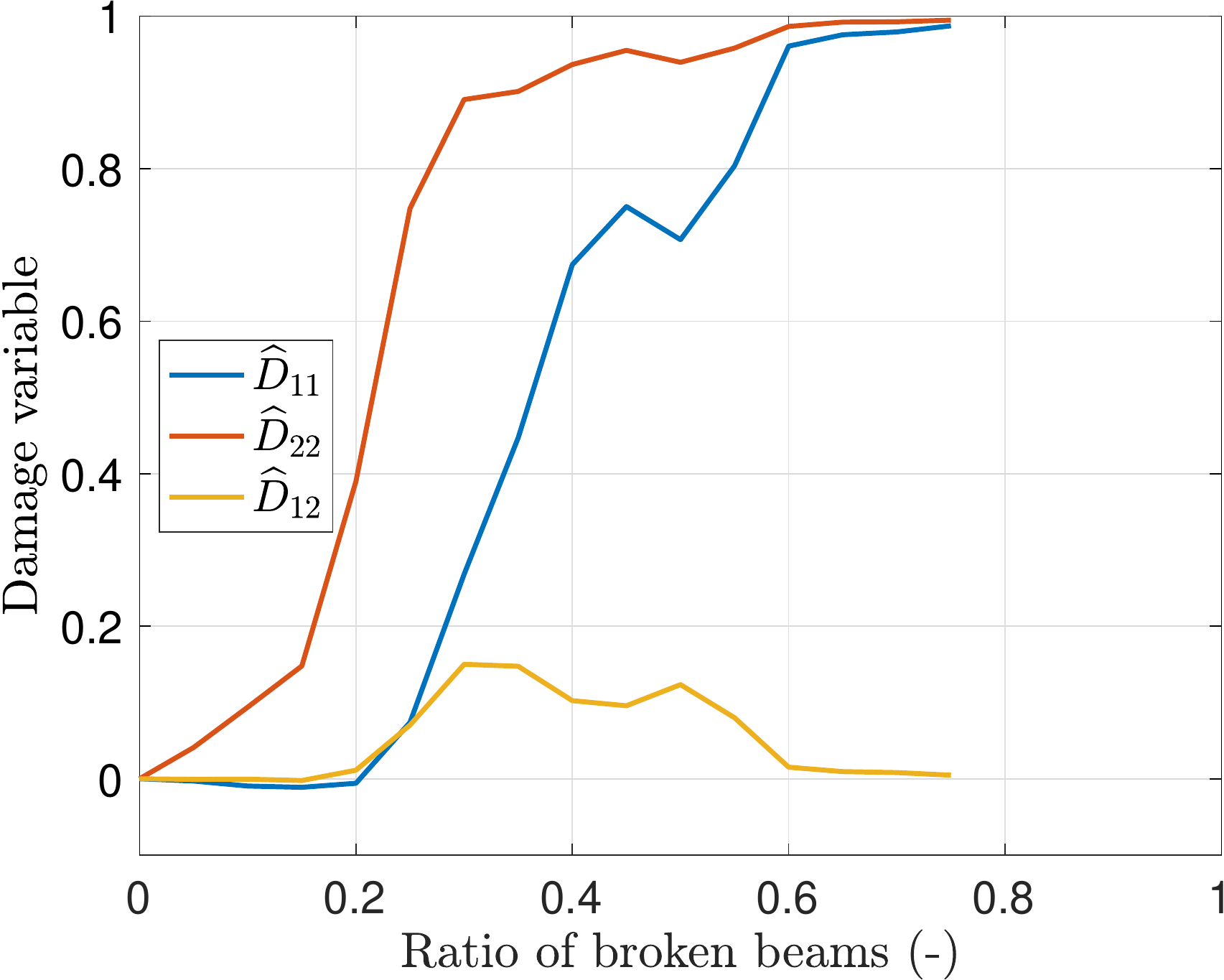}
  \caption{Willam loading: Evolution of the components of the damage tensor $\widehat \bD$ with the ratio of broken beams.}
  \label{fig:endo_souplesse_willam_Dhat}
\end{figure}

The figures~\ref{fig:endo_souplesse_traction_Dhat}
to~\ref{fig:endo_souplesse_willam_Dhat} show the evolutions of the components of the damage tensor $\widehat \bD$ with the ratio of broken beams for the studied loadings. One can see that the values of the components of $\widehat \bD$ are not bounded by 0 and 1. In view of the values obtained, this is in a non surprising manner also the case for the eigenvalues. Therefore, this definition based solely on the invert of the initial bulk modulus is is not compatible with a thermodynamics framework. Indeed, it does not ensure the positivity of $\bOmega$. The most problematic cases are the compressive loading (Fig.~\ref{fig:endo_souplesse_compression_Dhat}) and the simple shear loading (Fig.~\ref{fig:endo_souplesse_cisaillement_Dhat}), for which the evolution of the shear modulus might certainly not be easily linked to the evolution of the bulk modulus. A solution would be to introduce this shear modulus in the definition of the damage variable itself. However, this solution would be difficult to implement because one would need to introduce some parameters to combine both the bulk modulus part and  the shear modulus part, parameters that will have to be identified later.

\subsection{From stiffness tensor}
\label{S:DfromC}

A second -- alternative and, one will see, preferred -- anisotropic damage variable can be defined as related to the effective bulk modulus instead of its inverse, more precisely as related to the difference
\begin{equation*}
  \kappa-\tilde \kappa=\frac{1}{4}\tr\bd(\bC-\tilde\bC)
  =\frac{1}{4}\tr\tr_{12}\left( \bC-\tilde \bC\right)
\end{equation*}
We propose to define the damage variable as the dimensionless symmetric second order tensor:
\begin{align*}
  \bD & :=\bd(\bC)^{-1}\left(\bd(\bC)-\bd(\tilde\bC)\right)
  \\
      & =\left(\bd(\bC)-\bd(\tilde\bC)\right)\bd(\bC)^{-1}
\end{align*}
as $\bd(\bC)=\tr_{12} \bC=2\kappa\, \Idd$ is spherical due to initial isotropy, so that the previous commutativity property holds, with
\begin{equation}
  \bD=\frac{1}{2\kappa}\left(\bd(\bC)-\bd(\tilde\bC)\right)=\frac{1}{2\kappa}\tr_{12}\left( \bC- \tilde \bC\right)
\end{equation}

\begin{figure}[htp]
  \centering
  \includegraphics[align=c,width=0.8\linewidth]{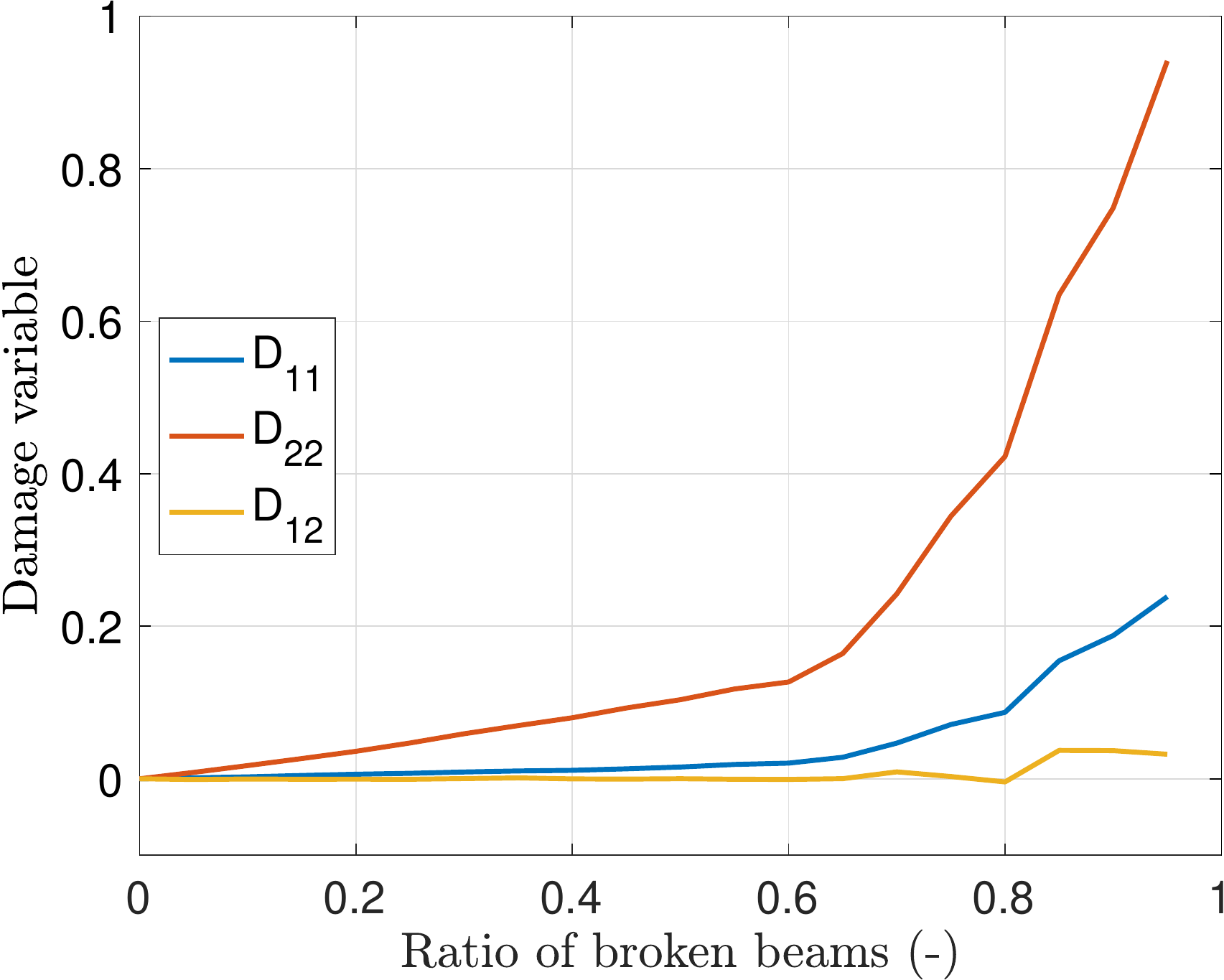}
  \caption{Tensile loading: Evolution of the components of the damage tensor $\bD$ with the ratio of broken beams.}
  \label{fig:endo_raideur_traction}
\end{figure}

\begin{figure}[htp]
  \centering
  \includegraphics[align=c,width=0.8\linewidth]{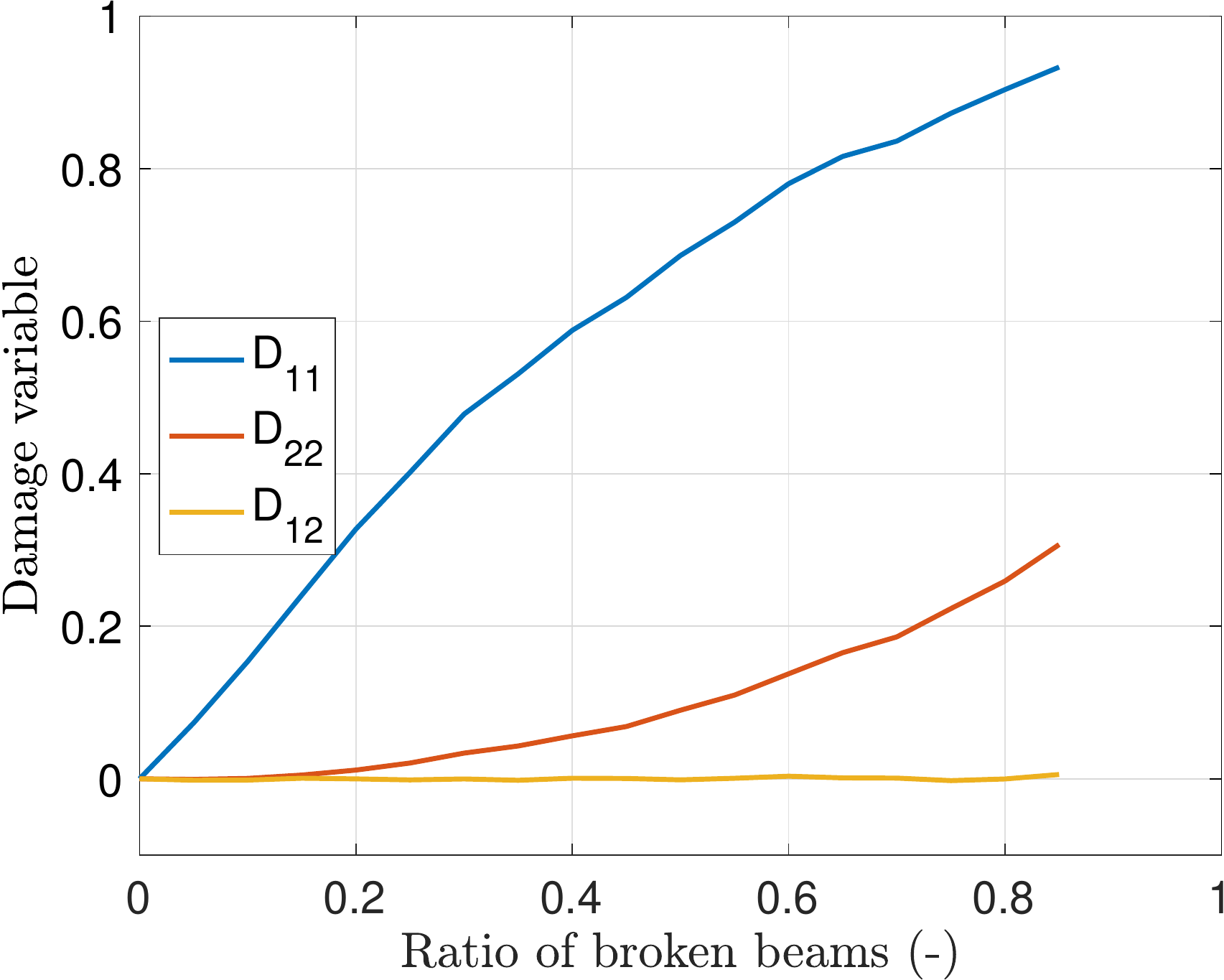}
  \caption{Compressive loading: Evolution of the components of the damage tensor $\bD$ with the ratio of broken beams.}
  \label{fig:endo_raideur_compression}
\end{figure}

\begin{figure}[htp]
  \centering
  \includegraphics[align=c,width=0.8\linewidth]{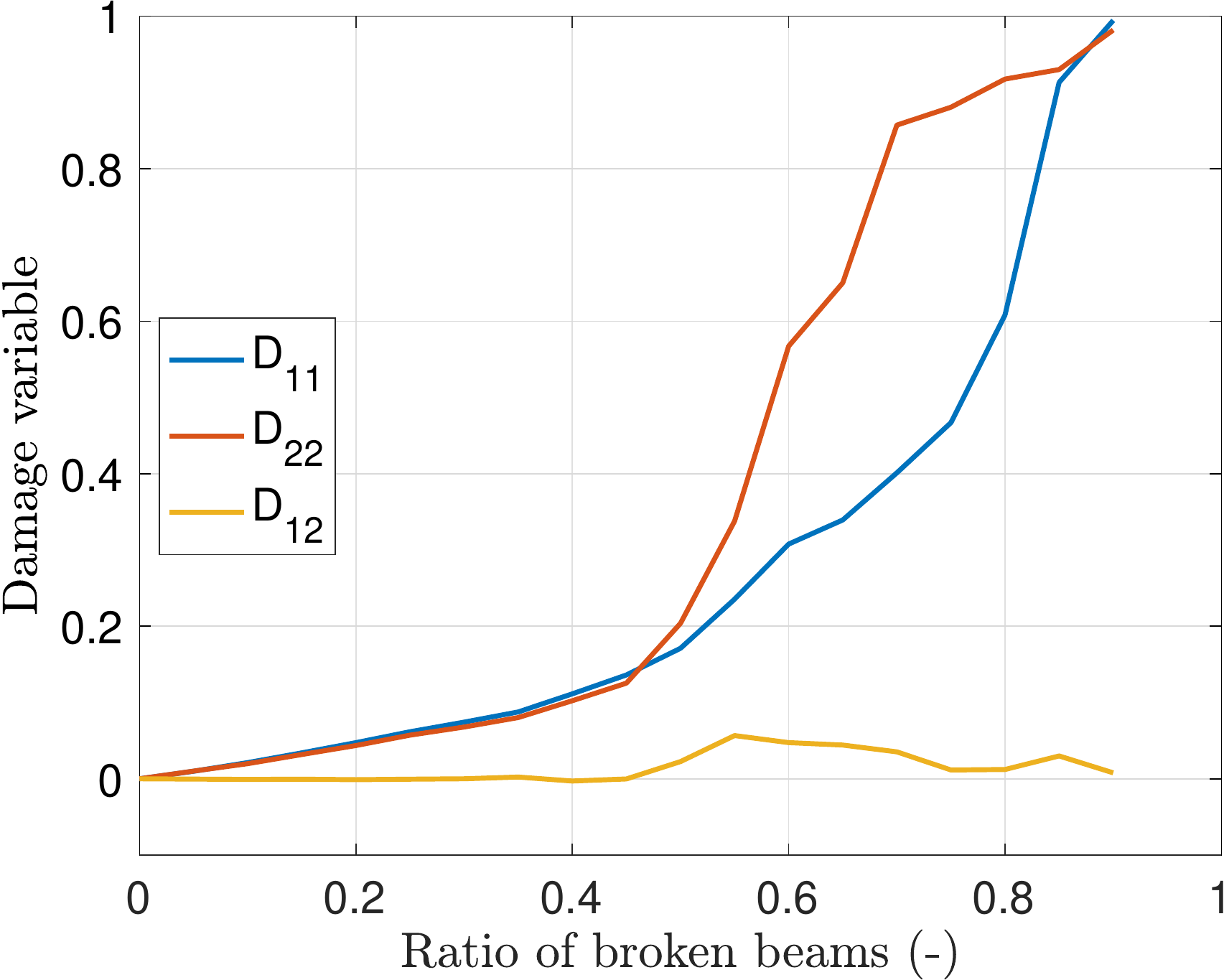}
  \caption{Bi-Tensile loading: Evolution of the components of the damage tensor $\bD$ with the ratio of broken beams.}
  \label{fig:endo_raideur_bitraction}
\end{figure}

\begin{figure}[htp]
  \centering
  \includegraphics[align=c,width=0.8\linewidth]{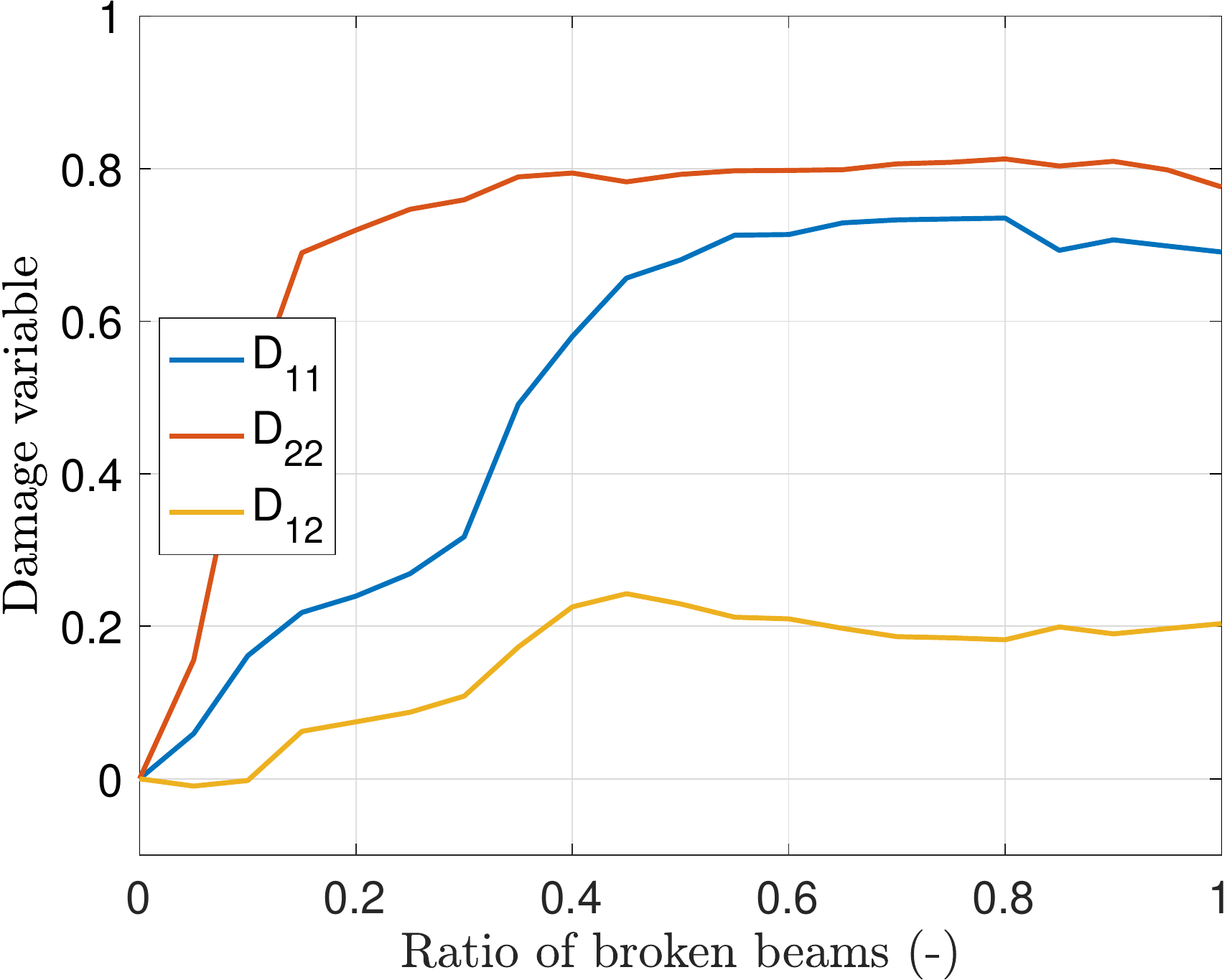}
  \caption{Simple shear loading: Evolution of the components of the damage tensor $\bD$ with the ratio of broken beams.}
  \label{fig:endo_raideur_cisaillement}
\end{figure}

\begin{figure}[htp]
  \centering
  \includegraphics[align=c,width=0.8\linewidth]{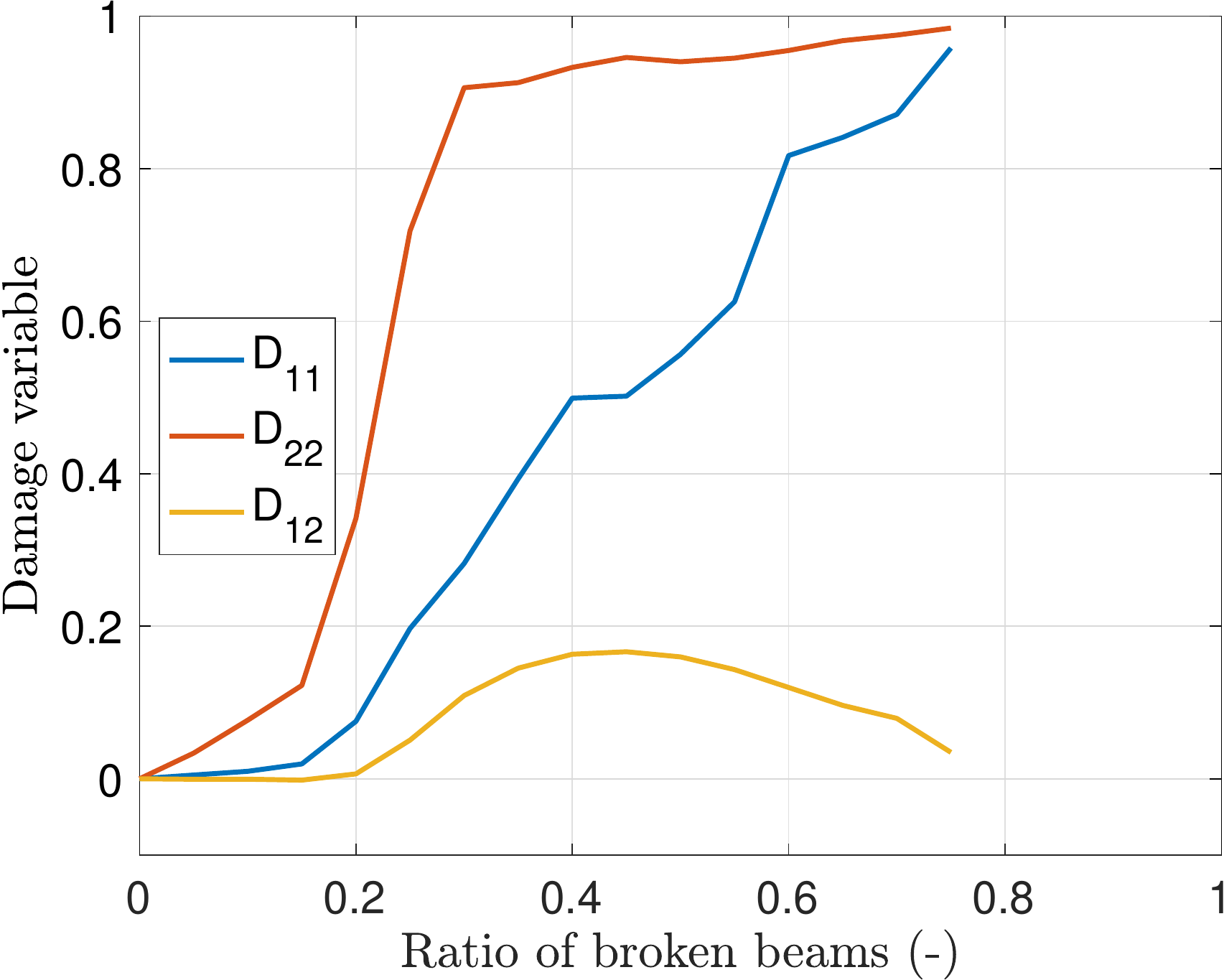}
  \caption{Willam loading: Evolution of the components of the damage tensor $\bD$ with the ratio of broken beams.}
  \label{fig:endo_raideur_willam}
\end{figure}

One must check that the damage tensor $\bD$ thus defined has positive eigenvalues bounded by 1 ($0\leq D_i\leq 1$).
The figures~\ref{fig:endo_raideur_traction}
to~\ref{fig:endo_raideur_willam} show the evolutions of the components of the damage tensor $\bD$ with the ratio of broken beams for the studied loadings.

We can observe that, this time, the components of the second-order tensorial damage variable remain postive and bounded by 1.
In most cases, a maximum damage close to 1 is reached on at least one of the tensor components before the end of the loading. This may explain why it is no longer possible to extract the effective elasticity tensors even though the ratio of broken beams has not yet reached 1.

\begin{figure}[htp]
  \centering
  \includegraphics[align=c,width=0.8\linewidth]{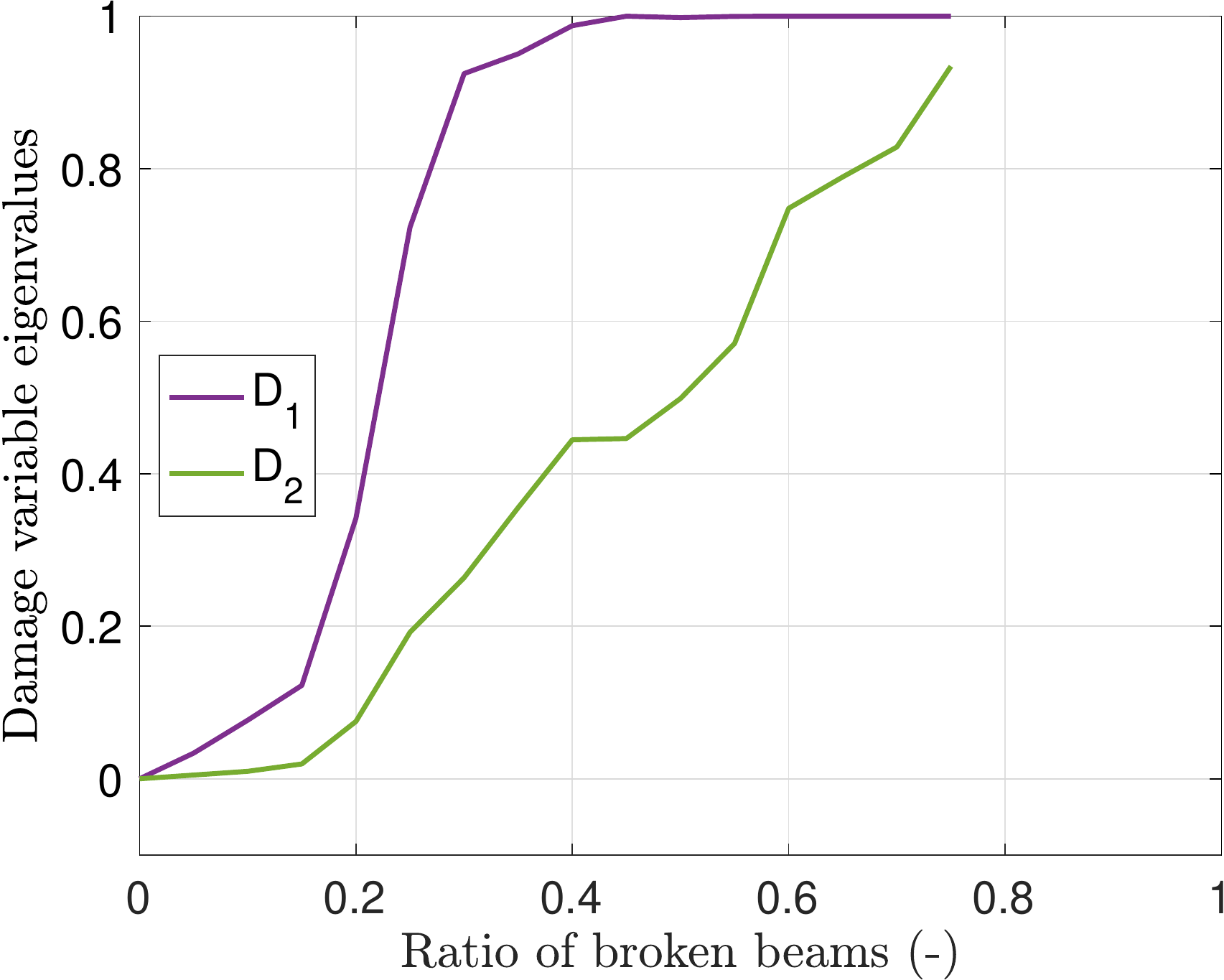}
  \caption{Willam loading: Evolution of the eigenvalues of the damage tensor $\bD$ with the ratio of broken beams.}
  \label{fig:endo_raideur_willam_vp}
\end{figure}

In pure proportional loading we expect a strict increase in the components of the tensor damage variable. This is not always the case. Indeed, in the case of bi-tension (Fig.~\ref{fig:endo_raideur_bitraction}) and simple shear (Fig.~\ref{fig:endo_raideur_cisaillement}), the slight decreases of the shear component $D_{12}$ is mainly related to local non proportionality (and rotation of the principal axes of tensor $\bD$).
The drop is clear in the Willam loading case (see figure~\ref{fig:endo_raideur_willam}). However, the damage eigenvalues $D_1, D_2$ of $\bD$ always increase for this loading 
as illustrated in figure~\ref{fig:endo_raideur_willam_vp}.

\section{Conclusion}
\label{sec:conclusion}

An upper bound to the distance to orthotropy has been obtained for bi-dimensional elasticity tensors. It naturally introduces a second --instead of fourth-- order tensor which models the medium orthotropy. This has allowed us to propose and measure second order tensorial damages variables fully representative of the effective anisotropic degradation due to complex cracking patterns. We have performed 2D discrete simulations with a beam-particle model of initially isotropic Representative Area Elements, allowing for strong interactions between cracks (up to their coalescence and complete failure). We have analyzed these simulations in a systematic manner.

The  results associated with the discrete computations confirm the accuracy of the orthotropic approximation of the elasticity tensor for a 2D cracked medium, under both proportional or non-proportional loading cases. They fully justify the use of a single second order damage tensor variable, instead of a fourth order one \citep{Cha1979,LO1980} or of two second order damage variables \citep{desmorat2016second}, even in the strong crack interaction case.

Two definitions, in stiffness and in compliance, for a symmetric second-order damage tensor have been proposed and studied in this paper. They are both based on the evolution of the bulk modulus (or of its inverse). It has been observed that the damage variable derived from the stiffness tensor has the necessary properties to be a suitable candidate for the formulation of a continuous anisotropic damage model in 2D from discrete simulations.

The methodology followed does not fix \emph{a priori} a particular working basis, it makes accessible all the tensorial (multiaxial) components measurement of the damage variable are any time step, therefore of its evolution.

\appendix

\section{Harmonic square roots of bi-dimensional harmonic tensors}
\label{sec:harmonic-square-root}

The goal of this appendix is to provide a simple proof of the following theorem, using results in \citep{desmorat2020computation}.

\begin{thm}
  Any fourth-order harmonic bi-dimensional tensor $\bH$ can be written either as an harmonic product $\bh \ast \bh$ (an harmonic square), or as $-\bk \ast \bk$ (the opposite of an harmonic square), where $\bh$ and $\bk$ are second-order bi-dimensional harmonic tensors.
\end{thm}

To prove this result, we will recall first that there is an isomorphism $\bH \cong \rh$ between bi-dimensional harmonic tensors $\bH$ of order $n$ and homogeneous harmonic polynomials $\rh$ of degree $n$ in two variables $x$ and $y$. Rather than $x$ and $y$, one can use the complex variables $z=x +i y$ and $\bar z=x -i y$. Then, any homogeneous harmonic polynomial of degree $n$ writes as $\rh_{1} = \Re(\bar z_1 z^n)$, where $z_1\in \CC$ and the harmonic product between two homogeneous harmonic polynomials $\rh_{1}$ and $\rh_{2}$, of respective degree $n_{1}$ and $n_{1}$ translates into
\begin{equation*}
  \rh_{1} \ast \rh_{2} = \frac{1}{2}\Re(\overline{z_1z_2}z^{n_{1}+n_{2}}).
\end{equation*}

For instance, the deviatoric second order tensors
\begin{equation*}
  \bh =
  \begin{pmatrix}
    a_1 & b_1  \\
    b_1 & -a_1
  \end{pmatrix}
  \quad \text{and} \quad
  \bk =
  \begin{pmatrix}
    a_2 & b_2   \\
    b_2 & - a_2
  \end{pmatrix}
\end{equation*}
are represented respectively by the harmonic polynomials
\begin{equation*}
  \rh_1 = \Re(\bar z_1 z^2), \quad \text{and} \quad \rh_2 = \Re(\bar z_2 z^2),
\end{equation*}
where $z_1=a_1+i b_1$ and $z_2=a_2+i b_2$, and an harmonic fourth order tensor $\bH$ (represented by its Kelvin matrix~\eqref{eq:KelvH}) corresponds to the homogeneous harmonic polynomial $\rh_3 = \Re(\bar z_3 z^4)$, where
\begin{equation*}
  z_3=H_{1111} +i H_{1112}.
\end{equation*}

Thus, the harmonic square tensorial equations
\begin{equation*}
  \bH = \bh \ast \bh = -\bk \ast \bk
\end{equation*}
translate, in terms of harmonic homogeneous polynomials, as
\begin{equation*}
  \rh_3 = \Re(\bar z_3 z^4) = \Re(\bar z_1^2 z^4) = -\Re(\bar z_2^2 z^4).
\end{equation*}
and the solutions are provided by roots of the algebraic equations
\begin{equation}\label{eq:z3z12}
   z_1^2 = - z_2^2 = z_3.
\end{equation}

\begin{rem}
  Note that both equations $\bH = \bh \ast \bh$ and $\bH = -\bk \ast \bk$ have exactly two opposite solutions, when $\bH \ne 0$.
\end{rem}




\begin{thebibliography}{48}
\expandafter\ifx\csname natexlab\endcsname\relax\def\natexlab#1{#1}\fi
\providecommand{\url}[1]{\texttt{#1}}
\providecommand{\href}[2]{#2}
\providecommand{\path}[1]{#1}
\providecommand{\DOIprefix}{doi:}
\providecommand{\ArXivprefix}{arXiv:}
\providecommand{\URLprefix}{URL: }
\providecommand{\Pubmedprefix}{pmid:}
\providecommand{\doi}[1]{\href{http://dx.doi.org/#1}{\path{#1}}}
\providecommand{\Pubmed}[1]{\href{pmid:#1}{\path{#1}}}
\providecommand{\bibinfo}[2]{#2}
\ifx\xfnm\relax \def\xfnm[#1]{\unskip,\space#1}\fi
\bibitem[{André et~al.(2012)André, Iordanoff, Charles and
  Néauport}]{andre2012discrete}
\bibinfo{author}{André, D.}, \bibinfo{author}{Iordanoff, I.},
  \bibinfo{author}{Charles, J.l.}, \bibinfo{author}{Néauport, J.},
  \bibinfo{year}{2012}.
\newblock \bibinfo{title}{Discrete element method to simulate continuous
  material by using the cohesive beam model}.
\newblock \bibinfo{journal}{Computer methods in applied mechanics and
  engineering} \bibinfo{volume}{213}, \bibinfo{pages}{113--125}.
\bibitem[{Backus(1970)}]{backus1970geometrical}
\bibinfo{author}{Backus, G.}, \bibinfo{year}{1970}.
\newblock \bibinfo{title}{A geometrical picture of anisotropic elastic
  tensors}.
\newblock \bibinfo{journal}{Reviews of geophysics} \bibinfo{volume}{8},
  \bibinfo{pages}{633--671}.
\bibitem[{Bagi(1996)}]{bagi1996stress}
\bibinfo{author}{Bagi, K.}, \bibinfo{year}{1996}.
\newblock \bibinfo{title}{Stress and strain in granular assemblies}.
\newblock \bibinfo{journal}{Mechanics of materials} \bibinfo{volume}{22},
  \bibinfo{pages}{165--177}.
\bibitem[{Chaboche(1979)}]{Cha1979}
\bibinfo{author}{Chaboche, J.L.}, \bibinfo{year}{1979}.
\newblock \bibinfo{title}{Le concept de contrainte effective appliqu\'e \`a
  l'\'elasticit\'e et \`a la viscoplasticit\'e en pr\'esence d'un endommagement
  anisotrope}, in: \bibinfo{editor}{Boehler, J.P.} (Ed.),
  \bibinfo{booktitle}{Colloque Int. CNRS 295, Villard de Lans},
  \bibinfo{publisher}{Martinus Nijhoff Publishers and Editions du CNRS, 1982}.
  pp. \bibinfo{pages}{737--760}.
\bibitem[{Challamel et~al.(2015)Challamel, Picandet and
  Pijaudier-Cabot}]{challamel2015discrete}
\bibinfo{author}{Challamel, N.}, \bibinfo{author}{Picandet, V.},
  \bibinfo{author}{Pijaudier-Cabot, G.}, \bibinfo{year}{2015}.
\newblock \bibinfo{title}{From discrete to nonlocal continuum damage mechanics:
  Analysis of a lattice system in bending using a continualized approach}.
\newblock \bibinfo{journal}{International Journal of Damage Mechanics}
  \bibinfo{volume}{24}, \bibinfo{pages}{983--1012}.
\bibitem[{Cordebois and Sidoroff(1982)}]{CS1982}
\bibinfo{author}{Cordebois, J.}, \bibinfo{author}{Sidoroff, F.},
  \bibinfo{year}{1982}.
\newblock \bibinfo{title}{Endommagement anisotrope en \'elasticit\'e et
  plasticit\'e}.
\newblock \bibinfo{journal}{J. Meca. Th. Appl., Special Volume} ,
  \bibinfo{pages}{45--65}.
\bibitem[{Cundall and Strack(1979)}]{cundall1979discrete}
\bibinfo{author}{Cundall, P.A.}, \bibinfo{author}{Strack, O.D.},
  \bibinfo{year}{1979}.
\newblock \bibinfo{title}{A discrete numerical model for granular assemblies}.
\newblock \bibinfo{journal}{geotechnique} \bibinfo{volume}{29},
  \bibinfo{pages}{47--65}.
\bibitem[{D'Addetta et~al.(2002)D'Addetta, Kun and
  Ramm}]{daddetta2002application}
\bibinfo{author}{D'Addetta, G.A.}, \bibinfo{author}{Kun, F.},
  \bibinfo{author}{Ramm, E.}, \bibinfo{year}{2002}.
\newblock \bibinfo{title}{On the application of a discrete model to the
  fracture process of cohesive granular materials}.
\newblock \bibinfo{journal}{Granular Matter} \bibinfo{volume}{4},
  \bibinfo{pages}{77--90}.
\bibitem[{Delaplace(2008)}]{delaplace2008modelisation}
\bibinfo{author}{Delaplace, A.}, \bibinfo{year}{2008}.
\newblock \bibinfo{title}{Modélisation discrète appliquée au
  comportement des matériaux et des structures}.
\newblock \bibinfo{journal}{Mémoire d'habilitation à diriger des
  recherches de l'Ecole Normale Supérieure de Cachan} .
\bibitem[{Delaplace and Desmorat(2007)}]{delaplace2007discrete}
\bibinfo{author}{Delaplace, A.}, \bibinfo{author}{Desmorat, R.},
  \bibinfo{year}{2007}.
\newblock \bibinfo{title}{Discrete 3d model as complimentary numerical testing
  for anisotropic damage}.
\newblock \bibinfo{journal}{Int J Fract} \bibinfo{volume}{148},
  \bibinfo{pages}{115--128}.
\bibitem[{Desmorat and Desmorat(2015)}]{desmorat2015tensorial}
\bibinfo{author}{Desmorat, B.}, \bibinfo{author}{Desmorat, R.},
  \bibinfo{year}{2015}.
\newblock \bibinfo{title}{Tensorial polar decomposition of 2d fourth-order
  tensors}.
\newblock \bibinfo{journal}{Comptes Rendus Mécanique}
  \bibinfo{volume}{343}, \bibinfo{pages}{471--475}.
\bibitem[{Desmorat and Desmorat(2016)}]{desmorat2016second}
\bibinfo{author}{Desmorat, B.}, \bibinfo{author}{Desmorat, R.},
  \bibinfo{year}{2016}.
\newblock \bibinfo{title}{Second order tensorial framework for 2d medium with
  open and closed cracks}.
\newblock \bibinfo{journal}{European Journal of Mechanics-A/Solids}
  \bibinfo{volume}{58}, \bibinfo{pages}{262--277}.
\bibitem[{Desmorat et~al.(2020)Desmorat, Olive, Auffray, Desmorat and
  Kolev}]{desmorat2020computation}
\bibinfo{author}{Desmorat, B.}, \bibinfo{author}{Olive, M.},
  \bibinfo{author}{Auffray, N.}, \bibinfo{author}{Desmorat, R.},
  \bibinfo{author}{Kolev, B.}, \bibinfo{year}{2020}.
\newblock \bibinfo{title}{Computation of minimal covariants bases for 2d
  coupled constitutive laws}.
\newblock \bibinfo{journal}{arXiv preprint arXiv:2007.01576} .
\bibitem[{Desmorat(2006)}]{Des2006}
\bibinfo{author}{Desmorat, R.}, \bibinfo{year}{2006}.
\newblock \bibinfo{title}{Positivity of intrinsic dissipation of a class of
  nonstandard anisotropic damage models}.
\newblock \bibinfo{journal}{C. R. Mecanique} \bibinfo{volume}{334},
  \bibinfo{pages}{587--592}.
\bibitem[{Desmorat(2016)}]{desmorat2016anisotropic}
\bibinfo{author}{Desmorat, R.}, \bibinfo{year}{2016}.
\newblock \bibinfo{title}{Anisotropic damage modeling of concrete materials}.
\newblock \bibinfo{journal}{International Journal of Damage Mechanics}
  \bibinfo{volume}{25}, \bibinfo{pages}{818--852}.
\bibitem[{Desmorat et~al.(2007)Desmorat, Gatuingt and
  Ragueneau}]{desmorat2007nonlocal}
\bibinfo{author}{Desmorat, R.}, \bibinfo{author}{Gatuingt, F.},
  \bibinfo{author}{Ragueneau, F.}, \bibinfo{year}{2007}.
\newblock \bibinfo{title}{Nonlocal anisotropic damage model and related
  computational aspects for quasi-brittle materials}.
\newblock \bibinfo{journal}{Engineering Fracture Mechanics}
  \bibinfo{volume}{74}, \bibinfo{pages}{1539--1560}.
\bibitem[{Halm and Dragon(1998)}]{HD1998}
\bibinfo{author}{Halm, D.}, \bibinfo{author}{Dragon, A.}, \bibinfo{year}{1998}.
\newblock \bibinfo{title}{An anisotropic model of damage and frictional sliding
  for brittle materials}.
\newblock \bibinfo{journal}{European Journal of Mechanics-A/Solids}
  \bibinfo{volume}{17}, \bibinfo{pages}{439--460}.
\bibitem[{Herrmann et~al.(1989)Herrmann, Hansen and
  Roux}]{herrmann1989fracture}
\bibinfo{author}{Herrmann, H.J.}, \bibinfo{author}{Hansen, A.},
  \bibinfo{author}{Roux, S.}, \bibinfo{year}{1989}.
\newblock \bibinfo{title}{Fracture of disordered, elastic lattices in two
  dimensions}.
\newblock \bibinfo{journal}{Physical Review B} \bibinfo{volume}{39},
  \bibinfo{pages}{637}.
\bibitem[{Hrennikoff(1941)}]{hrennikoff1941solution}
\bibinfo{author}{Hrennikoff, A.}, \bibinfo{year}{1941}.
\newblock \bibinfo{title}{Solution of problems of elasticity by the framework
  method}.
\newblock \bibinfo{journal}{J. appl. Mech.} .
\bibitem[{Jivkov(2014)}]{jivkov2014structure}
\bibinfo{author}{Jivkov, A.}, \bibinfo{year}{2014}.
\newblock \bibinfo{title}{Structure of micro-crack population and damage
  evolution in quasi-brittle media}.
\newblock \bibinfo{journal}{Theoretical and Applied Fracture Mechanics}
  \bibinfo{volume}{70}, \bibinfo{pages}{1--9}.
\bibitem[{Kachanov(1992)}]{kachanov1992effective}
\bibinfo{author}{Kachanov, M.}, \bibinfo{year}{1992}.
\newblock \bibinfo{title}{Effective elastic properties of cracked solids:
  critical review of some basic concepts}.
\newblock \bibinfo{journal}{Appl. Mech. Rev.} \bibinfo{volume}{45},
  \bibinfo{pages}{304--335}.
\bibitem[{Leckie and Onat(1981)}]{leckie1981tensorial}
\bibinfo{author}{Leckie, F.}, \bibinfo{author}{Onat, E.}, \bibinfo{year}{1981}.
\newblock \bibinfo{title}{Tensorial nature of damage measuring internal
  variables}, in: \bibinfo{booktitle}{Physical non-linearities in structural
  analysis}. \bibinfo{publisher}{Springer}, pp. \bibinfo{pages}{140--155}.
\bibitem[{Leckie and Onat(1980)}]{LO1980}
\bibinfo{author}{Leckie, F.A.}, \bibinfo{author}{Onat, E.T.},
  \bibinfo{year}{1980}.
\newblock \bibinfo{title}{Tensorial nature of damage measuring internal
  variables}. \bibinfo{publisher}{J. Hult and J. Lemaitre eds, Springer
  Berlin}. chapter \bibinfo{chapter}{Physical Non-Linearities in Structural
  Analysis}.
\newblock pp. \bibinfo{pages}{140--155}.
\bibitem[{Lemaitre and Chaboche(1985)}]{LC1985}
\bibinfo{author}{Lemaitre, J.}, \bibinfo{author}{Chaboche, J.L.},
  \bibinfo{year}{1985}.
\newblock \bibinfo{title}{M\'ecanique des mat\'eriaux solides}.
\newblock \bibinfo{publisher}{Dunod, english translation 1990 'Mechanics of
  Solid Materials' Cambridge University Press}.
\bibitem[{Lemaitre and Desmorat(2005)}]{lemaitre2005engineering}
\bibinfo{author}{Lemaitre, J.}, \bibinfo{author}{Desmorat, R.},
  \bibinfo{year}{2005}.
\newblock \bibinfo{title}{Engineering damage mechanics: ductile, creep, fatigue
  and brittle failures}.
\newblock \bibinfo{publisher}{Springer Science \& Business Media}.
\bibitem[{Mazars(1984)}]{mazars1984application}
\bibinfo{author}{Mazars, J.}, \bibinfo{year}{1984}.
\newblock \bibinfo{title}{Application de la mécanique de l'endommagement au
  comportement non linéaire et à la rupture du béton de structure}.
\newblock Ph.D. thesis. Université Pierre et Marie Curie - Paris 6.
\bibitem[{Mazars et~al.(1990)Mazars, Berthaud and
  Ramtani}]{mazars1990unilateral}
\bibinfo{author}{Mazars, J.}, \bibinfo{author}{Berthaud, Y.},
  \bibinfo{author}{Ramtani, S.}, \bibinfo{year}{1990}.
\newblock \bibinfo{title}{The unilateral behaviour of damaged concrete}.
\newblock \bibinfo{journal}{Engineering Fracture Mechanics}
  \bibinfo{volume}{35}, \bibinfo{pages}{629--635}.
\bibitem[{Meguro and Hakuno(1989)}]{meguro1989fracture}
\bibinfo{author}{Meguro, K.}, \bibinfo{author}{Hakuno, M.},
  \bibinfo{year}{1989}.
\newblock \bibinfo{title}{Fracture analyses of concrete structures by the
  modified distinct element method}.
\newblock \bibinfo{journal}{Doboku Gakkai Ronbunshu} \bibinfo{volume}{1989},
  \bibinfo{pages}{113--124}.
\bibitem[{Murakami(1988)}]{Mur1988}
\bibinfo{author}{Murakami, S.}, \bibinfo{year}{1988}.
\newblock \bibinfo{title}{Mechanical modeling of material damage}.
\newblock \bibinfo{journal}{ASME J. Appl. Mech.} \bibinfo{volume}{55},
  \bibinfo{pages}{280--286}.
\newblock \DOIprefix\doi{10.1115/1.3173673}.
\bibitem[{Olive et~al.(2018a)Olive, Kolev, Desmorat and
  Desmorat}]{olive2018harmonic}
\bibinfo{author}{Olive, M.}, \bibinfo{author}{Kolev, B.},
  \bibinfo{author}{Desmorat, B.}, \bibinfo{author}{Desmorat, R.},
  \bibinfo{year}{2018}a.
\newblock \bibinfo{title}{Harmonic factorization and reconstruction of the
  elasticity tensor}.
\newblock \bibinfo{journal}{Journal of Elasticity} \bibinfo{volume}{132},
  \bibinfo{pages}{67--101}.
\bibitem[{Olive et~al.(2018b)Olive, Kolev, Desmorat and
  Desmorat}]{kolev2018characterization}
\bibinfo{author}{Olive, M.}, \bibinfo{author}{Kolev, B.},
  \bibinfo{author}{Desmorat, R.}, \bibinfo{author}{Desmorat, B.},
  \bibinfo{year}{2018}b.
\newblock \bibinfo{title}{Characterization of the symmetry class of an
  elasticity tensor using polynomial covariants}.
\newblock \bibinfo{journal}{arXiv:1807.08996 [math.RT]} .
\bibitem[{Oliver-Leblond(2019)}]{oliver2019discontinuous}
\bibinfo{author}{Oliver-Leblond, C.}, \bibinfo{year}{2019}.
\newblock \bibinfo{title}{Discontinuous crack growth and toughening mechanisms
  in concrete: A numerical study based on the beam-particle approach}.
\newblock \bibinfo{journal}{Engineering Fracture Mechanics}
  \bibinfo{volume}{207}, \bibinfo{pages}{1--22}.
\bibitem[{Poisson(1828)}]{poisson1828memoire}
\bibinfo{author}{Poisson, S.D.}, \bibinfo{year}{1828}.
\newblock \bibinfo{title}{Mémoire sur l'équilibre et le mouvement des
  corps élastiques}.
\newblock \bibinfo{publisher}{F. Didot}.
\bibitem[{Ramtani et~al.(1992)Ramtani, Berthaud and
  Mazars}]{ramtani1992orthotropic}
\bibinfo{author}{Ramtani, S.}, \bibinfo{author}{Berthaud, Y.},
  \bibinfo{author}{Mazars, J.}, \bibinfo{year}{1992}.
\newblock \bibinfo{title}{Orthotropic behavior of concrete with directional
  aspects: modelling and experiments}.
\newblock \bibinfo{journal}{Nuclear Engineering and design}
  \bibinfo{volume}{133}, \bibinfo{pages}{97--111}.
\bibitem[{Rinaldi(2013)}]{rinaldi2013bottom}
\bibinfo{author}{Rinaldi, A.}, \bibinfo{year}{2013}.
\newblock \bibinfo{title}{Bottom-up modeling of damage in heterogeneous
  quasi-brittle solids}.
\newblock \bibinfo{journal}{Continuum Mechanics and Thermodynamics}
  \bibinfo{volume}{25}, \bibinfo{pages}{359--373}.
\bibitem[{Rinaldi and Lai(2007)}]{rinaldi2007statistical}
\bibinfo{author}{Rinaldi, A.}, \bibinfo{author}{Lai, Y.C.},
  \bibinfo{year}{2007}.
\newblock \bibinfo{title}{Statistical damage theory of 2d lattices: Energetics
  and physical foundations of damage parameter}.
\newblock \bibinfo{journal}{International Journal of Plasticity}
  \bibinfo{volume}{23}, \bibinfo{pages}{1796--1825}.
\bibitem[{Schlangen and Van~Mier(1992)}]{schlangen1992simple}
\bibinfo{author}{Schlangen, E.}, \bibinfo{author}{Van~Mier, J.G.M.},
  \bibinfo{year}{1992}.
\newblock \bibinfo{title}{Simple lattice model for numerical simulation of
  fracture of concrete materials and structures}.
\newblock \bibinfo{journal}{Materials and Structures} \bibinfo{volume}{25},
  \bibinfo{pages}{534--542}.
\bibitem[{Schouten(1989)}]{schouten1989tensor}
\bibinfo{author}{Schouten, J.A.}, \bibinfo{year}{1989}.
\newblock \bibinfo{title}{Tensor analysis for physicists}.
\newblock \bibinfo{publisher}{Courier Corporation}.
\bibitem[{Spencer(1970)}]{spencer1970note}
\bibinfo{author}{Spencer, A.}, \bibinfo{year}{1970}.
\newblock \bibinfo{title}{A note on the decomposition of tensors into traceless
  symmetric tensors}.
\newblock \bibinfo{journal}{International Journal of Engineering Science}
  \bibinfo{volume}{8}, \bibinfo{pages}{475--481}.
\bibitem[{Tillemans and Herrmann(1995)}]{tillemans1995simulating}
\bibinfo{author}{Tillemans, H.J.}, \bibinfo{author}{Herrmann, H.J.},
  \bibinfo{year}{1995}.
\newblock \bibinfo{title}{Simulating deformations of granular solids under
  shear}.
\newblock \bibinfo{journal}{Physica A: Statistical Mechanics and its
  Applications} \bibinfo{volume}{217}, \bibinfo{pages}{261--288}.
\bibitem[{Vannucci(2005)}]{vannucci2005plane}
\bibinfo{author}{Vannucci, P.}, \bibinfo{year}{2005}.
\newblock \bibinfo{title}{Plane anisotropy by the polar method}.
\newblock \bibinfo{journal}{Meccanica} \bibinfo{volume}{40},
  \bibinfo{pages}{437--454}.
\bibitem[{Vannucci and Verchery(2001)}]{VV2001}
\bibinfo{author}{Vannucci, P.}, \bibinfo{author}{Verchery, G.},
  \bibinfo{year}{2001}.
\newblock \bibinfo{title}{Stiffness design of laminates using the polar
  method}.
\newblock \bibinfo{journal}{International Journal of Solids and Structures}
  \bibinfo{volume}{38}, \bibinfo{pages}{9281--9894}.
\newblock \DOIprefix\doi{10.1016/S0020-7683(01)00177-9}.
\bibitem[{Vassaux et~al.(2016)Vassaux, Oliver-Leblond, Richard and
  Ragueneau}]{vassaux2016beam}
\bibinfo{author}{Vassaux, M.}, \bibinfo{author}{Oliver-Leblond, C.},
  \bibinfo{author}{Richard, B.}, \bibinfo{author}{Ragueneau, F.},
  \bibinfo{year}{2016}.
\newblock \bibinfo{title}{Beam-particle approach to model cracking and energy
  dissipation in concrete: Identification strategy and validation}.
\newblock \bibinfo{journal}{ICement and Concrete Composites}
  \bibinfo{volume}{70}, \bibinfo{pages}{1--14}.
\bibitem[{Vassaux et~al.(2015a)Vassaux, Richard, Ragueneau and
  Millard}]{vassaux2015regularised}
\bibinfo{author}{Vassaux, M.}, \bibinfo{author}{Richard, B.},
  \bibinfo{author}{Ragueneau, F.}, \bibinfo{author}{Millard, A.},
  \bibinfo{year}{2015}a.
\newblock \bibinfo{title}{Regularised crack behaviour effects on continuum
  modelling of quasi-brittle materials under cyclic loading}.
\newblock \bibinfo{journal}{Engineering Fracture Mechanics}
  \bibinfo{volume}{149}, \bibinfo{pages}{18--36}.
\bibitem[{Vassaux et~al.(2015b)Vassaux, Richard, Ragueneau, Millard and
  Delaplace}]{vassaux2015lattice}
\bibinfo{author}{Vassaux, M.}, \bibinfo{author}{Richard, B.},
  \bibinfo{author}{Ragueneau, F.}, \bibinfo{author}{Millard, A.},
  \bibinfo{author}{Delaplace, A.}, \bibinfo{year}{2015}b.
\newblock \bibinfo{title}{Lattice models applied to cyclic behavior description
  of quasi-brittle materials: advantages of implicit integration}.
\newblock \bibinfo{journal}{International Journal for Numerical and Analytical
  Methods in Geomechanics} \bibinfo{volume}{39}, \bibinfo{pages}{775--798}.
\bibitem[{Verchery(1982)}]{verchery1982invariants}
\bibinfo{author}{Verchery, G.}, \bibinfo{year}{1982}.
\newblock \bibinfo{title}{Les invariants des tenseurs d'ordre 4 du type de
  l'élasticité}, in: \bibinfo{booktitle}{Mechanical behavior of
  anisotropic solids/comportment Méchanique des Solides Anisotropes}.
  \bibinfo{publisher}{Springer}, pp. \bibinfo{pages}{93--104}.
\bibitem[{Vianello(1997)}]{vianello1997integrity}
\bibinfo{author}{Vianello, M.}, \bibinfo{year}{1997}.
\newblock \bibinfo{title}{An integrity basis for plane elasticity tensors}.
\newblock \bibinfo{journal}{Archives of Mechanics} \bibinfo{volume}{49},
  \bibinfo{pages}{197--208}.
\bibitem[{Willam et~al.(1989)Willam, Pramono and Sture}]{willam1989fundamental}
\bibinfo{author}{Willam, K.}, \bibinfo{author}{Pramono, E.},
  \bibinfo{author}{Sture, S.}, \bibinfo{year}{1989}.
\newblock \bibinfo{title}{Fundamental issues of smeared crack models}, in:
  \bibinfo{booktitle}{Fracture of concrete and rock}.
  \bibinfo{publisher}{Springer}, pp. \bibinfo{pages}{142--157}.

\end{thebibliography}
\end{document}